\documentclass[12pt]{article}
\usepackage{epsfig,amsmath,amssymb,graphics,color,calc} 
\usepackage{cite}

\usepackage{amsmath} 
\usepackage{amsthm}
\usepackage{amssymb}
\usepackage{amsopn}
\newcommand{\sezione}[2]{ 
\refstepcounter{section}\label{#2} 
\setcounter{equation}{0} 
\setcounter{subsection}{0} 
\addcontentsline{toc}{section} 
      {\normalsize\textbf{\thesection.\ #1}} 
\bigskip\bigskip\noindent 
\normalsize\textbf{\thesection.\ #1}\nopagebreak\smallskip\nopagebreak} 
\def\thesection{{\normalsize\arabic{section}}} 
\newcommand{\subsec}[2]{ 
\refstepcounter{subsection}\label{#2} 
\addcontentsline{toc}{subsection} 
      {\normalsize\normalfont\textit{\thesubsection.\ #1}} 
\medskip\medskip\noindent 
\normalsize\normalfont
\textit{\thesubsection. \ #1}\nopagebreak\smallskip\nopagebreak} 
\def\thesubsection{{\normalsize
{\arabic{section}.\arabic{subsection}}}} 
\newcounter{appendice}

\newtheorem{teo}{Theorem}[section]	\newtheorem{pro}[teo]{Proposition}
\newtheorem{defi}[teo]{Definition}	\newtheorem{lem}[teo]{Lemma}
\newtheorem{cor}[teo]{Corollary}	\newtheorem{rem}[teo]{Remark}
\newtheorem{con}[teo]{Condition}

\newcommand{\bteo}[1]{\begin{teo}\label{#1}}
\newcommand{\bpro}[1]{\begin{pro}\label{#1}}
\newcommand{\bdefi}[1]{\begin{defi}\label{#1}}
\newcommand{\blem}[1]{\begin{lem}\label{#1}}
\newcommand{\bcor}[1]{\begin{cor}\label{#1}}
\newcommand{\brem}[1]{\begin{rem}\label{#1}}
\newcommand{\bcon}[1]{\begin{con}\label{#1}}

\newcommand{\eteo}{\end{teo}}	\newcommand{\epro}{\end{pro}}
\newcommand{\edefi}{\end{defi}}	\newcommand{\elem}{\end{lem}}
\newcommand{\ecor}{\end{cor}}	\newcommand{\erem}{\end{rem}}
\newcommand{\econ}{\end{con}}

\newcommand{\Pro}{\noindent{\it Proof.\/}\ \ }

\renewcommand{\eqref}[1]{(\ref{#1})}

\newcommand{\be}[1]{\begin{equation}\label{#1}}
\newcommand{\bea}[1]{\begin{eqnarray}\label{#1}}
\newcommand{\besn}{\begin{equation*}}
\newcommand{\beasn}{\begin{eqnarray*}}
\newcommand{\nn}{\nonumber}
\renewcommand{\(}{\left(}		\renewcommand{\)}{\right)}
\renewcommand{\[}{\left[}		\renewcommand{\]}{\right]}
\renewcommand{\lg}{\left\{}		\newcommand{\rg}{\right\}}

\newcommand{\su}{\subset}	\newcommand{\ssu}{\subset\subset}
\newcommand{\sm}{\setminus}	\newcommand{\es}{\emptyset}

\newcommand{\diam}{\mathop{\rm diam}\nolimits}
\newcommand{\dis}{\mathop{\rm d}\nolimits}
\newcommand{\supp}{\mathop{\rm supp}\nolimits}

\newcommand{\ten}{\rightarrow}	
	
	\newcommand{\id}{{1 \mskip -5mu {\rm I}}}
 
\newcommand{\noi}{\noindent}

\newcommand{\ul}[1]{\underline{#1}}

\renewcommand{\a}{\alpha}	\renewcommand{\b}{\beta}	
			
	\newcommand{\f}{\varphi} 
\newcommand{\g}{\gamma}		\newcommand{\h}{\eta}
	
\newcommand{\m}{\mu}		
\renewcommand{\o}{\omega}	
	\renewcommand{\r}{\rho}
\newcommand{\s}{\sigma}		
\renewcommand{\t}{\tau}		
	
\newcommand{\z}{\zeta}

\newcommand{\G}{\Gamma}		\renewcommand{\L}{\Lambda}
	\renewcommand{\O}{\Omega} 
		\newcommand{\Ps}{\Psi} 
	\newcommand{\X}{\Xi}

\newcommand{\cA}{\mathcal A}	\newcommand{\cB}{\mathcal B} 
\newcommand{\cC}{\mathcal C}	\newcommand{\cD}{\mathcal D} 
\newcommand{\cE}{\mathcal E}	\newcommand{\cF}{\mathcal F} 
\newcommand{\cG}{\mathcal G}	\newcommand{\cH}{\mathcal H} 
	 
	\newcommand{\cL}{\mathcal L} 
\newcommand{\cM}{\mathcal M}	 
\newcommand{\cO}{\mathcal O}	 
\newcommand{\cQ}{\mathcal Q}	\newcommand{\cR}{\mathcal R} 
\newcommand{\cS}{\mathcal S}	 
	 
	\newcommand{\cX}{\mathcal X} 
\newcommand{\cY}{\mathcal Y}	 
	\newcommand{\bB}{\mathbb B}

\newcommand{\bG}{\mathbb G}	 
	 
	\newcommand{\bL}{\mathbb L} 
	\newcommand{\bN}{\mathbb N} 
	 
\newcommand{\bQ}{\mathbb Q}	\newcommand{\bR}{\mathbb R} 
	\newcommand{\bT}{\mathbb T} 
	\newcommand{\bV}{\mathbb V} 
	 
	\newcommand{\bZ}{\mathbb Z} 
	\newcommand{\Z}{\mathbb Z}
\newcommand{\newatop}[2]{\genfrac{}{}{0pt}{}{#1}{#2}}

\newcommand{\Es}{Y}
\newcommand{\dEs}{y}

\newcommand{\rest}{\!\restriction\!}
\newcommand{\env}{\cQ}

\newcommand{\sclos}[2]{\overline{#2}^{(#1)}}
\newcommand{\srclos}[3]{\overline{#3}^{(#1),#2}}
\newcommand{\proj}[1]{\widehat{#1}}
\newcommand{\tree}{\bT}

\renewcommand{\tilde}{\widetilde}
\renewcommand{\dis}{\mathrm d}
\newcommand{\disuno}{\mathrm D}
\newcommand{\diamuno}{\mathrm{Diam}}
\renewcommand{\complement}{\mathrm{c}}

\newcommand{\sopra}{\textrm{u}}
\newcommand{\sotto}{\textrm{d}}

\newcommand{\cionc}{}
\newcommand{\bonta}{\ell^{d(1/6-\delta)}\chi^{-1/2}}

\definecolor{light}{gray}{.9}

\begin{document}
\begin{titlepage}
%
%
%
\par\vskip 1cm\vskip 2em

\begin{center}
{\LARGE Renormalization Group in the uniqueness region: \\ 
$\phantom .$ \\ 
weak Gibbsianity and convergence}
\par
\vskip 2.5em \lineskip .5em
{\large
\begin{tabular}[t]{c}
$\mbox{Lorenzo Bertini}^{1} \phantom{m} \mbox{Emilio N.M.\ Cirillo}^{2}
\phantom{m} \mbox{Enzo Olivieri}^{3}$ 
\\
\end{tabular}
\par
}

\medskip
{\small
\begin{tabular}[t]{ll}
{\bf 1} & {\it 
Dipartimento di Matematica, Universit\`a di Roma La Sapienza}\\
&  Piazzale Aldo Moro 2, 00185 Roma, Italy\\
&  E--mail: {\tt bertini@mat.uniroma1.it}\\
\\
{\bf 2} & {\it
Dipartimento Me.\ Mo.\ Mat., Universit\`a di Roma La Sapienza}\\
&  Via A.\ Scarpa 16, 00161 Roma, Italy\\
&  E--mail: {\tt cirillo@dmmm.uniroma1.it}\\
\\
{\bf 3} & {\it
Dipartimento di Matematica, Universit\`a di Roma Tor Vergata}\\
& Via della Ricerca Scientifica, 00133 Roma, Italy\\ 
& E--mail: {\tt olivieri@mat.uniroma2.it}\\
\end{tabular}
}
\bigskip
\end{center}
\par\noindent
{\bf Communicating author:} Enzo Olivieri\\
{\bf E\_mail:} \texttt{olivieri@mat.uniroma2.it}\\
{\bf Telephone number:} +39--06--72594686\\
{\bf Fax number:} +39--06--72594699
\vskip 1 em

\centerline{\bf Abstract} 
\smallskip
We analyze the block averaging transformation applied to lattice gas 
models with short range interaction in the uniqueness region below 
the critical temperature. 
We prove weak Gibbsianity of the 
renormalized measure and convergence of the renormalized
potential in a weak sense.
Since we are arbitrarily close to the coexistence region we have a
diverging characteristic length of the system: the correlation length or the
critical length for metastability, or both. Thus, to perturbatively treat
the problem we have to use a scale--adapted expansion. Moreover, such a model 
below the critical temperature resembles a disordered system in presence of 
Griffiths' singularity. Then the 
cluster expansion that we use must be graded with its minimal scale length 
diverging when the coexistence line is approached. 

\vskip 0.8 em

\vfill


\bigskip\bigskip
\footnoterule
\vskip 1.0em
{\small 
\noindent
The authors acknowledge the support of Cofinanziamento MIUR.
\vskip 1.0em
\noindent
}
\end{titlepage}
\vfill\eject
 
\sezione{Introduction}{s.int} 
\par\noindent
In this paper we  analyze, from a rigorous point of view, the well known
Renormalization Group (RG) map called Block Averaging Transformation (BAT).
Following \cite{[DS5]} we say that a
stochastic field is {\it strongly} resp.\ {\it weakly }  
Gibbsian if its family of conditional
probabilities has the Gibbsian form with respect to a potential absolutely 
{\it uniformly} resp.\ {\it pointwise almost surely}  converging. 
Thus in both cases the DLR equations are satisfied but with different 
notions on the summability properties of the potential. We refer to 
\cite{[EFS]} for a general description of Gibbs formalism especially in
connection with renormalization--group maps and to 
\cite{[DS5],[MRSvM],[BKL]} for a discussion of the weak Gibbs property.

Under suitable strong mixing conditions i.e., exponential decay of 
truncated expectations, for the original ({\it object})
system we establish the weak Gibbs property of the renormalized 
({\it image}) measure and the convergence, in a suitable sense, of the 
renormalized potential under iteration of BAT.
A relevant application will be the standard two--dimensional Ising model 
in the uniqueness region. For this case, when  the temperature is higher 
than the critical value $T_c$, actually we have strong Gibbsianity  
of the renormalized measure for all large enough scales of the transformation 
as shown in \cite{[BCO]}. 
On the other hand for $T<T_c$ violation of strong Gibbsianity 
is expected and actually proven for $T\ll T_c$, see \cite{[EFS]}. 
In the present paper we prove the weak 
Gibbsianity of the renormalized measure. 

Let us focus, for the moment, on the two--dimensional 
Ising model. A more general setup will be 
introduced in the following section. We give here some specific definitions.
The state space of  the object  system  is 
$\cS=\otimes_{x \in \cL}\cS_x,$ with $\cS_x= \{-1,+1\},$ $\cL=\bZ^2$;
for $\Lambda\subset\cL$ we set 
$\cS_\Lambda=\otimes_{x\in\Lambda}\cS_x$.
The (negative) Hamiltonian in a finite volume $\Lambda$ with 
boundary condition 
$\tau\in\cS_{\cL\setminus\Lambda}$ is
$$
H_{\Lambda}(\sigma\tau)=
\beta\sum_{\newatop{\{x,y\}\subset\Lambda:}
                   {|x-y|=1}
    }\sigma_x\sigma_y+
\beta\sum_{\newatop{x\in \Lambda, y\not \in \Lambda:}
                      {|x-y| =1}
             }
 \sigma_x\tau_y +\beta h\sum_{x\in \Lambda}
\sigma_x
$$
with $\beta ={1/T}>0 $ the inverse temperature, $h\in\bR$   
the magnetic field and $\sigma\in\cS_\Lambda$. 
Notice that in this section we use the magnetic language whereas in the 
following we will use the equivalent lattice gas formulation.
The corresponding finite volume Gibbs measure is
$$
\mu^\t_{\beta,h,\Lambda}(\sigma)= 
\frac{\exp\{H_{\Lambda}(\sigma\tau)\}}
     {\displaystyle\sum_{\sigma'\in\cS_\Lambda}
       \exp\{H_{\Lambda}(\sigma'\tau)\}}
$$
We denote by $\m =\m_{\beta,h}$ the unique infinite volume Gibbs measure
in the uniqueness region deprived of the critical point given by
$\{\b<\b_c\}\cup \{\b>\b_c, \,h \neq 0\}$, where
$\beta_c =1/T_c=\log(1+\sqrt{2})/2$ is the inverse critical temperature,
see for instance \cite{[Ga]}.

Let $\cL^{(\ell)}=(\ell \bZ)^2, \ell\in\bN$ and partition
$\cL$ as the disjoint union of $\ell$--block $Q_\ell(i)  = Q_\ell(0)+i$,
where $i\in\cL^{(\ell)}$, and $Q_\ell(0)$ is the square of side
$\ell$ with the origin the site with smallest coordinates.
We associate with each $i\in \cL^{(\ell)}$ a {\it renormalized spin} 
$m_i$ taking values in
$$
\cS^{(\ell)}_i=\left\{\frac{-\ell^d-\ell^d\bar m}{\sqrt{\ell^d \chi}}, 
                      \frac{-\ell^d+2-\ell^d\bar m}{\sqrt{\ell^d \chi}}, 
                      \dots, 
                      \frac{\ell^d-\ell^d\bar m}{\sqrt{\ell^d \chi}}
               \right\}
$$
where $\bar m=\bar m_{\beta, h}=\m_{\b,h}(\sigma_0)$ is the equilibrium
magnetization and
$\chi=\chi(\beta,h)=\sum_{x\in\cL}
 [\mu_{\b,h}(\sigma_0\sigma_x)-\mu_{\beta,h}(\sigma_0)\mu_{\beta,h}(\sigma_x)]$
is the susceptibility. For $I\subset\cL^{(\ell)}$ we write 
$\cS^{(\ell)}_I=\otimes_{i\in I}\cS^{(\ell)}_i$; we also set 
$\cS^{(\ell)}=\otimes_{i\in\cL^{(\ell)}}\cS^{(\ell)}_i$. 

The renormalized measure $\nu^{(\ell)}=\nu^{(\ell)}_{\beta,h}$ on the
renormalized space $\cS^{(\ell)}$ 
is defined via its finite dimensional distributions; 
let $I\subset\subset\cL^{(\ell)}$, where $\subset\subset$ means finite 
subset, and pick $\tilde m\in\cS^{(\ell)}_I$, then set 
\begin{equation} 
\label{rinmes}
\nu^{(\ell)}_{\b,h}\big(\{m\in\cS^{(\ell)}:\,m_I=\tilde m\}\big)=  
 \int_\cS 
  {\rm d}\mu_{\beta,h}(\sigma) 
   \prod_{i\in I} \delta(M_i(\sigma_{Q_\ell(i)})-\tilde m_i)
\end{equation}
where for all $i\in\cL^{(\ell)}$ and $\eta\in\cS_{Q_\ell(i)}$ we have
introduced the empirical magnetization, centered and normalized,
\begin{equation}
\label{rinspi}
M_i(\eta)=\frac{1}{\sqrt{\ell^d\chi}}
           \sum_{x\in Q_\ell(i)}[\eta_x-\bar m]
\end{equation}
We write $\nu_{\beta,h}^{(\ell)}=T^{(\ell)}\mu_{\beta,h}$ and note that 
the semigroup property holds namely, $T^{(\ell)}\,T^{(\ell')}=T^{(\ell\ell')}$. 
The image measure $\nu^{(\ell)}_{\b,h}$ represents the distribution of 
the empirical block magnetization 
under the object measure $\m_{\b,h}$. 

We would like to analyze the map on the
potentials induced by the map $T^{(\ell)}$ that was defined on  
(infinite volume) measures.
A preliminary condition for this program is that the 
renormalized measure is strongly or weakly Gibbsian  with respect to the
renormalized potential, see \cite{[EFS]}.

We introduce, now, the finite volume setup.
Let $I\subset\subset\cL^{(\ell)}$ be a finite box in $\cL^{(\ell)}$ and 
consider the corresponding box  $\Lambda=Q_\ell(I)\subset\cL$. 
We introduce the {\it renormalized Hamiltonian} 
$\cH^{(\ell),\tau}_I$ with boundary condition 
$\tau\in\cS_{\cL\setminus\Lambda}$ by setting 
\begin{equation}
\label{rinham}
e^{\cH^{(\ell),\tau}_I(m)}=
  \sum_{\sigma\in\cS_\Lambda} 
    e^{H^\t_\L(\sigma)}
    \prod_{i\in I}\delta(M_i(\s_{Q_\ell(i)})-m_i)
\end{equation}
for each $m\in\cS^{(\ell)}_I$.
In the computation of the {\it renormalized potential} associated with the 
renormalized Hamiltonian $\cH^{(\ell),\tau}_I$,
a crucial role is played by the {\it constrained systems} obtained 
by conditioning the object system to a fixed renormalized spin configuration,
see the pioneering paper \cite{[CasG]}.
More precisely, the equilibrium probability measure of the constrained 
model associated with the renormalized configuration $m\in\cS^{(\ell)}_I$
on the finite volume $\Lambda=Q_\ell(I)\subset\subset\cL$ is given by
\begin{equation}
\label{conmis}
\m^{(\ell),\tau}_{m,\Lambda}(\sigma)=
  \frac{e^{H^\tau_\Lambda(\sigma)}
        \prod_{i\in I}\delta(M_i(\s_{Q_\ell(i)})-m_i)}
       {\sum_{\eta\in\cS_\Lambda}e^{H^\t_\Lambda(\eta)} 
          \prod_{i\in I}\delta(M_i(\eta_{Q_\ell(i)})-m_i)}
\end{equation}
for all $\sigma\in\cS_\Lambda$. 
Notice that from (\ref{rinham}) it follows that the renormalized Hamiltonian 
$\cH^{(\ell),\tau}_I(m)$ is equal to the logarithm of the partition 
function of the corresponding constrained system which is defined as 
\begin{equation}
\label{ham}
Z^{(\ell),\t}_{m,\Lambda}=
 \sum_{\sigma\in\cS_\Lambda} 
  e^{H^\tau_\Lambda(\sigma)} 
    \prod_{i\in I}\delta(M_i(\s_{Q_\ell(i)})-m_i)
\end{equation}
The measure $\m^{(\ell),\t}_{m,I}$ can be called 
{\it multi--canonical}, because it is nothing but the original measure 
{\it constrained} to the assigned magnetizations in the
$\ell$--blocks contained in $\Lambda$. 
Of course $\m^{(\ell),\t}_{m,I}$ does not depend at all on the magnetic
field $h$.

It has been shown in \cite{[EFS]} that for any $\ell\in\bN$ 
even there exists $\beta_0=\beta_0(\ell)$ such that the renormalized 
measure $\nu^{(\ell)}_{\beta,h}$, arising from the application of BAT map 
to the Ising measure
$\m_{\beta,h}$, is non--Gibbsian at any $h$ and $\beta>\beta_0$. This
pathological behavior is a consequence of violation of {\it quasi--locality}, 
a continuity property of its conditional
probabilities  which constitutes a necessary condition for Gibbsianity,
see \cite{[EFS],[E1],[E2]}.
This, in turn, is  a consequence of a first order
phase transition with long range order of a 
particular constrained model: the one
corresponding to  $m_i =0$ for all $i\in\cL^{(\ell)}$.
It is clear that this pathology is completely 
independent of the value of the magnetic field
$h$ acting on the object system. 
On the other hand it is also clear that this ``bad"
configuration, inducing non--Gibbsianity, is very atypical with respect to
$\nu^{(\ell)}_{\b,h}$ for $h\neq0$.
Thus it is reasonable to expect the validity of a weaker 
property of Gibbsianity.

Before discussing this point let us recall the main result of 
\cite{[BCO]} on strong Gibbsianity
above $T_c$ in two dimensions which will be 
useful for a comparison with the results
obtained in the present paper for the case with $T<T_c$. 
The case $d>2$ will be discussed later on.

\bteo{hight}
Consider the two--dimensional Ising system with 
$\b <\b_c$ and $h \in \bR$ given.
Then there exists $\ell_0\in\bN$ such that for any
$\ell$ large enough multiple of $\ell_0$ the measure 
$\nu^{(\ell)}=\nu^{(\ell)}_{\b,h}$ is Gibbsian
in the sense that for each $Y\subset\subset\cL^{(\ell)}$
and for each local function $f:\cS_Y^{(\ell)}\to\bR$ we have
\begin{equation}
\label{i:dlr}
\begin{array}{l}
{\displaystyle
\nu^{(\ell)}(f)=
 \int_{\cS^{(\ell)}}
  \nu^{(\ell)}(dm')\, \frac{1}{Z_Y(m')}\,
  \sum_{m\in\cS_Y^{(\ell)}}f(m)}\\
{\displaystyle
\;\;\;\;\;\;\;\;\;\;\;\;\;\;\;\;\;\;\;\;\;\;\;\;\;\;
\times\exp\Big\{\sum_{X\cap Y\neq\emptyset}
               [\psi_X^{(\ell)}(m_{Y \cap X }m'_{Y^\complement\cap X})+
               \phi_X^{(\ell)}(m_{Y \cap X }m'_{Y^\complement \cap X})]\Big\}}
\end{array}
\end{equation}
where
\begin{equation}
\label{i:fpart}
 Z_Y(m') =\sum_{m\in\cS_Y^{(\ell)}}
       \exp\Big\{\sum_{X\cap Y\neq\emptyset}
       [\psi_X(m_{Y \cap X }m'_{Y^\complement\cap X})+\phi_X(m_{Y \cap X
}m'_{Y^\complement\cap X})]\Big\}
\end{equation}
The family
$\{\phi^{(\ell)}_X+\psi^{(\ell)}_X,\, X\su\su\cL^{(\ell)}\}$,
with $\phi^{(\ell)}_X,\psi^{(\ell)}_X:\cS^{(\ell)}_X\to\bR$,
is translationally invariant and satisfies the uniform bound
\begin{equation}
\label{i:finnorm}
\sum_{X\ni 0} e^{\alpha|X|} 
\sup_{m\in\cS^{(\ell)}_X}  
  \Big(
    \big|\psi^{(\ell)}_X(m)\big|+\big|\phi^{(\ell)}_X(m)\big|\Big)
 <\infty
\end{equation}
for a suitable $\a>0$.
Moreover, there exists
$\kappa\in\bN$ such that $\Psi_{X}^{(\ell)}= 0$ if $\diam(X)\ge\kappa$.
Finally we have that for the same $\alpha$ as in \eqref{i:finnorm}
\begin{equation}
\label{i:conver}
\lim_{\ell\ten\infty}
\sum_{X\ni 0} e^{\alpha|X|}
\sup_{m\in\cS^{(\ell)}_X}
  \big|\phi^{(\ell)}_X(m)\big|=0
\end{equation}
$\psi^{(\ell)}_{\{i\}}(m_i)=-m_i^2/2$, for $i\in\cL^{(\ell)}$,
and there exists $a>0$ such that
$$
\lim_{\ell\ten\infty} 
\sup_{\newatop{m\in \cS^{(\ell)}_X}
              {|m_i|\le\ell^a,i\in X}}
    \big|\psi^{(\ell)}_X(m)\big|=0
\quad \quad {\rm for }~~ |X|\ge 2
$$
\eteo
The crucial point to obtain the above result is the validity of a strong mixing condition
for the object system {\it uniformly} in the magnetic field $h$. 
This fails below $T_c$, because of the phase transition at $h=0$.  
By only assuming strong mixing of the object
system, without uniformity in $h$, we can expect only weak Gibbsianity since, 
as we said before, for $T<T_c$ it has been proven violation 
of strong Gibbsianity in \cite{[EFS]}.
Let us now state our main results on weak Gibbsianity and convergence of 
the renormalized potential as $\ell\to\infty$; this theorem is an immediate
consequence of the more general results that will be stated in 
Theorems~\ref{t:wgib} and \ref{t:conv}.

\bteo{t:inwgib}
Consider the two--dimensional Ising model. Given 
$(\b,h)\in\{\b<\b_c\}\cup \{\b>\b_c, \,h \neq 0\}$,
there exists $\ell_0$ such that for any large enough
$\ell$ multiple of $\ell_0$, the measure 
$\nu^{(\ell)}$ is weakly Gibbsian in the sense 
that it satisfies the DLR equations \eqref{i:dlr} with respect to a potential
$\{\psi_X^{(\ell)}+\phi_X^{(\ell)},\,X\subset\subset\cL^{(\ell)}\}$,
  $\psi_X^{(\ell)},\phi_X^{(\ell)}:\, \cS^{(\ell)}_X\mapsto\bR$,
satisfying the following.

There exists a measurable set $\bar\cS^{(\ell)}\subset\cS^{(\ell)}$, such that
$\nu^{(\ell)}(\bar\cS^{(\ell)})=1$, and functions
$r^{(\ell)}_i:\bar\cS^{(\ell)}\mapsto\bN\setminus\{0\}$,
for all $i\in\cL^{(\ell)}$, such that
for each $m\in\bar\cS^{(\ell)}$, if
$X\ni i$ and $\diam(X)>r^{(\ell)}_i(m)$ then $\psi_X^{(\ell)}(m)=0$.
Furthermore, for each $i\in\cL^{(\ell)}$ and $m\in\bar\cS^{(\ell)}$
there exists a real $c_i^{(\ell)}(m)\in[0,\infty)$ such that
\begin{equation}
\label{in:mtb5}
\sum_{X\ni i}
  |\psi_X^{(\ell)}(m_X)|\le c_i^{(\ell)}(m)
\end{equation}
There exists $C$ independent of $\ell$ such that
\begin{equation}
\label{i:eee}
\sup_{m\in\cS^{(\ell)}}
\sup_{i\in\cL^{(\ell)}}\,
\sum_{X\ni i}\,
 |\phi_X^{(\ell)}(m)|<C
\end{equation}
For each $i\in\cL^{(\ell)}$ we have $\psi_{\{i\}}^{(\ell)}(m)=-m_i^2/2$ and
for each $q\in[1,+\infty)$
\begin{equation}
\label{i:ee2}
\lim_{\ell\to\infty}\,
\sup_{i\in\cL^{(\ell)}}
  \nu^{(\ell)}\Big(\Big|
              \sum_{X\ni i:\,|X|\ge2}\psi_X^{(\ell)}
              \Big|^q\Big)
=0
\end{equation}
and
\begin{equation}
\label{i:ee1}
\lim_{\ell\to\infty}
\sup_{m\in\cS^{(\ell)}}
\sup_{i\in\cL^{(\ell)}}
\sum_{X\ni i} \,
 |\phi_X^{(\ell)}(m)|
  =0
\end{equation}
\end{teo}
\par\noindent
{\it Remark}. From the more general result stated in Theorems~\ref{t:wgib}
and \ref{t:conv} below, we get that Theorem~\ref{t:inwgib} extends to
the case $d>2, h\neq 0, \b >\b_0(d,|h|)$ for a suitable function 
$\b_0:\bN\times \bR^+\to\bR^+$. 
Indeed in this case the required strong mixing condition is satisfied.

In this low--temperature case we have a diverging scale even when we are far 
away from the
critical point. It is not the correlation length but, rather, the 
``critical length for
metastability", diverging when $h \to 0$ as $1/h$, representing the scale 
for which the
magnetic field decides the phase; this is given as the scale for which the 
boundary conditions
are ``screened" by the magnetic field $h$.
Notice that in the region $\{\b<\b_c\}\cup \{\b>\b_c, \,h
\neq 0\}$, where for $d=2$ the strong mixing is satisfied
\cite{[MOS],[ScS]}, both the 
critical length and
the correlation length can diverge even simultaneously.

\medskip
Let us discuss, now, the result of Theorem~\ref{t:inwgib}. 
As we said above it is sufficient that
there exists one ``bad" renormalized configuration giving rise to 
long range order for the
corresponding constrained system, to induce violation of Gibbsianity. 
For BAT it has been
shown in \cite{[EFS]} that for any magnetic field $h$, 
a bad configuration is $m_i=0$ for all $i\in\cL^{(\ell)}$.
For $h\neq0$ this is a very atypical configuration; however with small but 
positive probability we have arbitrarily large bad regions with $m_i$
close to zero.
To be more precise we shall call ``good" a block magnetization $m_i$ 
belonging to a suitable interval such that: inside such interval 
the system has a good behavior in the strong mixing sense and 
the probability to be bad (not good) is sufficiently small, 
see Subsection~\ref{s:fs}.
In order to prove weak Gibbsianity the key property is that bad regions are 
far apart: larger
and larger bad regions are sparser and sparser. 

As we discussed in 
\cite{[BCOabs]} this situation is
similar to that of disordered systems in presence of the Griffiths' 
singularity. A multi--scale analysis is needed. 
The natural approach, quite complicated from the technical
point of view, is to use a graded cluster expansion.
For disordered systems there are clever methods,
see \cite{[Berretti],[vDKP]},
avoiding cluster expansion, that enable to prove results 
like exponential decay of correlations for almost all realizations
of the disorder.
In the case of BAT, in order to compute renormalized potentials in 
the weakly Gibbsian case,
the use of the full theory of graded cluster expansion 
(like the one in \cite{[FI]})
appears to be unavoidable.
Since we want to study a region of parameters arbitrarily close 
to the critical point ($h\ne0$, $0<T\le T_c$) the distinctive character of our 
graded cluster expansion is that the
minimal scale may be chosen arbitrarily large and diverging as 
$T\to T_c$ and/or $h\to 0$.
The minimal scale involved in our discussion being divergent, we need to 
use a scale--adapted cluster expansion,
see \cite{[O],[OP],[BCO]}, based on a finite size mixing condition. 

In this case, contrary to low and high temperature expansion or high magnetic 
field expansion, the small parameter is the ratio between the diverging 
length and the suitably large finite size where the mixing condition holds.
We want to stress again that in our approach, according to the general 
renormalization group ideology, we first fix the values of the 
thermodynamic parameters of the object system and, subsequently, the value of 
the scale of BAT. In other words we {\it take advantage} from choosing the 
scale $\ell$ of the transformation large enough. On the other side we cannot 
exclude that, for given values of $\b$ and $h$, if
$\ell$ is not sufficiently large, weak Gibbsianity ceases to be valid.
In \cite{[MRSvM],[BKL]} the authors study decimation transformation,
see \cite{[EFS]}, at large $\b$ and {\it arbitrary h}. 
They first fix the scale of the transformation and, subsequently, 
choose the temperature below which they get weak Gibbsianity.

In the present paper, in the context of weak Gibbsianity, we give also 
results of convergence
of renormalized potentials when iterating BAT, which, 
by the semigroup property, corresponds
to taking the limit as $\ell \to \infty$.
It appears clear that for that purpose one has to use a 
perturbative theory based 
on scale--adapted cluster expansion. Even far away from the critical point, 
in order to prove directly convergence, one needs to take advantage from 
choosing large $\ell$. 
In \cite{[Cam]} the author uses a high temperature expansion giving rise to 
a polymer system whose activity is small uniformly in $\ell$; he then
proves convergence by making use of the
general result \cite{[Hu]} according to which, to get
convergence in a suitable sense, one needs only to prove uniform boundedness 
in a suitable norm.
This situation is similar to the 
one of \cite{[S1]} where the author uses a low
temperature expansion that converges uniformly in $\ell$.

The paper is organized as follows:
in Section~\ref{s:not} we introduce the basic notation and state our main 
results on the weak Gibbsianity and convergence of the renormalized 
potentials as $\ell\to\infty$. 
In Section~\ref{s:cattivi} we prove the required probability estimates on 
the configuration of ``bad" magnetizations. Then, in Section~\ref{s:con} 
we construct the full measure set where the conditional probabilities have the 
Gibbs form. In Section~\ref{s:cegp}, following \cite{[O],[OP],[BCO]},
we perform the scale--adapted cluster expansion on the ``good" part of the 
lattice. In Section~\ref{s:diso} we apply the theory of the 
graded expansion developed in \cite{[BCOabs]} to prove the main results. 

\sezione{Notation and results}{s:not} 
\par\noindent
In this section we give the basic definitions, 
introduce the general setup, and state 
our main results. 

\subsec{The lattice}{s:lat} 
\par\noindent
For $x=(x_1,\cdots,x_d)\in\bR^d$ we let $|x|:=\sup_{k=1,\cdots,d} |x_k|$.
The spatial structure is modeled by the $d$--dimensional lattice $\cL:=\Z^d$, 
in which we let $e_i$, $i=1,\dots,d$, be the coordinate unit vectors. 
For each strictly positive integer $s$, we introduce the $s$--rescaled lattice
$\cL^{(s)}:=(s\bZ)^d$ which is embedded in $\cL$ namely, 
points in $\cL^{(s)}$ are also points in $\cL$, see Fig.~\ref{f:lat0}.
Given an integer $s\ge 1$ we next define some geometrical notions on the 
$s$--rescaled lattice $\cL^{(s)}$. If $s=1$ they refer to the original lattice
$\cL$ and in such a case we drop $s$ from the notation.

We set $e_i^{(s)}:=s\,e_i$, $i=1,\dots,d$ and
use $\Lambda^\complement:=\cL^{(s)}\setminus \Lambda$ 
to denote the complement of $\Lambda\subset\cL^{(s)}$. 
For $\Lambda$ a finite subset of $\cL^{(s)}$ 
(we use $\Lambda\subset\subset\cL^{(s)}$ to 
indicate that $\Lambda$ is finite), 
$|\Lambda|$ denotes the cardinality of $\Lambda$. 
We consider $\cL^{(s)}$ endowed with the distance 
$\dis_{s}(x,y):=|x-y|/s$. 
As usual for $\Lambda,\Delta\subset\cL^{(s)}$ we set 
$\dis_{s}(\Lambda,\Delta):=\inf\{\dis_{s}(x,y),\, x\in\Lambda,\, 
                                                      y\in\Delta\}$
and
$\diam_{s}(\Lambda):=\sup\{\dis_{s}(x,y),\; x,y\in\Lambda\}$.
Moreover, 
for each $\Lambda\subset\subset\cL^{(s)}$ we denote by 
$\env^{(s)}(\Lambda)\subset\subset\cL^{(s)}$ the smallest parallelepiped,
with axes parallel to the coordinate directions, containing $\Lambda$.
We say that $x,y\in\cL^{(s)}$ are {\it nearest neighbors} iff 
$\dis_s(x,y)=1$; we say that $\Lambda\subset\cL^{(s)}$ is $s$--{\it connected}
iff for each $x,y\in\Lambda$ there exists a path of pairwise nearest neighbor
sites of $\Lambda$ joining $x$ and $y$. 

For $x\in\cL^{(s)}$ and $m$ a strictly positive real we set 
$Q_m^{(s)}(x):=\{y\in\cL^{(s)}:\,x_k\le y_k\le x_k+s(m-1),\,
                      \forall k=1,\dots,d\}$.
For $X\subset\subset\cL^{(s)}$ and $m>0$ we set 
$B_m^{(s)}(X):=\{y\in\cL^{(s)}:\,\dis_{s}(X,y)\le m\}$;
if $x\in\cL^{(s)}$ we write 
$B_m^{(s)}(x)$ for $B_m^{(s)}(\{x\})$.
Note that $Q_m^{(s)}(x)$ is the cube of $s$--side length 
$[m]$ with $x$ the site with smallest coordinates, 
while $B_m^{(s)}(x)$ is the ball of $s$--radius $[m]$ centered at $x$, 
hence it is a cube of $s$--side length $2[m]+1$.  
We remark, also, that $|Q_m^{(s)}(x)|=[m]^d$ and $|B_m^{(s)}(x)|=(2[m]+1)^d$.
We shall denote $Q_m^{(s)}(0)$, resp.\ $B_m^{(s)}(0)$, simply by 
$Q_m^{(s)}$, resp.\ $B_m^{(s)}$. 

For $r>0$ and $\Lambda\subset\cL^{(s)}$ we set
$\partial^{(s),r}\Lambda:=\{x\in\Lambda^\complement:\,
                          \dis_{s}(x,\Lambda)\le r\}$; 
finally we set 
$\srclos{s}{r}{\Lambda}:=\Lambda\cup\partial^{(s),r}\Lambda$. 
If $r=1$ we drop it from the notation, i.e.\ 
$\partial^{(s)}\Lambda:=\partial^{(s),1}\Lambda$ and
$\srclos{s}{1}{\Lambda}=:\sclos{s}{\Lambda}$.

Let 
$\cE^{(s)}:=\lg\{x,y\},\, x,y\in\cL^{(s)}:\, \dis_{s}(x,y)=1 \rg$
be the collection of {\em edges} in $\cL^{(s)}$.
Note that, according to our
definitions, the edges can be also diagonal.
We say that two edges $e,e'\in\cE^{(s)}$
are connected if and only if $e\cap e'\neq\emptyset$.
A subset $(V,E)\su(\cL^{(s)},\cE^{(s)})$ is said to be connected iff
for each pair $x,y\in V$, with $x\neq y$,
there exists in $E$ a path of connected
edges joining them. For $\Lambda\subset\subset\cL^{(s)}$ we then set
\be{treedec0}
\tree_s(\Lambda):=\inf\lg |E| \,, \  (V,E) \su (\cL^{(s)},\cE^{(s)})
                \textrm{ is connected and } V \supset\Lambda 
\rg
\end{equation}
We agree that $\tree_s(\Lambda)=0$ if $|\Lambda|=1$ and 
remark that for each $x,y\in\cL^{(s)}$ we have
$\tree_s(\{x,y\})=\dis_{s}(x,y)$.

Let $u$ be a multiple of $s$, 
we define the {\it unpacking} and the {\it packing} 
operators which associate subsets of the $u$--rescaled lattice to subsets
of the $s$--rescaled lattice and vice versa. More precisely, 
the {\it unpacking} operator $\cO_u^s$ maps a set $\Lambda\subset\cL^{(u)}$ 
to 
$$ 
\cO_u^{s}\Lambda:=\bigcup_{x\in\Lambda}Q_{u/s}^{(s)}(x)
$$
Note that 
the cubes $Q_{u/s}^{(s)}(x)$ appearing above are disjoint 
namely, $Q_{u/s}^{(s)}(x)\cap Q_{u/s}^{(s)}(y)=\emptyset$ for any 
$x,y\in\Lambda$ such that $x\neq y$. 
The {\it packing} operator $\cO_s^u$ maps a set $\Lambda\subset\cL^{(s)}$ 
to $\cO_s^u\Lambda:=\{x\in\cL^{(u)}:\,
           \Lambda\cap Q^{(s)}_{u/s}(x)\neq\emptyset\}$.
We note that the restriction of 
$\cO_s^u$ to the range of $\cO_u^s$ is the inverse operator of $\cO_u^s$ 
namely, $\cO_s^u\cO_u^s\Lambda=\Lambda$ for all $\Lambda\subset\cL^{(u)}$. 
Note that, as mentioned before, we let $\cO_u:=\cO_u^1$ and $\cO^u:=\cO_1^u$. 

\setlength{\unitlength}{1.pt}
\begin{figure}
\begin{picture}(200,150)(-125,50)
\thinlines
\multiput(-5,0)(0,12){19}{\line(1,0){226}}
\multiput(0,-5)(12,0){19}{\line(0,1){226}}
\multiput(0,0)(0,36){7}{
 \multiput(0,0)(36,0){7}{\circle{8}}}
\multiput(0,0)(0,24){10}{
 \multiput(0,0)(24,0){10}{\circle*{4}}}
\multiput(-6,0)(0,72){4}{
 \multiput(0,-6)(72,0){4}{\line(0,1){12}}}
\multiput(-6,0)(0,72){4}{
 \multiput(0,6)(72,0){4}{\line(1,0){12}}}
\multiput(6,0)(0,72){4}{
 \multiput(0,6)(72,0){4}{\line(0,-1){12}}}
\multiput(6,0)(0,72){4}{
 \multiput(0,-6)(72,0){4}{\line(-1,0){12}}}
\end{picture}
\vskip 2. cm
\caption{The lattices $\cL$, $\cL^{(2)}$, $\cL^{3)}$ and $\cL^{(6)}$ are 
depicted in the two--dimensional case.
Sites in $\cL$ are represented by the intersections of the lines,
solid circles represent sites belonging to $\cL^{(2)}$,
open circles represent sites belonging to $\cL^{(3)}$, 
open squares represent sites of $\cL^{(6)}$.
}
\label{f:lat0}
\end{figure}
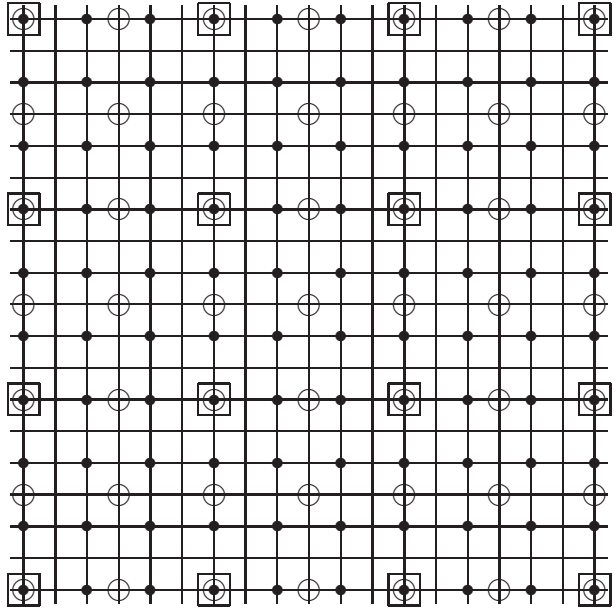

\subsec{The configuration space}{s:conf} 
\par\noindent
We deal with lattice systems whose single spin space is not
translationally invariant and labelled by points in the 
$s$--rescaled lattice $\cL^{(s)}$.
As usual for $s=1$ we recover the notation for 
the original lattice and drop $s$ from the notation.  
Given a collection of strictly 
positive integers $S_x^{(s)}$, $x\in\cL^{(s)}$, such that 
$S^{(s)}:=\sup_{x\in\cL^{(s)}}S_x^{(s)}<+\infty$,
the configuration space associated to $x\in\cL^{(s)}$ 
is a finite set $\cS_x^{(s)}$, with $|\cS_x^{(s)}|=S_x^{(s)}+1$
which we consider endowed with the discrete topology, the 
associated Borel $\sigma$--algebra is denoted by 
$\cF^{(s)}_{\{x\}}$. 

The configuration space in $\Lambda\subset\cL^{(s)}$ is defined as 
$\cS^{(s)}_\Lambda:=\otimes_{x\in\Lambda}\cS^{(s)}_x$ and
equipped with the product topology; we will 
let $\cS^{(s)}_{\cL^{(s)}}=:\cS^{(s)}$. 
We denote by $\cF^{(s)}$ the Borel $\sigma$--algebra on $\cS^{(s)}$ and
for each $\Lambda\subset\cL^{(s)}$ we set
$\cF^{(s)}_{\Lambda}:=\otimes_{x\in\Lambda}\cF^{(s)}_{\{x\}}\subset\cF^{(s)}$.

Given $\Delta\subset\Lambda\subset\cL^{(s)}$ and 
$\sigma:=\{\sigma_x\in\cS^{(s)}_x,\,x\in\Lambda\}\in\cS^{(s)}_\Lambda$, 
we denote by $\sigma_\Delta$ the {\it restriction} of $\sigma$ 
to $\Delta$ namely, $\sigma_\Delta:=\{\sigma_x,\,x\in\Delta\}$.
Let $m$ be a positive integer and 
let $\L_1,\dots,\L_m\subset\cL^{(s)}$ be pairwise disjoint subsets of 
$\cL^{(s)}$; for each  
$\sigma_k\in\cS^{(s)}_{\Lambda_k}$, with $k=1,\dots,m$, we denote by  
$\sigma_1\sigma_2\cdots\sigma_m$
the configuration in $\cS^{(s)}_{\L_1\cup\cdots\cup\L_m}$ such that 
$(\sigma_1\sigma_2\cdots\sigma_m)_{\Lambda_k}=\sigma_k$ 
for all $k\in\{1,\dots,m\}$. 
For $x\in\cL^{(s)}$ we define the shift $\Theta_x$
acting on $\cS^{(s)}$ by setting $(\Theta_x\sigma)_y:=\sigma_{y+x}$, 
for all $y\in\cL^{(s)}$ and $\sigma\in\cS^{(s)}$.

A function $f:\cS^{(s)}\rightarrow\bR$ is called a {\em local} function
if and only if there exists $\Lambda\subset\subset\cL^{(s)}$ such that
$f\in\cF^{(s)}_\Lambda$ namely, $f$ is 
$\cF^{(s)}_\Lambda$--measurable for some bounded set $\Lambda$.
For $f$ a local function we shall denote by $\supp(f)$,
the so--called support of $f$, the smallest $\Lambda\subset\subset\cL^{(s)}$
such that $f\in\cF^{(s)}_\Lambda$. If $f\in\cF^{(s)}_\Lambda$ we shall
sometimes misuse the notation by writing $f(\sigma_\Lambda)$ for $f(\sigma)$.
We also introduce $C(\cS^{(s)})$ the space of continuous functions on 
$\cS^{(s)}$ which becomes a Banach space under the norm 
$\|f\|_\infty:=\sup_{\sigma\in\cS^{(s)}}|f(\sigma)|$; note that the 
local functions are dense in $C(\cS^{(s)})$. 

\subsec{The potential}{s:pot} 
\par\noindent
Consider the integer $s\ge1$, a {\it potential} $\Phi^{(s)}$, for a lattice 
model on $\cL^{(s)}$ with configuration space $\cS^{(s)}$ as above, 
is a collection of local functions labelled by finite subsets of $\cL^{(s)}$
namely, 
$\Phi^{(s)}:=\{\Phi_X^{(s)}\in\cF_X^{(s)},\,X\subset\subset\cL^{(s)}\}$.
We say that $\Phi^{(s)}$ is 
{\it finite range} iff there exists $r>0$ such that 
$\Phi_X^{(s)}=0$ if $\diam_{s}(X)>r$; we say it is 
{\it translationally invariant} iff for each $x\in\cL^{(s)}$, 
$\Phi_X^{(s)}(\sigma)=\Phi_{X-x}^{(s)}(\Theta_x\sigma)$.
We note that the potentials, which do not need to satisfy 
the conditions above, form a linear space in which,
given $a\ge 0$, we introduce the norm $\|\cdot\|_a$ defined by
\begin{equation}
\label{nnnorm}
\|\Phi^{(s)}\|_a:=\sup_{x\in\cL^{(s)}}\sum_{X\ni x}e^{a\diam_{s}(X)}
\|\Phi_X^{(s)}\|_\infty
\end{equation}
We also note that in the translation invariant case we can omit the 
supremum above. 
Note that the Banach space defined by the norm above is too large to have
a satisfactory theory of high temperature phases. Indeed in 
\cite{[DM1], [DM2]} Dobrushin and Martirosyan have shown the 
following: let $h:\bN\to\bR_+$ and set 
$$
\|\Phi^{(s)}\|_{\textrm{DM}}:=
\sum_{X\ni 0} h(|X|)\,\|\Phi^{(s)}_X\|_\infty
$$
If $\exp\{-\gamma n\}h(n)\to 0$ in the limit $n\to\infty$ for all $\gamma>0$,
there exist complex interactions with arbitrarily small norm 
$\|\cdot\|_{\textrm{DM}}$, giving rise to phase transition in the sense
that the corresponding partition function vanishes for a sequence
of cubes $\Lambda_n\to\cL^{(s)}$, see also \cite{[EF]}. 

Given $\Lambda\su\su\cL^{(s)}$ 
and a potential $\Phi^{(s)}$ with bounded $\|\cdot\|_0$ norm,
the {\it finite volume Hamiltonian} 
associated to a configuration $\sigma\in\cS^{(s)}$ in 
$\Lambda$ is given by:
\begin{equation}
\label{1.2}
H_\Lambda^{(s)}(\sigma):=
 \sum_{\newatop{X\subset\subset\cL^{(s)}}
               {X\cap\Lambda\neq\emptyset}}\Phi_X^{(s)}(\sigma)
\end{equation}
Note that the sum on the r.h.s.\ of (\ref{1.2}) is absolutely convergent
(uniformly in $\sigma$) 
by the boundedness of $\|\Phi^{(s)}\|_0$. We also remark that
for a potential of range $r$ the Hamiltonian depends only on 
$\sigma_{\srclos{s}{r}{\Lambda}}$, namely 
$H_\Lambda^{(s)}\in\cF^{(s)}_{\srclos{s}{r}{\Lambda}}$.
We also let $E_{\Lambda}^{(s)}(\sigma)$ be the self--interaction 
associated to the volume $\Lambda$ i.e., the Hamiltonian 
with free boundary conditions namely, 
\begin{equation}
\label{fbc}
E_\Lambda^{(s)}(\sigma):=\sum_{X\subset\Lambda}\Phi_X^{(s)}(\sigma)
\end{equation}
We have that the map $E_\Lambda^{(s)}:\cS^{(s)}\to\bR$ 
depends only on the spins inside $\Lambda$ namely,
$E_\Lambda^{(s)}\in\cF_{\Lambda}^{(s)}$.

\subsec{The Gibbs measures}{s:gibbs} 
\par\noindent
Pick $s\ge 1$ and consider a 
potential $\Phi^{(s)}$ of bounded $\|\cdot\|_0$ norm. For each 
$\Lambda\subset\subset\cL^{(s)}$ we define the (finite volume) Gibbs measure 
in $\Lambda$, with boundary condition $\tau\in\cS^{(s)}$, as
the following measure on $\cS^{(s)}_\Lambda$:
$$
\mu^{(s),\tau}_{\Lambda}(\sigma):=
  \frac{1}{Z^{(s)}_\Lambda(\tau)} 
  \exp\left\{H^{(s)}_\Lambda\(\sigma\tau_{\Lambda^\complement}\)\right\}
$$  
for any $\sigma\in\cS^{(s)}_{\Lambda}$, 
where $ Z^{(s)}_\Lambda(\tau)$, called {\it partition function}, 
is the normalization constant, i.e.
\begin{equation}
\label{poert}
Z^{(s)}_\Lambda(\tau):=
\sum_{\sigma\in\cS^{(s)}_\Lambda} 
e^{H^{(s)}_\Lambda\(\sigma\tau_{\Lambda^\complement}\)}
\end{equation}
Note that we defined the Gibbs measure with a sign convention opposite to 
the usual one and include the inverse temperature in the definition
of the Hamiltonian; in fact it will be kept fixed in our analysis.  

We regard $\mu_\Lambda^{(s),\tau}$ also as a measure on the whole 
$\cS^{(s)}$ by giving zero mass to the configurations $\sigma$ which do 
not agree with $\tau$ on $\Lambda^\complement$. 
The (infinite volume) Gibbs states associated to the potential
$\Phi^{(s)}$ are then the probability measures $\mu^{(s)}$ on $\cS^{(s)}$ 
which satisfy the DLR equations
\begin{equation}
\label{dlr0}
\int\! \mu^{(s)}(d\tau)\, \mu_\Lambda^{(s),\tau}(f) 
= 
\mu^{(s)}(f) \quad\quad  \textrm{ for any } 
\Lambda\subset\subset\cL^{(s)}, \,\, f\in C(\cS^{(s)})
\end{equation}
where $\mu^{(s)}(f)$ denotes the expectation of $f$ w.r.t.\ $\mu^{(s)}$.

Given two local functions $f,g:\cS^{(s)}\to\bR$ we denote, finally, by 
$\mu^{(s)}(f;g):= \mu^{(s)}(fg)-\mu^{(s)}(f)\mu^{(s)}(g)$ the covariance
between $f$ and $g$.

\smallskip 
\noindent{\bf Condition SM$^{(s)}(\ell_0,b,B)$} ({\it Strong Mixing})
\hfill\break
{\sl Given a positive integer $\ell_0$ and two strictly positive reals $b,B$ 
we say that the potential 
$\Phi^{(s)}$ satisfies SM$^{(s)}(\ell_0,b,B)$ if and only if for any volume 
$I\su\su\cL^{(s\ell_0)}$ by setting 
$\Lambda:=\cO^s_{s\ell_0}I= \bigcup_{i\in I} Q^{(s)}_{\ell_0}(i)$,
the following bound holds.
For each pair of local functions $f,g$ such that
$\supp(f)\subset\Lambda$, $\supp(g)\subset\Lambda$, and  
$|\supp(f)|\wedge|\supp(g)|\,\exp\{-b\dis_s(\supp(f),\supp(g))\}\le 1$
we have
\begin{equation}
\label{e:sm}
\sup_{\tau\in\cS^{(s)}}
|\mu_\Lambda^{(s),\tau}(f;g)|\le 
B \|f\|_\infty \|g\|_\infty 
  |\supp(f)|\wedge|\supp(g)|\,e^{-b\,\dis_s(\supp(f),\supp(g))}
\end{equation}
}

It is a standard result that if there exist $\ell_0,b,B$ such that 
Condition~SM$^{(s)}(\ell_0,b,B)$ is satisfied then there exists a 
unique infinite volume Gibbs state namely, the DLR equations (\ref{dlr0}) 
admit a unique solution. 

\subsec{Lattice gas potential}{s:lg}
\par\noindent
A lattice gas is a translational invariant 
Gibbs field in the case $s=1$, which
is then dropped from the notation, and $S_x=1$ for each $x\in\cL$;
in such a case the single site configuration space associated to 
$x\in\cL$ is $\cX_x:=\{0,1\}$.
With the notation introduced in Subsection~\ref{s:conf}, for each 
$\Lambda\subset\cL$, we denote by $\cX_\Lambda$ the configuration 
space on $\Lambda$ equipped with the product topology 
and, as usual, we let $\cX:=\cX_{\cL}$.
For
$\eta\in\cX$ the value $\eta_x\in\{0,1\}$ is interpreted as the occupation 
number in $x\in\cL$. 
Moreover, we denote by $\cF$ the Borel $\sigma$--algebra on $\cX$ and set
$\cF_\Lambda:=\{\eta_x\in\cX_x,\,x\in\cL\}\subset\cF$. 

For a translationally invariant lattice gas 
we denote by $U$ the potential and observe that
$U_{\{x\}}(\eta)=\lambda\eta_x+a$ for some 
constants $\lambda,a\in\bR$. We neglect
the constant $a$, which do not affect the definition of the Gibbs
measure, and note that $\lambda$ is interpreted as the {\it
chemical potential}. We also introduce the {\it activity} $z\in\bR_+$
by $z:=e^\lambda$ which we use to parameterize lattice gases with different
chemical potentials. In such a case we write $U=(z,U_{>1})$ where
$U_{>1}:=\{U_X\in\cF_X,\,X\subset\subset\cL,|X|>1\}$ and call $U_{>1}$ 
the interaction. 

Coherently with the notation introduced in Subsection~\ref{s:gibbs}
the infinite volume Gibbs measure is denoted by $\mu$. 
We shall sometimes write $\mu_z$ for the infinite volume Gibbs measure, 
$\mu_{\Lambda,z}^\tau$ for the finite volume Gibbs measure 
on $\Lambda\subset\subset\cL$, 
and $Z_{\Lambda,z}(\tau)$ for the partition function of the lattice gas 
in order to explicitly indicate the dependence on the activity $z$.

\subsec{Block averaging transformation (BAT)}{s:bat}
\par\noindent
Let $\mu$ be the (unique) infinite volume Gibbs measure of a finite
range translationally invariant lattice gas satisfying Condition
SM$(\ell_0,b,B)$. 
Let $\rho:=\mu(\h_0)$ be the equilibrium density and
let us denote the compressibility by 
\begin{equation}
\label{chi}
\chi:=\sum_{x\in\cL}\mu(\eta_0;\h_x)
\end{equation}
Note that SM$(\ell_0,b,B)$ implies that there exists a real number
$C\in(0,+\infty)$ such that $C^{-1}\le\chi\le C$.

We consider a positive integer $\ell$ and the {\it renormalized lattice} 
$\cL^{(\ell)}$. For $I\subset\cL^{(\ell)}$ we define 
the function $N_I^{(\ell)}:\cX_I\to\{0,1,\dots,\ell^d\}^I$, which counts
the total number of particles in each block $Q_\ell(i)$, as follows
\begin{equation}
\label{dNi}
\big(N^{(\ell)}(\eta)\big)_i:=\sum_{x\in Q_\ell(i)} \eta_x 
\end{equation}
for all $i\in I$ and $\eta\in\cX_I$. As usual we will let 
$N^{(\ell)}_{\cL^{(\ell)}}=:N^{(\ell)}$ and 
$N^{(\ell)}_{\{i\}}=:N_i^{(\ell)}$ for all $i\in\cL^{(\ell)}$. 

For any $i\in\cL^{(\ell)}$ we define, moreover, the set 
\begin{equation}
\cM^{(\ell)}_i := \left \{\frac{-\r|Q_\ell|}
                                   {\sqrt{|Q_\ell|\chi}}, 
                              \frac{1-\r|Q_\ell|}{\sqrt{|Q_\ell| \chi}}, 
                              \dots,
                              \frac{|Q_\ell|(1-\r)}
                                   {\sqrt{|Q_\ell| \chi}}\right \}
\label{Ombi}
\end{equation}
that we consider equipped with the discrete topology. 
For $I\subset\cL^{(\ell)}$
we introduce, following the notation in Subsection~\ref{s:conf},
the {\it renormalized configuration space} 
$\cM_I^{(\ell)}:=\otimes_{i\in I}\cM_i^{(\ell)}$;
we set $\cM_{\cL^{(\ell)}}=:\cM^{(\ell)}$ and 
denote by $\cB^{(\ell)}$ its Borel $\sigma$--algebra. 
For $I\subset\cL^{(\ell)}$ we also set 
$\cB^{(\ell)}_I:=\sigma\{m_i\in\cM^{(\ell)}_i,\, i\in I\}\subset\cB^{(\ell)}$.
Moreover we define the measurable function 
$M_I^{(\ell)}:\,(\cX_{\cO_\ell I},\cF_{\cO_\ell I})\longrightarrow
       (\cM^{(\ell)}_I,\cB^{(\ell)}_I)$
by setting 
\begin{equation}
\label{Centr}
M_I^{(\ell)}(\eta):=\frac{N^{(\ell)}_I(\eta)-\r|Q_\ell|}{\sqrt{|Q_\ell|\chi}}
\end{equation}
for $\eta\in\cX_{\cO_\ell I}$. We let also
$M_{\cL^{(\ell)}}^{(\ell)}=:M^{(\ell)}$ and 
$M_{\{i\}}^{(\ell)}=:M_i^{(\ell)}$ for all $i\in\cL^{(\ell)}$. 

Finally, we define the {\it renormalized measure} 
$\nu^{(\ell)}:=\mu\circ(M^{(\ell)})^{-1}$, which 
is naturally induced by $M^{(\ell)}$ on $\cM^{(\ell)}$, and 
for $I\subset\cL^{(\ell)}$ and $\tau\in\cX$ we let
$\nu^{(\ell),\tau}_I:=\mu^\tau_{\cO_\ell I}\circ(M_I^{(\ell)})^{-1}$.
We avoid the troublesome issue of describing Gibbs measures on non--compact 
single spin space, see \cite{[EFS]} for a discussion, and consider
$\nu^{(\ell)}$ only for finite $\ell$. 

\subsec{Main results}{s:main}
\par\noindent
In this subsection we state the main theorems on the weak Gibbsianity
of the renormalized measure and the corresponding convergence 
as $\ell\to\infty$.

\bteo{t:wgib}
Let $U$ be a lattice gas potential 
satisfying SM($\ell_0,b,B$). Then for any large enough 
$\ell$ multiple of $\ell_0$ 
there exists a family of functions 
$\{\psi_I^{(\ell)},\phi_I^{(\ell)},\,I\subset\subset\cL^{(\ell)}\}$,
with $\psi_I^{(\ell)},\phi_I^{(\ell)}:\, \cM^{(\ell)}\mapsto\bR$, 
such that 
\begin{enumerate}
\item\label{i:mtb0}
for each $I\subset\subset\cL^{(\ell)}$ we have 
$\psi_I^{(\ell)},\phi_I^{(\ell)}\in\cB^{(\ell)}_I$.
\item\label{i:mtb1}
For each $I\subset\subset\cL^{(\ell)}$ we have 
$\psi_I^{(\ell)},\phi_I^{(\ell)}\equiv 0$ if $I$ is not $\ell$--connected.
\item\label{i:mtb2}
The functions 
$\psi_I^{(\ell)},\phi_I^{(\ell)}$ are translationally invariant in the 
sense specified in Subsection~\ref{s:pot} namely, 
$\psi_I^{(\ell)}(m)=\psi_{I-i}^{(\ell)}(\Theta_i m)$
and 
$\phi_I^{(\ell)}(m)=\phi_{I-i}^{(\ell)}(\Theta_i m)$
for any $i\in\cL^{(\ell)}$,
$I\subset\subset\cL^{(\ell)}$, and $m\in\cM^{(\ell)}$.
\item\label{i:mtb5}
There exist a measurable set $\bar\cM^{(\ell)}\subset\cM^{(\ell)}$, such that 
$\nu^{(\ell)}(\bar\cM^{(\ell)})=1$, and functions 
$r^{(\ell)}_i:\bar\cM^{(\ell)}\mapsto\bN\setminus\{0\}$, 
for all $i\in\cL^{(\ell)}$, such that 
for each $m\in\bar\cM^{(\ell)}$ if 
$I\ni i$ and $\diam_\ell(I)>r^{(\ell)}_i(m)$ then $\psi_I^{(\ell)}(m)=0$.
In particular for each $i\in\cL^{(\ell)}$ and $m\in\bar\cM^{(\ell)}$
there exists a real $c_i^{(\ell)}(m)\in[0,\infty)$ such that 
\begin{equation}
\label{mtb5}
\sum_{I\ni i}|\psi_I^{(\ell)}(m)|\le c_i^{(\ell)}(m)
\end{equation} 
\item\label{i:mtb4}
For each $q\in [1,\infty)$ 
\begin{equation}
\label{ee3f}
\sup_{i\in\cL^{(\ell)}}
  \nu^{(\ell)}\Big(\Big|
              \sum_{I\ni i}\psi_I^{(\ell)}
              \Big|^q\Big)<\infty
\end{equation} 
\item\label{i:mtb3}
There exists $\alpha'>0$ such that 
\begin{equation}
\label{eee}
\sup_{m\in\cM^{(\ell)}}
\sup_{i\in\cL^{(\ell)}}
\sum_{I\ni i} e^{\alpha'\diam_\ell(I)} 
 |\phi_I^{(\ell)}(m)|<\infty
\end{equation} 
\item\label{i:mtb6}
The DLR equations hold, namely for each $J\subset\subset\cL^{(\ell)}$
and for each local function $f\in\cB^{(\ell)}_J$ we have
\begin{equation}
\label{dlr}
\nu^{(\ell)}(f)=
 \int_{\cM^{(\ell)}} 
  \nu^{(\ell)}(dm')\, \frac{1}{Z_J(m')}\,
  \sum_{m\in\cM_J^{(\ell)}}
   \!\!\!f(m)\exp\Big\{
                  \sum_{I\cap J\neq\emptyset}
                 [\psi_I^{(\ell)}(mm'_{J^\complement})+
                  \phi_I^{(\ell)}(mm'_{J^\complement})]\Big\}
\end{equation}
where 
$$
 Z_J(m'):=\sum_{m\in\cM_J^{(\ell)}}
       \exp\Big\{\sum_{I\cap J\neq\emptyset}
       [\psi^{(\ell)}_I(mm'_{J^\complement})
        +\phi^{(\ell)}_I(mm'_{J^\complement})]\Big\}
$$
\end{enumerate}
\eteo

In the next theorem we state that in a suitable sense the renormalized 
potential converges to the one of independent harmonic oscillators. 

\bteo{t:conv}
In the same hypotheses as in Theorem \ref{t:wgib}, the family
$\{\psi_I^{(\ell)},\phi_I^{(\ell)},\,I\subset\subset\cL^{(\ell)}\}$ 
is such that 
\begin{enumerate}
\item\label{imtc2}
for each $i\in\cL^{(\ell)}$ we have $\psi_{\{i\}}^{(\ell)}(m)=-m_i^2/2$ and 
for each $q\in[1,+\infty)$ 
\begin{equation}
\label{ee2}
\lim_{\ell\to\infty}\,
\sup_{i\in\cL^{(\ell)}}
  \nu^{(\ell)}\Big(\Big|
              \sum_{I\ni i:\,|I|\ge2}\psi_I^{(\ell)}
              \Big|^q\Big)
=0
\end{equation} 
\item\label{imtc1}
there exists $\alpha'>0$ such that 
\begin{equation}
\label{ee1}
\lim_{\ell\to\infty} 
\sup_{m\in\cM^{(\ell)}}
\sup_{i\in\cL^{(\ell)}}
\sum_{I\ni i} e^{\alpha'\diam_\ell(I)}
|\phi_I^{(\ell)}(m)| 
  =0  
\end{equation} 
\end{enumerate}
where the limits $\ell\to\infty$ are taken along multiples 
of $\ell_0$.
\eteo

In order to compute the renormalized potential we compute the
partition function of the constrained system (\ref{ham}). Since we are
in the low temperature regime, the constrained system will
not have good mixing properties {\em for all\/}  possible values of the
image variable $m\in\cM^{(\ell)}$. We thus look at the constrained model as a
disordered system and look for properties which hold for
$\nu^{(\ell)}$--{\em almost all} image variables, that 
is we have to construct the set $\bar\cM^{(\ell)}$ properly. 
More precisely, we enclose the constrained models in a huge volume and
we try to compute their partition function via a uniform convergent
cluster expansion. We then face the typical problem of the Griffiths'
phase in disordered systems: anomalous values of the image
variables, which do occur somewhere in our volume, might produce
arbitrarily large regions of strong interaction. 

To overcome the above difficulty we follow a classical strategy in
disordered systems. Let us fix a configuration of the image variables.
We first perform a cluster expansion in the domains where the
constrained model verifies a uniform strong mixing condition implying
an effective weak interaction on a proper scale.
We are then left with an
effective residual interaction between the domains of strong interaction.
Since anomalous values of the image variables have small probability,
in the set $\bar\cM^{(\ell)}$ the strong interacting domains are well
separated on the lattice; we can thus use the graded cluster expansion 
developed in \cite{[BCOabs]} to treat the effective interaction.

\subsec{Synopsis}{s:str}
\par\noindent
In Section~\ref{s:cattivi} we construct the full measure set $\bar\cM^{(\ell)}$
algorithmically.
The required probability estimates are proven in a general setting, not
necessarily Gibbsian, of the underlying distribution of the disorder.
The analysis is based on the exponential decay of correlations and the
key recursive estimate in Lemma~\ref{t:recatt} is inspired by the approach
to the Anderson localization in \cite{[vDK]}.

In Section~\ref{s:con} we define the constrained models. We also 
introduce the condition, see (\ref{igood}),
on the image variable $m$ to identify
the good part of the lattice where the constrained systems satisfy 
the uniform strong mixing condition. We finally prove, in 
Theorem~\ref{t:partD}, that the general theory developed in 
Section~\ref{s:cattivi} can be applied. 

We then fix a value $m\in\bar\cM^{(\ell)}$ and compute the partition
function of the corresponding constrained model for a sequence 
of volumes invading the whole lattice. More precisely in 
Section~\ref{s:cegp} we use a procedure similar to 
\cite{[O],[OP],[BCO]} to integrate over the good part of the 
lattice and get the expansion in Theorem~\ref{t:ip3}.
In Section~\ref{s:diso} we feed the effective potential of 
Theorem~\ref{t:ip3} to the general theory developed 
in \cite{[BCOabs]} to integrate over the bad part of the lattice. 

To complete the proof of Theorem~\ref{t:wgib} we need to express
the output of \cite{[BCOabs]}, see Theorem~\ref{t:yteo}, 
as the sum of local functions. This is not a trivial point since,
a priori, the partition function of the constrained models depends
on the whole infinite volume image variable $m$. Nevertheless,
the graded cluster expansion in \cite{[BCOabs]} has been developed 
with a volume cutoff, see (\ref{supol}) and (\ref{cjp->Ys}), 
at each step of the iteration, 
so that the recursive construction allows us to prove locality 
of the renormalized potentials, see Theorem~\ref{t:mep.1}. 
The convergence stated in Theorem~\ref{t:conv} is an easy byproduct 
of the whole analysis.

\sezione{Bounds on the badness probability}{s:cattivi} 
\par\noindent
To compute the partition function of the {\em
constrained models},
see Section~\ref{s:con} below, we face the typical problem of the Griffiths'
phase in disordered systems: anomalous values of the image
variables, which do occur somewhere, might produce
arbitrarily large regions of strong correlation \cite{[BCOabs]}. 
In the present section we obtain some probability estimates on the
multi--scale geometry of these regions, based on the hypotheses 
that such anomalous values have small probability.

In this section we denote the lattice $\bZ^d$ by $\bL$.
In order to use a setup compatible with the one in \cite{[BCOabs]},
we use the distance $\disuno(x,y):=\sum_{i=1}^d|x_i-y_i|$ for all
$x=(x_1,\dots,x_d),y=(y_1,\dots,y_d)\in\bL$;
on the other hand we recall that $\dis(x,y)=\sup_{i=1,\dots,d}|x_i-y_i|$
as defined in Subsection~\ref{s:lat}.
Accordingly for $X\subset\cL$ we set 
$\diamuno(X):=\sup\{\disuno(x,y):\,x,y\in X\}$.
Moreover, given $X\subset\bL$ and $m\ge0$ we set 
$O_m(X):=\{y\in\bL:\,\disuno(y,X)\le m\}$ be the $m$--{\it neighborhood} of
$X$ w.r.t\ the metric $D$. 
If $x\in\bL$ we write $O_m(x)$ instead of $O_m(\{x\})$. 
We also recall that $B_m(X)$, see Subsection~\ref{s:lat}, is
the $m$--neighborhood of $X$ w.r.t.\ the metric $\dis$; of course 
$B_m(X)\supset O_m(X)$ for all $X\subset\bL$. 

We describe the strength of the disorder at the site $x$ in terms of a
binary variable $\o_x\in\{0,1\}$. We denote by $\o\in\O:=\{0,1\}^\bL$
the random field $\{\o_x,\:x\in\bL\}$; we consider $\O$ endowed with
its Borel $\s$--algebra $\cA$ and we let $\bQ$, a probability
in $\O$, be the distribution of $\o$. We also introduce the family of
$\s$--algebras on $\O$ defined by $\cA_\L:=\s\{\o_x,\,x\in\L\}$. 
We measure the diluteness of the system via the parameter 
\be{ddil} 
p:=\sup_{x\in\bL}\bQ(\omega_x=1) 
\end{equation}
which, in our analysis, will be sufficiently small.
We also assume that the correlations under $\bQ$ 
are exponentially decaying; more precisely we assume there exist 
reals $b''>0$ and $B''<\infty$ such that for each pair of local functions 
$f,g$ such that 
$|\supp(f)|\wedge|\supp(g)|\,\exp\{-b''\disuno(\supp(f),\supp(g))\}\le 1$
we have
\begin{equation}
\label{smin}
|\bQ(f;g)|\le 
B'' \|f\|_\infty \|g\|_\infty 
  |\supp(f)|\wedge|\supp(g)|\,e^{-b''\disuno(\supp(f),\supp(g))}
\end{equation}

We have a first classification of sites in {\em good\/} (where
$\o_x=0$) and {\em bad\/} (where $\o_x=1$). We strengthen the notion 
of {\it steep scales} introduced in \cite{[BCOabs]}.

\bdefi{t:seq}
We say that two 
strictly increasing sequences $\G=\{\G_j\}_{j\ge 0}$ and
$\g=\{\g_j\}_{j\ge 0}$ are 
{\em moderately steep scales} iff they
satisfy the following conditions:
\begin{enumerate}
\item\label{seq:<} 
$\Gamma_0=0$, $\gamma_0\ge0$, 
$\G_1\ge 2$, and $\G_j < \g_j / 2 $ for any $j\ge 1$.
\item\label{seq:con-dist} 
For $j\ge 0$ set
${\displaystyle \vartheta_j:=\sum_{i=0}^j (\G_i+ \g_i)}$ and
$\lambda:=\inf_{j\ge0}(\Gamma_{j+1}/\vartheta_j)$; then $\lambda\ge10$.
\item\label{seq:covol}
We have
${\displaystyle \sum_{j=1}^\infty\frac{\G_j}{\g_j}\le\frac{1}{2}}$
where we understand $\Gamma_0/\gamma_0=0$ even in the case 
$\gamma_0$. 
\item\label{seq:con-ano} 
There exist reals $a>0$ and $\varepsilon\in(0,1)$ such that 
${\displaystyle 
  2\cdot3^d\vartheta_{k+1}^d\le\exp\{a(1+\varepsilon)^k\}}$
for all $k\ge0$.
\item\label{seq:con-icf} 
For $a$ and $\varepsilon$ as above we have that 
${\displaystyle 
\sum_{k=1}^\infty\vartheta_k^s\exp\{-a(1+\varepsilon)^k/q\}<\infty
}$
for all $s\ge0$ and $q>1$.
\end{enumerate}
\edefi

We remark that with respect to the definition of steep scales given 
in \cite{[BCOabs]} we have added the conditions
\ref{seq:con-ano} and \ref{seq:con-icf}, and 
strengthened item~\ref{seq:con-dist} to $\lambda\ge10$. 
An explicit example of moderately steep scales is given in (\ref{okscale}) 
below. 

\bdefi{t:gentle} 
We say that $\cG:=\{\cG_j\}_{j\ge 0}$, where each $\cG_j$ is a collection of 
finite subsets of $\bL$ is a {\em graded disintegration of} $\bL$ iff
\begin{enumerate}
\item\label{i:gd1}
for each $g\in\bigcup_{j\ge0}\cG_j$ 
there exists a unique $j\ge0$, which 
is called the {\em grade} of $g$, such that $g\in\cG_j$;
\item\label{i:gd2}
the collection $\bigcup_{j\ge0}\cG_j$ 
of finite subsets of $\bL$ is a partition 
of the lattice $\bL$ namely, it is a collection of not empty pairwise disjoint 
finite subsets of $\bL$ such that 
\begin{equation}
\label{part}
\bigcup_{j\ge 0}\,\bigcup_{g\in\cG_j}g=\bL.
\end{equation}
\end{enumerate}
\hfill\break
Given $\bG_0\subset\bL$ and $\Gamma,\gamma$ steep scales, we say that 
a graded disintegration $\cG$ is a {\em gentle disintegration} of $\bL$ 
with respect to $\bG_0,\Gamma,\gamma$ iff
the following recursive conditions hold:
\begin{enumerate}
\setcounter{enumi}{2}
\item
\label{gent:null} 
$\cG_0=\big\{\{x\},\,x\in\bG_0\big\}$;
\item
\label{gent:diam} 
if $g\in\cG_j$ then $\diamuno(g) \le \G_j$ for any $j\ge 1$;
\item
\label{gent:dist}
set $\bG_j:=\bigcup_{g\in\cG_j}g\subset\bL$, $\bB_0:=\bL\setminus\bG_0$ and 
$\bB_j:=\bB_{j-1}\setminus\bG_j$, then for 
any $g\in\cG_j$ we have $\disuno\(g,\bB_{j-1}\setminus g\)>\g_j$ for
any $j\ge 1$;
\item
\label{gent:cas}
$\forall x\in\bL$ we have 
$k_x:=\sup\big\{j\ge 1:\,\exists g\in\cG_j\textrm{ such that }
                \dis(x,\env(g))\le \vartheta_j\big\}<\infty$,
where we recall $\env(g)$ has been defined in Subsection~\ref{s:lat}.
\end{enumerate}
Sites in $\bG_0$ (resp.\ $\bB_0$) are called {\em good} (resp.\ {\em bad});
similarly we call $j$--gentle (resp.\ $j$--bad) 
the sites in $\bG_j$ (resp.\ $\bB_j$).
Elements of $\cG_j$, with $j\ge 1$, are called $j$--gentle atoms.
Finally, we set 
$\cG_{\ge j}:= \bigcup_{i\ge j} \cG_i$.
\edefi

The results of the present section are summarized in the following
Theorem. 

\bteo{t:fm}
Let the sequences $\G,\g$ be moderately steep scales 
in the sens of Definition \ref{t:seq}. 
Assume also that (\ref{smin}) holds,
$p\le\exp\{-a/(1-\varepsilon)\}$ 
and the sequences $\Gamma,\gamma$ are such that 
\begin{equation}
\label{gag}
2\cdot9^dB''\vartheta_{k+1}^{2d} \exp\{-b''\Gamma_{k+1}/20\}
\le
\exp\Big\{-\frac{a}{1-\varepsilon}(1+\varepsilon)^{k+1}\Big\}
\end{equation}
for any $k\ge0$. Set finally $\bG_0 (\o) := \{x\in\bL\,:\ \o_x=0\}$. 
Then there exists a set $\bar\O\in\cA$ with 
$\bQ(\bar\O)=1$ such that
\begin{enumerate}
\item\label{gent:pa}
for each $\o\in\bar\O$ there exists a gentle disintegration 
$\cG=\cG(\o)$ in the sense of Definition \ref{t:gentle}; 
\item\label{gent:mis}
for each $x\in\bL$ and  $X\subset\subset\bL$ we have that 
$\left\{\omega:\; \bG_k(\omega)\ni x\right\}\in\cA_{O_{\vartheta_k}(x)}$ 
and also 
$\left\{\omega:\; \cG_k(\omega)\ni X\right\}\in
\cA_{O_{\vartheta_k}(X)}$.
\end{enumerate} 
\eteo

Let us first describe an algorithm to construct the families $\cG_k$ for 
$k\ge 1$;
from this it will follow item \ref{gent:mis} in Theorem \ref{t:fm}.  
Given a configuration $\o\in\O$ and $\G$, $\g$ 
{\em moderately steep scales}, 
we define the following inductive procedure in a finite volume
$\Lambda\subset\subset\bL$ which constructs the $k$--{\em gentle} sites
in $\Lambda$.

\smallskip
\par\noindent
\texttt{
\begin{enumerate}
\item\label{a0}
set $k=1$;
\item\label{a1}
set $i=1$ and $V=\emptyset$;
\item\label{a2}    
if 
$(\bB_{k-1}\cap\Lambda)\setminus V=\emptyset$ 
then goto \ref{a6};
\item\label{a3} 
pick a point 
$x\in(\bB_{k-1}\cap\Lambda)\setminus V$,   
set $A=O_{\Gamma_k}(x)\cap\bB_{k-1}$ and $V=V\cup A$;
\item\label{a4} 
if $\diamuno(A)\le\Gamma_k$ and 
$\disuno\(A,\bB_{k-1}\setminus A\)> \gamma_k$
then $g_k^{i}=A$ and $i=i+1$;
\item\label{a5}
goto \ref{a2}; 
\item\label{a6}
set $\cG_k :=\{g_k^m,\; m=1,\dots,i-1\}$,  
with the convention $\cG_k=\emptyset$ if $i=1$, 
$\bG_k :=\bigcup_{m=1}^{i-1}g_k^m$, and $\bB_{k}:=\bB_{k-1}\setminus\bG_k$;
\item\label{a7}
set $k=k+1$, if $\Gamma_k\le\diamuno(\Lambda)$ goto \ref{a1} else exit;
\end{enumerate} 
}

\smallskip
Let us briefly describe what the above algorithm does. At step $k$ 
we have inductively constructed $\bB_{k-1}$, the set of $(k-1)$--bad 
sites; we stress that sites in $\bL\setminus\Lambda$ may belong to 
$\bB_{k-1}$. Among the sites in $\bB_{k-1}\cap\Lambda$ we are now looking
for the $k$--gentle ones. The set $V$ is used to keep track of the sites
tested for $k$--gentleness. At step \ref{a3} we pick a new site 
$x\in\bB_{k-1}\cap\Lambda$ and test it, at step \ref{a4}, for 
$k$--gentleness w.r.t.\ $\bB_{k-1}$, i.e. including also bad sites 
in $\bL\setminus\Lambda$.  
Note that the families $\cG_k$ for any $k\ge 1$ do not depend on the way
in which $x$ is chosen at step \ref{a3} of the algorithm. Suppose, indeed,
to choose $x\in(\bB_{k-1}\cap\Lambda)\setminus V$ at step \ref{a3} and to
find that $A=O_{\Gamma_k}(x)\cap\bB_{k-1}$ is a $k$--gentle atom.
Consider $x'\in A$ such that $x'\neq x$ and set 
$A':=O_{\Gamma_k}(x')\cap\bB_{k-1}$: since $A$ satisfies the test 
for $k$--gentleness at step \ref{a4} of the algorithm, we have
$A\subset A'$. By changing the role of $x$ and $x'$ we get $A=A'$.

After a finite number of operations, the algorithm stops and outputs the family
$\cG_k(\Lambda)$ (note we wrote explicitly the dependence on $\L$)
with the following property. If $g\in\cG_k(\L)$ then $\diamuno(g)\le\Gamma_k$
and $\disuno\(g,\bB_{k-1}(\Lambda)\setminus g\) >\gamma_k$. 
We call a set $g\in\cG_k(\L)$ an atom of $k$--{\it gentle} sites;
note however that $g$ is not necessarily connected. 

We finally take an increasing sequence of sets
$\Lambda_i\subset\subset\bL$, invading $\bL$ and we sequentially
perform the above algorithm.
This means the algorithm for $\Lambda_i$ is performed 
independently of the outputs prevoiusly obtained, i.e., for $\Lambda_j$ 
$j<i$. 
It is easy to show that if $g\in\cG_k(\Lambda_i)$ then   
$g\in\cG_k(\Lambda_{i+1})$; therefore  
$\cG_k(\Lambda_i)$ is increasing in $i\ge 1$, so
that we can define $\cG_k:=\lim_{i\to\infty}\cG_k(\Lambda_i)
=\bigcup_{i}\cG_k(\Lambda_i)$ and 
$\bG_k:=\lim_{i\to\infty}\bG_k(\Lambda_i)=\bigcup_{g\in\cG_k}g$. 
Hence, $\bB_k(\Lambda_i)=\bB_{k-1}(\Lambda_i)\setminus\bG_k(\Lambda_i)=
\bL\setminus\cup_{j=0}^{k-1}\bG_j(\Lambda_i)$ is decreasing in $i\ge 1$, 
so that $\bB_k:=\lim_{i\to\infty}\bB_k(\Lambda_i)=
\bigcap_i\bB_k(\Lambda_i)$. We also remark that, by construction, 
$\{\bB_k,\; k\ge 0\}$ is a decreasing sequence.    
We say $x\in\bL$ is $k$--{\it gentle} (resp. $k$--{\it bad}) 
iff $x\in\bG_k$  (resp. $x\in\bB_k$).

Note that it follows from the construction that it is possible to decide 
whether a site $x$ at step $k$ 
is $k$--gentle by looking only at the $\omega$'s inside 
a ball centered at $x$ of radius $\vartheta_k$ 
(as defined in item \ref{seq:con-dist}
of Definition \ref{t:seq}).  

\begin{lem}\label{t:mis}
Let $\bG_k$ and $\cG_k$, $k=0,1,\dots$, as constructed above. Then
item \ref{gent:mis} in Theorem \ref{t:fm} holds, i.e.\ 
for each $x\in\bL$
\be{e:mis} 
\left\{\omega:\; x\in\bG_k(\omega)\right\}
\in\cA_{O_{\vartheta_k}(x)}
\end{equation} 
and for each $X\subset\bL$
\be{e:mis2} 
\left\{\omega:\; X\in\cG_k(\omega)\right\}
\in\cA_{O_{\vartheta_k}(X)}
\end{equation} 
\end{lem}

\noi{\it Proof.}\  
We first prove (\ref{e:mis}). We proceed by induction. 
For $k=0$ \eqref{e:mis} holds trivially. 
Let $O:=O_{\Gamma_k}(x)$. 
{}From the algorithmic construction above we have 
\begin{equation}
\label{e:aaa}
\{x\in\bG_{k}\}= 
\left\{x\in\bB_{k-1}\right\}\cap
\left\{\diamuno\left(O\cap\bB_{k-1}\right)\le\Gamma_k\right\}  
\cap\left\{
\disuno\left(\bB_{k-1}\cap O,\bB_{k-1}
         \setminus O\right)>\gamma_k\right\}
\end{equation}
Since $\vartheta_k$ is increasing, by the inductive hypotheses
\begin{equation}
\label{e:aac}
\{x\in\bB_{k-1}\}=\bigcap_{h=0}^{k-1}\{x\notin\bG_h\}
\in\bigvee_{h=0}^{k-1}\cA_{O_{\vartheta_h}(x)}
=
\cA_{O_{\vartheta_{k-1}}(x)}
\end{equation}
On the other hand 
\begin{equation}
\label{e:aab}
\left\{\diamuno\left(O\cap\bB_{k-1} \right)\le\Gamma_k\right\}  
=
\bigcap_{\newatop{y,z\in O}
                 {\disuno(y,z)>\Gamma_k}}
\left(\left\{y\notin\bB_{k-1}\right\}\cup\left\{z\notin\bB_{k-1}\right\}
\right)
\in
\bigvee_{y\in O}\cA_{O_{\vartheta_{k-1}}(y)}
\end{equation}
where we used (\ref{e:aac}).  
Finally
\begin{eqnarray}
\left\{\disuno\left(\bB_{k-1}\cap O,\bB_{k-1}\setminus O\right)>\gamma_k\right\}
&=&
\left\{\nexists (y,z)\in\left(O\cap\bB_{k-1}\right)\times 
                           \left(\bB_{k-1}\setminus O\right): 
                           \disuno(y,z)\le\gamma_k\right\}
\nn\\
&=&\bigcap_{\newatop{y\in O,z\in\bL\setminus O:}
                    {\disuno(y,z)\le\gamma_k}}
\left(
\left\{y\notin\bB_{k-1}\right\}\cup\left\{z\notin\bB_{k-1}\right\} 
\right)\nn
\end{eqnarray} 
hence
\begin{equation}
\label{e:aad}
\left\{\disuno\left(\bB_{k-1}\cap O,\bB_{k-1}\setminus O\right)>\gamma_k\right\}
\in
\bigvee_{\genfrac{}{}{0pt}{}{y\in\bL:}{\disuno(x,y)\le\Gamma_k+\gamma_k}}
\cA_{O_{\vartheta_{k-1}}(y)}   
\end{equation}
where we used again \eqref{e:aac}. 
Recalling $\vartheta_k=\vartheta_{k-1}+\Gamma_k+\gamma_k$, 
equation (\ref{e:mis}) follows from 
\eqref{e:aaa}--\eqref{e:aad}. 

Similar arguments can be used to prove (\ref{e:mis2}).
If $\diamuno(X)>\Gamma_k$ then $X\notin\cG_k$. We consider, now, the case 
$\diamuno(X)\le\Gamma_k$; we have:
\begin{eqnarray}
\left\{X\in\cG_k\right\} & = &
\left\{x\in\bB_{k-1},\,\forall x\in X\right\}
\cap
\left\{\disuno\left(X,\bB_{k-1}\setminus X\right)>\gamma_k\right\}
\nn\\ 
& = & 
\bigcap_{x\in X}\left\{x\in\bB_{k-1}\right\} 
\cap
\left\{\disuno\left(X,\bB_{k-1}\setminus X\right)>\gamma_k\right\}
\label{e:abb} 
\end{eqnarray}
Now, from (\ref{e:aac}) we have that 
\begin{equation}
\label{e:aba}
\bigcap_{x\in X}
\left\{x\in\bB_{k-1}\right\}\in
\cA_{O_{\vartheta_{k-1}}(X)}
\end{equation}
Moreover,
\begin{eqnarray}
\left\{\disuno\left(X,\bB_{k-1}\setminus X\right)>\gamma_k\right\}
& = &
\left\{\nexists(x,y)\in X\times\left(\bB_{k-1}\setminus X\right):\,
       \disuno(x,y)\le\gamma_k\right\}
\nn\\ 
& = & 
\bigcap_{y\in\bL\setminus X:\,\disuno(y,X)\le\gamma_k}
     \left\{y\notin\bB_{k-1}\right\} 
     \in
     \cA_{O_{\vartheta_{k-1}+\gamma_k}(X)}
\label{e:abc} 
\end{eqnarray}
and from (\ref{e:abb})--(\ref{e:abc}) we finally get (\ref{e:mis2}).
\qed 

\bteo{t:sexpb}
Let the hypotheses of Theorem~\ref{t:fm} be satisfied; recall 
$a$ and $\varepsilon$ have been defined 
in item~\ref{seq:con-ano} of Definition~\ref{t:seq}.
Then
\be{sexpb}
\sup_{x\in\bL} \bQ\(\o\::\:x\in\bB_k\)
  \le\exp\Big\{-\frac{a}{1-\varepsilon}(1+\varepsilon)^k\Big\}
\end{equation}
\eteo

Postponing the proof of the above bound, let us show how it implies,
via a straightforward application of Borel--Cantelli lemma, Theorem \ref{t:fm}.

\medskip
\par\noindent
{\it Proof of Theorem \ref{t:fm}.} \/
Proof of item~\ref{gent:pa}:
for each $\omega\in\Omega$ let $\cG\equiv\cG(\omega)=\{\cG_j(\omega)\}_{j\ge0}$ 
be constructed by the algorithm described below Theorem~\ref{t:fm}.
We have to show that $\cG$ satisfies 
items~\ref{i:gd1}--\ref{gent:cas} in Definition~\ref{t:gentle} $\bQ$--a.s.: 
items~\ref{i:gd1}, \ref{gent:null}, \ref{gent:diam} and \ref{gent:dist}
hold by construction.
We prove first that there exists a set $\bar\Omega\subset\Omega$ of 
full $\bQ$--measure such that item~\ref{i:gd2} (of Definition~\ref{t:gentle})
holds, namely such that $\bigcup_{j\ge0}\cG_j(\omega)$ is a 
partition of the lattice $\bL$.
Let $\bB_{\infty}$ be the random subset of the lattice given by 
$\bB_{\infty}:=\lim_{n\to\infty}\bB_n=\bigcap_{n=0}^{\infty}\bB_n$.  
{}From Theorem \ref{t:sexpb} and the Borel--Cantelli lemma we get 
\begin{equation}
0=\bQ\left(\bigcap_{n=0}^{\infty}\bigcup_{k=n}^{\infty}\{x\in\bB_k\}\right)
=\bQ\left(\bigcap_{n=0}^{\infty}\{x\in\bB_n\}\right)
=\bQ\left(\{x\in\bB_{\infty}\}\right)
\end{equation} 
where we have used that $\bB_n,\; n\in\bN$, is a decreasing family 
of subsets of the lattice.  
Whence, by taking a countable union, we get
\begin{equation*}
0=\bQ\left(\bigcup_{x\in\bL}\{x\in\bB_{\infty}\}\right)
=\bQ \left(\bB_{\infty}\neq\emptyset\right)
=1-\bQ\left(\bL=\bigcup_{j=0}^{\infty}\bG_j\right)
\end{equation*}
We prove, finally, that also item~\ref{gent:cas} of Definition~\ref{t:gentle}
is satisfied $\bQ$--a.s.: it is enough to note that for $x\in\bL$ we have
\begin{equation}
\label{ppppp}
\begin{array}{l}
{\displaystyle
\sum_{k=1}^\infty 
\bQ \(\o:\,\exists g\in \cG_k(\o):\, \dis(x,\env(g))\le\vartheta_k \) 
}\\
\phantom{merdonemerdonemerda}
\le 
{\displaystyle
\sum_{k=1}^\infty 
\bQ \(\o:\,\exists y \in \bB_{k-1}(\o):\, \dis(x,y) \le \vartheta_k+\G_k  \) 
}
\\
\phantom{merdonemerdonemerda}
\le
{\displaystyle
\sum_{k=1}^\infty \[2 (\vartheta_k+\G_k) +1\]^d
    \sup_{y\in\bL}\bQ\(\o:\,y\in\bB_{k-1}(\o) \)
}
\\
\phantom{merdonemerdonemerda}
\le
{\displaystyle
\sum_{k=1}^\infty \[2 (\vartheta_k+\G_k) +1\]^d 
\exp\Big\{-\frac{a}{1-\varepsilon}(1+\varepsilon)^{k-1}\Big\}
}
\\
\phantom{merdonemerdonemerda}
\le
{\displaystyle
\sum_{k=1}^\infty \[2 (\vartheta_k+\G_k) +1\]^d 
\exp\Big\{-a(1+\varepsilon)^k\Big\}
< \infty
}
\end{array}
\end{equation}
where we used the bound \eqref{sexpb} and item \ref{seq:con-icf} in 
Definition \ref{t:seq}. The proof is
completed by applying again Borel--Cantelli. 

Proof of item~\ref{gent:mis} of the Theorem: it has already been proven 
in Lemma~\ref{t:mis} for the graded disintegration constructed via 
the algorithm described below Theorem~\ref{t:fm}.
\qed

\medskip
The key step in proving Theorem \ref{t:sexpb} is the following
recursive estimate on the probability of the degree of badness.

\blem{t:recatt} 
Let the hypotheses of Theorem~\ref{t:fm} be satisfied. Set 
$p_k:=\sup_{x\in\bL}\bQ\(x\in\bB_k\)$, for $k=0,1,\dots$; set also  
$A_k(x):=O_{\g_k+\G_k}(x)\setminus O_{(\G_k-1)/2}(x)$ and 
$|A_k|=|A_k(x)|$. Then for each $k=0,1,\dots$ we have
\be{recatt} 
p_{k+1}\le|A_{k+1}|\Big(p_k^2+B''|O_{\vartheta_k}|
\exp\{-b''\Gamma_{k+1}/20\}\Big)
\end{equation}
\elem

\Pro 
Recalling the definition of the $k$--bad set $\bB_k$ we have 
\be{e:1}
\left\{x\in\bB_{k+1}\right\}=
\left\{x\in\bB_k\right\}\cap 
\left\{x\notin\bG_{k+1}\right\}
\end{equation}  
On the other hand, by the construction of the ($k+1$)--gentle sites,  
\be{e:2}
\left\{x\in\bB_k\right\}\cap 
\left\{x\notin\bG_{k+1}\right\}
\subset
\left\{x\in\bB_k\right\}\cap 
\left\{\exists y\in A_{k+1}(x):\; y\in\bB_k\right\}
\end{equation}
indeed, given $\bB_k$, if there were no $k$--bad site in the annulus
$A_{k+1}(x)$ then $x$ would have been $(k+1)$--gentle. {}From
\eqref{e:1} and \eqref{e:2}
\begin{equation}
\label{e:3}
\begin{array}{rl}
{\displaystyle
\bQ\left(x\in\bB_{k+1}\right)\le
}
&
{\displaystyle
\bQ
\bigg(
\bigcup_{y\in A_{k+1}(x)}
\left\{x\in\bB_k\right\}
\cap
\left\{y\in\bB_k\right\}
\bigg)
}
\\
&\\
{\displaystyle
\le
}
&
{\displaystyle
\sum_{y\in A_{k+1}(x)}
\bQ
\big(
\left\{x\in\bB_k\right\}
\cap
\left\{y\in\bB_k\right\}
\big)
}
\\
&\\
{\displaystyle
=
}
&
{\displaystyle
\sum_{y\in A_{k+1}(x)}
\Big[
\bQ
\left(
\left\{x\in\bB_k\right\}
\right)
\bQ
\left(
\left\{y\in\bB_k\right\}
\right)
+\bQ\left(\id_{\{x\in\bB_k\}};\id_{\{y\in\bB_k\}}\right)\Big]
}
\\
\end{array}
\end{equation} 
We note, now, that for $x\in\bL$ and $y\in A_{k+1}(x)$ we have 
\begin{equation}
\label{geom}
\begin{array}{lcl}
\disuno(O_{\vartheta_k}(x),O_{\vartheta_k}(y)) 
& \ge & {\displaystyle
         \Big[\frac{\Gamma_{k+1}-1}{2}\Big]-2\big[\vartheta_k\big]+1
         \ge 
         \frac{\Gamma_{k+1}}{2}-2\vartheta_k-\frac{1}{2}}\\
&&\\
& \ge & {\displaystyle
         \frac{\lambda-8}{4\lambda}\Gamma_{k+1}\ge
         \frac{1}{20}\Gamma_{k+1}}\\
\end{array}
\end{equation}
recall we assumed $\lambda\ge10$ in item~\ref{seq:con-dist} of 
Definition~\ref{t:seq}. 
By Lemma~\ref{t:mis}, (\ref{smin}), and (\ref{e:3}), we finally get 
the bound (\ref{recatt}).
\qed

\medskip
\noi{\it Proof of Theorem \ref{t:sexpb}.}\
The thesis follows by induction from $p_0:=p\le\exp\{-a/(1-\varepsilon)\}$,
Lemma~\ref{t:recatt}, item~\ref{seq:con-ano} in Definition~\ref{t:seq},
equation (\ref{gag}), $|A_{k+1}|\le3^d\vartheta_{k+1}^d$, and 
$|O_{\vartheta_k}|\le 3^d\vartheta_{k+1}^d$.
\qed 

\sezione{The constrained models}{s:con}
\par\noindent 
In dealing with the renormalization group transformation it is 
necessary to pack spins associated to different sites of the  
lattice so that a new variable, often called {\it block spin}, is obtained.

\subsec{The block spin models}{s:spm}
\par\noindent
Recall the general setup introduced in Subsections~\ref{s:conf}--\ref{s:gibbs}
for a spin model on lattice $\cL^{(s)}$, with $s\ge1$ integer, 
with potential $\Phi^{(s)}$, and Gibbs measure $\mu^{(s)}$. 
For $u$ a positive multiple of $s$ we consider the 
lattice $\cL^{(u)}$ and associate to each site $i\in\cL^{(u)}$
the {\it single site block spin configuration space} 
\begin{equation}
\label{sinspi}
\cS^{(s),u}_i:=\bigotimes_{x\in Q_{u/s}^{(s)}(i)}\cS_x^{(s)}
              =\cS^{(s)}_{Q_{u/s}^{(s)}(i)}
\end{equation}
We can then consider the {\it block spin configuration space} 
$\cS^{(s),u}_I:=\otimes_{i\in I}\cS_i^{(s),u}$,
for any $I\subset\cL^{(u)}$, equipped with the product topology.
As usual we let $\cS^{(s),u}_{\cL^{(u)}}=:\cS^{(s),u}$ and 
denote its Borel $\sigma$--algebra by $\cF^{(s),u}$. 
Moreover, for each $I\subset\cL^{(u)}$ we set 
$\cF^{(s),u}_I:=\sigma\{\zeta_i\in\cS^{(s),u}_i,\, i\in I\}\subset\cS^{(s),u}$.

As for the lattices we introduce operators 
which allow to pack spins and unpack block spins. With an abuse of notation
we shall use the same symbol as in Subsection~\ref{s:lat}. 
We define the 
{\it packing} operator $\cO_s^u:\cS^{(s)}\to\cS^{(s),u}$ associating to each 
spin configuration $\sigma\in\cS^{(s)}$ the unique block spin configuration 
$\cO_s^u\sigma\in\cS^{(s),u}$ such that 
$(\cO_s^u\sigma)_i:=\{\sigma_x,\,x\in Q_{u/s}^{(s)}(i)\}$ for all 
$i\in\cL^{(u)}$. 
The {\it unpacking} operator $\cO_u^s:\cS^{(s),u}\to\cS^{(s)}$ associates 
to each block spin configuration $\zeta\in\cS^{(s),u}$ the 
unique spin configuration $\cO_u^s\zeta\in\cS^{(s)}$ such that 
$\zeta_i=\{(\cO_u^s\zeta)_x,\,x\in Q_{u/s}^{(s)}(i)\}$ for all $i\in\cL^{(u)}$.
Note that in the case of infinite volume 
configurations the packing and the 
unpacking operators are one the inverse of the other. 

We remark also that the two operators allow the packing of the 
spin $\sigma$--algebra and the unpacking of the block spin one namely, 
for each $I\subset\cL^{(u)}$ and $\Lambda\subset\cL^{(s)}$ we have 
\begin{equation}
\label{spisigf}
\cO_u^s\big(\cF^{(s),u}_I\big)=\cF^{(s)}_{\cO_u^sI}
\;\;\;\;\textrm{ and }\;\;\;\;
\cO_s^u\big(\cF^{(s)}_\Lambda\big)\subset\cF^{(s),u}_{\cO_s^u\Lambda}
\end{equation}
Where in the last relation the equality between the two $\sigma$--algebras
stands if and only if $\cO_u^s\cO_s^u\Lambda=\Lambda$. 

To a block spin configuration we can naturally associate the potential 
$\Phi^{(s),u}$ defined as follows; for each $I\subset\subset\cL^{(u)}$ 
the function $\Phi^{(s),u}_I:\cS^{(s),u}\to\bR$ is defined as 
\begin{equation}
\label{spipot}
\Phi^{(s),u}_I:=\sum_{\newatop{X\subset\cL^{(s)}:}
                              {\cO_s^u X=I}}
                [\Phi^{(s)}_X\circ\cO_u^s]
\end{equation}
We remark that 
$\Phi^{(s),u}\in\cF^{(s),u}_I$.  
Given $I\subset\subset\cL^{(u)}$, 
we consider the {\em block spin Hamiltonian} 
$H^{(s),u}_I:\cS^{(s),u}\to\bR$ associating to each 
block spin configuration $\zeta\in\cS^{(s),u}$ the Hamiltonian 
\begin{equation}
\label{spiham}
H^{(s),u}_I(\zeta):=\sum_{J\cap I\neq\emptyset}\Phi^{(s),u}_J(\zeta)
\;\;\;\;\textrm{ and }\;\;\;\;
E^{(s),u}_I(\zeta):=\sum_{J\subset I}\Phi^{(s),u}_J(\zeta)
\end{equation}
It is easy to show that, given $I\subset\subset\cL^{(u)}$
and the block spin configuration $\zeta\in\cS^{(s),u}$, the Hamiltonian 
$H^{(s),u}_I$ is the Hamiltonian of the unique spin configuration
$\cO_u^s\zeta\in\cS^{(s)}$ obtained by unpacking $\zeta$; indeed
\begin{equation}
\label{spihams}
H^{(s),u}_I(\zeta)
=\sum_{J\cap I\neq\emptyset}\Phi^{(s),u}_J(\zeta)
=\sum_{J\cap I\neq\emptyset}
  \sum_{\newatop{X\subset\cL^{(s)}:}{\cO_s^u X=J}}\Phi^{(s)}_X(\cO_u^s\zeta)
=\sum_{\newatop{X\subset\cL^{(s)}:}{X\cap\cO_u^sI\neq\emptyset}}
 \Phi^{(s)}_X(\cO_u^s\zeta)
=H_{\cO_u^s I}^{(s)}(\cO_u^s\zeta)
\end{equation}

We can finally define a Gibbs measure 
on the block spin configuration space $\cS^{(s),u}$, with its 
$\sigma$--algebra $\cF^{(s),u}$, by considering the measure 
$\mu^{(s),u}:=\mu^{(s)}\circ\cO_u^s$
which is Gibbsian w.r.t.\ the potential (\ref{spipot}).

We note that it is possible to make block spins out of block spins namely, we 
can consider $\cS^{(s),u}$ as the starting configuration space, fix a
multiple $v$ of $u$, and 
construct the block spin configuration space $\cS^{(s),u,v}$. 
Exploiting the fact that $v$ is a multiple of $s$ it 
is possible to construct the block spin space $\cS^{(s),v}$; 
note that the spaces $\cS^{(s),u,v}$ and $\cS^{(s),v}$ are different because
they are produced by grouping the original spins, living on scale $s$,
in two different ways.

For the sake of clearness we list here the particular cases in which 
we will make use of the block spin setup introduced above. 
First of all, we fix the the renormalization scale $\ell$ and the 
rougher scale $\wp\equiv \wp(\ell):=\cionc d\ell$.
On one hand we consider as original lattice model the 
lattice gas $\mu$ on $\cL$ with 
configuration space $\cX$ and algebra of the events $\cF$, 
see Subsection~\ref{s:lg}, and construct the 
block spin space $\cX^{(1),\ell}\equiv\cX^{\ell}$,  
its $\sigma$--algebra $\cF^{\ell}$, and the Gibbs measure 
$\mu^{\ell}=\mu\circ\cO_\ell$,
with $\cO_\ell:\cX^{\ell}\to\cX$ the unpacking operator.
Then, on the rougher scale $\wp$, we  
construct the space $\cX^{(1),\ell,\wp}\equiv\cX^{\ell,\wp}$, its 
$\sigma$--algebra $\cF^{\ell,\wp}$, and the Gibbs measure 
$\mu^{\ell,\wp}=\mu^{\ell}\circ\cO_\wp^{\ell}$
with $\cO_\wp^{\ell}:\cX^{\ell,\wp}\to\cX^{\ell}$ the unpacking operator.
On the other hand, we consider as original 
lattice model the renormalized model $\nu^{(\ell)}$ on $\cL^{(\ell)}$ 
with configuration space $\cM^{(\ell)}$ and algebra of events $\cB^{(\ell)}$,
see Subsection~\ref{s:bat},
and we construct the block spin space $\cM^{(\ell),\wp}$, its 
$\sigma$--algebra $\cB^{(\ell),\wp}$ and the measure 
$\nu^{(\ell),\wp}=\nu^{(\ell)}\circ\cO_\wp^\ell$,
with $\cO_\wp^{\ell}:\cM^{(\ell),\wp}\to\cM^{(\ell)}$ the unpacking operator.
The elements of $\cM^{(\ell)}$ will be denoted by $m$, and by $n$ those of
$\cM^{(\ell),\wp}$.

\subsec{The constrained models}{s:sccm}
\par\noindent
Let $\ell$ be the size of the BAT transformation, see Subsection~\ref{s:bat},
and pick a configuration of renormalized variables and $m\in\cM^{(\ell)}$.
We define the {\it single site constrained configuration space}
\begin{equation}
\label{dOni} 
\cX^{(\ell)}_{m,i}:=
     \big\{\zeta\in\cX^{\ell}_i:\, M_i^{(\ell)}(\zeta)=m_i\big\}
            \subset\cX^{\ell}_i
\end{equation}
which will be equipped with the discrete topology. 
For $I\subset\cL^{(\ell)}$ we consider the 
{\it constrained configuration space} 
$\cX^{(\ell)}_{m,I}:=\bigotimes_{i\in I}\cX^{(\ell)}_{m,i}
                \subset\cX^{\ell}_I$
equipped with the product topology; 
we remark that $\bigcup_{m\in\cM^{(\ell)}}\cX^{(\ell)}_{m,I}=\cX^{(\ell)}_I$.
As usual we let 
$\cX^{(\ell)}_{m,\cL^{(\ell)}}=:\cX^{(\ell)}_m$ and denote by 
$\cF^{(\ell)}_m$ the Borel $\sigma$--algebra 
of $\cX^{(\ell)}_m$; for each $I\subset\subset\cL^{(\ell)}$ we set 
$\cF^{(\ell)}_{m,I}:=\sigma\{\zeta_i\in\cX^{(\ell)}_{m,i},\,i\in I\}
 \subset\cF^{(\ell)}_m$. 
Finally, we consider the block spin potential $U^{(1),\ell}\equiv U^\ell$
constructed as in (\ref{spipot}) starting from the lattice gas potential 
$U^{(1)}\equiv U=(z,U_{>1})$.

We consider, now, $\tau\in\cX^\ell$ and emphasize that $\tau$ does not
depend on the fixed $m$, in the sense that it is chosen arbitrarily in a set 
not depending on $m$. Let $I\subset\subset\cL^{(\ell)}$, 
we define the probability measure for the constrained model on $I$
with boundary condition $\tau$ as follows: for each 
$\zeta\in\cX^{(\ell)}_{m,I}$
$$
\mu^{(\ell),\tau}_{m,I}(\zeta):=
      \frac{1}{Z^{(\ell)}_{m,I}(\tau)}\, 
      e^{H^{\ell}_I(\zeta\tau_{I^\complement})}
$$
where the Hamiltonian $H_I^\ell$ is defined as in (\ref{spiham}) and the 
partition function $Z^{(\ell)}_{m,I}(\tau)$ is given by 
\begin{equation}
\label{epartf}
Z^{(\ell)}_{m,I}(\tau)
:=\sum _{\zeta\in\cX^{(\ell)}_{m,I}} e^{H^{\ell}_I(\zeta\tau_{I^\complement})}
\end{equation}
Note that the function $H^\ell_I:\cX^\ell\to\bR$ can be evaluated 
in $\zeta\tau_{I^\complement}$, indeed 
$\zeta\in\cX^{(\ell)}_{m,I}\subset\cX^{(\ell)}_I$ and 
$\tau\in\cX^\ell$ imply 
$\zeta\tau_{I^\complement}\in\cX^\ell$.

We remark that the function 
$Z^{(\ell)}_{m,I}(\cdot)\in\cF^\ell_{I^\complement}$
can be looked at as the partition function of a not
translationally invariant finite volume system which is the original lattice 
gas {\it constrained} to have fixed values 
$\rho|Q_{\ell}|+m_i\sqrt{|Q_{\ell}|\chi}$
of the total number of
particles in each block $Q_\ell(i)$ for all $i\in I$.
Its elementary 
variables are the original spin configurations in each block $Q_\ell(i)$
compatible with the assigned value $m_i$ namely, the 
set $\cX^{(\ell)}_{m,I}$ defined in (\ref{dOni}).
Finally we note that for each $\tau\in\cX^\ell$ we have 
$m\mapsto Z^{(\ell)}_{m,I}(\tau)\in\cB^{(\ell)}_I$.

The finite volume renormalized measure $\nu^{(\ell),\tau}_I$, introduced in 
Subsection~\ref{s:bat}, 
which is a probability measure on $\cM^{(\ell)}_I$, 
can be written in the Gibbsian form w.r.t.\ to the renormalized Hamiltonian given by
$\log Z^{(\ell)}_{m,I}(\tau)$.
Our aim will then be to compute the partition function 
$Z^{(\ell)}_{m,I}(\tau)$ for given $m\in\cM^{(\ell)}$. 
More precisely we are interested in finding an expression for 
$\log Z^{(\ell)}_{m,I}(\tau)$ that allows to extract the renormalized 
potential with a procedure having sense in the thermodynamics limit.

\begin{table}
\label{tab:def}
\begin{center}
\begin{tabular}{l|c|c|c}
\hline\hline
 & object model & constrained model & image model\\
\hline\hline
lattice & $\Lambda\subset\cL=\bZ^d$ 
        & $I\subset\cL^{(\ell)}=(\ell\bZ)^d$ 
        & $I\subset\cL^{(\ell)}=(\ell\bZ)^d$
        \\
\hline
configuration space & $\cX_\Lambda=\{0,1\}^\Lambda$
                 & $\cX^{(\ell)}_{m,I}=
                              \bigotimes_{i\in I}\cX^{(\ell)}_{m,i}$
                 & $\cM^{(\ell)}=\bigotimes_{i\in I}\cM_i^{(\ell)}$
                 \\
\hline
$\sigma$--algebra & $\cF_\Lambda$
                  & $\cF^{(\ell)}_{m,I}$
                  & $\cB^{(\ell)}_I$
                  \\
\hline
measure & $\mu_\Lambda^\sigma$
        & $\mu^{(\ell),\tau}_{m,I}$
        & $\nu^{(\ell),\tau}_I$
        \\
\hline\hline
\end{tabular}
\end{center}
\caption{Notation for the object, constrained and image model; in the table 
$m\in\cM^{(\ell)}$ is a given renormalized configuration, $\sigma\in\cX$ and 
$\tau\in\cX^{(\ell)}$ are fixed boundary conditions.}
\end{table}

\subsec{On goodness and badness}{s:fs}
\par\noindent
As mentioned at the end of Subsection~\ref{s:spm} technical reasons,
connected to the computation developed in Section~\ref{s:cegp} below, force
the introduction of the rougher scale $\wp=\cionc d\ell$.
We then pack the renormalized variables $m_i$ 
lying inside cubes of $\cL^{(\ell)}$ of side length $\cionc d$ to form 
a renormalized block spin $n_t$, with $t\in\cL^{(\wp)}$. More precisely 
we consider the block spin space $\cM^{(\ell),\wp}$, its 
$\sigma$--algebra $\cB^{(\ell),\wp}$ and the measure 
$\nu^{(\ell),\wp}=\nu^{(\ell)}\circ\cO_\wp^\ell$,
with $\cO_\wp^{\ell}:\cM^{(\ell),\wp}\to\cM^{(\ell)}$ the unpacking operator.

We define, now, the good part of the lattice $\cL^{(\wp)}$. 
We fix $\delta\in(0,1/6)$
and $n\in\cM^{(\ell),\wp}$; recall $\chi$ has been defined in 
(\ref{chi}), we set 
\begin{equation}
\label{igood}
\cL^{(\wp)}_\delta(n)\equiv\cL^{(\wp)}_\delta:=
\{t\in\cL^{(\wp)}:\,
               |(\cO_\wp^\ell n)_i|\le\bonta
               \textrm{ for all } i\in Q_{\wp/\ell}^{(\ell)}(t)
\}
\end{equation}
We say that a site $t\in\cL^{(\wp)}$ is {\it good} 
w.r.t\ $n\in\cM^{(\ell),\wp}$
if $t\in\cL^{(\wp)}_\delta(n)$; if $t$ is not good we say it is {\it bad}. 
Loosely speaking a cube of side $\cionc d\ell$ of the original lattice 
is good if the empirical density 
in {\it all} its $\cionc d^d$ sub--cubes of side 
$\ell$ differs from the infinite volume mean $\rho$ less than 
$\ell^{-d(1/3+\delta)}$; this choice ensures the validity of the 
central limit theorem inside the good blocks, see 
\cite[Theorem~4.5]{[BCO]}.

\subsec{On the goodness of good sites}{s:wgg}
\par\noindent
As we have already discussed in Subsection~\ref{s:str} our 
strategy of proof consists in performing a cluster expansion, 
similar to the one used in \cite{[BCO]}, in the good region of the lattice 
and to use the sparseness of the bad sites
to carry out the sum over the {\it bad} part of the lattice. 
In this subsection we deduce the property of the good blocks that 
will enable us to cluster expand the partition function of the constrained 
models in this region. 

We recall that $\ell$ is the scale of the renormalization 
transformation. Let $i\in\cL^{(\wp)}$ and $k\in\{1,\dots,d\}$,
we denote by $P^{i,k}$ the family of all not empty subsets 
$I\subset\cL^{(\wp)}$ such that for each 
$j\in I$ we have 
$j_k=i_k$ and $j_h\in\{i_h-\wp,i_h,i_h+\wp\}$ for all $h=1,\dots,d$ and
$h\neq k$. We set 
$$
I_\pm:=\partial^{(\wp)} I\cap\{j\in\cL^{(\wp)}:\, j_k=i_k\pm \wp\}
$$
and, for $m\in\cM^{(\ell)}$ and $\sigma\in\cX^\ell$, 
$\sigma_\pm:=\sigma_{\cO_{\wp}^\ell I_\pm}$ and 
$\sigma_0:=\sigma_{\cO_{\wp}^\ell(I_+\cup I_-)^\complement}$. 
Recall $\delta>0$ has been introduced in (\ref{igood}),
given $J\subset\cL^{(\ell)}$ we set 
$\cD^{(\ell)}_\delta(J):=\{m\in\cM^{(\ell)}:\,
                         |m_j|\le\bonta,\,j\in J\}$.

\begin{teo}
\label{t:bonta}
Let the lattice gas potential $U$ satisfy Condition SM$(\ell_0,b,B)$.
We have that there exists a real $C=C(\delta,\ell_0,b,B,\|U\|_0,r,d)<\infty$ 
such that for each $\ell$ multiple of $\ell_0$ and 
$i\in\cL^{(\wp)}$ we have
\begin{equation}
\label{CC}
\sup_{k=1,\dots,d}~
\sup_{I\in P^{i,k}}~
\sup_{m\in\cD^{(\ell)}_\delta(\cO_{\wp}^\ell I)}~
\sup_{\s,\z,\t\in\cX^\ell}~ 
\left| {\frac{Z_{m,\cO_{\wp}^\ell I}^{(\ell)}\(\s_+\s_-\t_0\) 
              Z_{m,\cO_{\wp}^\ell I}^{(\ell)}\(\z_+\z_-\t_0\)} 
             {Z_{m,\cO_{\wp}^\ell I}^{(\ell)}\(\s_+\z_-\t_0\)  
              Z_{m,\cO_{\wp}^\ell I}^{(\ell)}\(\z_+\s_-\t_0\)}
}
-1\right|
\le \frac{C}{\ell} 
\end{equation}
\end{teo}

To prove the above theorem we shall use 
Lemma~\ref{t:cruciale} below in which it is proven that 
the strong mixing condition holds uniformly in the activity. To state 
precisely such a property we introduce the notion of 
{\it lattice gas with not homogeneous activity}:
consider the configuration space $\cX:=\{0,1\}^{\cL}$ of the lattice gas,
the Borel $\sigma$--algebra $\cF$, see Subsection~\ref{s:lg}, and the 
family of local functions 
$U_{>1}=\{U_X\in\cF_X,\,X\subset\subset\cL,|X|>1\}$.
Let $\underline{z}:= \{z_x\in [0,\infty),\, x\in\cL\}$, the 
lattice gas potential with not homogeneous activity is the family of functions
$U^{\ul{z}}:=\{U_X\in\cF_X,\,X\subset\subset\cL\}$ with 
$$
U^{\underline{z}}_X(\eta):=\left\{
\begin{array}{ll}
\eta_x \log z_x &\textrm{ if there exists } x\in\cL\textrm{ such that }X=\{x\}\\
U_X(\eta) &\textrm{ if }|X|>1\\ 
\end{array}
\right.
$$
for all $\eta\in\cX$; we shall use the notation 
$U^{\ul{z}}:=(\ul{z},U_{>1})$. Given $\Lambda\subset\subset\cL^{(\ell)}$
and $\tau\in\cX$,
the finite volume Gibbs measure with boundary condition $\tau$ 
associated with the lattice gas potential with not homogeneous activity 
$U^{\ul{z}}$ is denoted by $\mu_{\Lambda,\ul{z}}^\tau$ and the corresponding 
partition function by $Z_{\Lambda,\ul{z}}^\tau$. Namely, we have 
\begin{equation}
\label{dmgcvf}
Z_{\Lambda,\ul{z}}^\tau:=
\sum_{\eta\in\cX_\Lambda}
 \exp\Big\{\sum_{\newatop{X\cap\L\neq\emptyset:}
                         {|X|>1}}
            U_X(\eta\tau_{\Lambda^\complement})
           +\sum_{x\in\Lambda}\eta_x\log z_x 
\Big\}
\end{equation}

It is easy to see that by \cite{[DS2],[DS3]}, see Remark 2 
in \cite[p.\ 849]{[BCO]}, the following lemma, stating that the strong
mixing condition (\ref{e:sm}) is satisfied uniformly in the activities, holds. 

\begin{lem}
\label{t:cruciale}
Let the lattice gas potential $U=(z,U_{>1})$ satisfy
Condition SM$(\ell_0,b,B)$.
Then there exist $\varepsilon>0$, $\ell'_0$ multiple of $\ell_0$, 
and two positive reals $b'=b'(\varepsilon,b,B,\ell_0)$ and 
$B'=B'(\varepsilon,b,B,\ell_0)<\infty$ 
such that for $\ul{z}=\{z_x\in[0,\infty):\,x\in\cL\}$ such that 
$|z_x-z|\le\varepsilon$, for all $x\in\cL$,
the lattice gas potential with not homogeneous activity 
$U^{\ul{z}}=(\ul{z},U_{>1})$ satisfies SM$(\ell'_0,b',B')$. 
\end{lem}

\noi{\it Proof of Theorem~\ref{t:bonta}.}\  
Let $\mu_z$ be the unique infinite volume Gibbs measure of the 
lattice gas with potential $U=(z,U_{>1})$ and set 
$\varrho:(0,+\infty)\ni z\longrightarrow\varrho(z):=\mu_z(\eta_0)\in(0,1)$.
Let $\varepsilon>0$ be as in Lemma~\ref{t:cruciale}, 
by the continuity of $\varrho$ we can choose 
$\varepsilon'>0$ such that 
$\varrho^{-1}\left(\varrho(z)-2\varepsilon',\varrho(z)+2\varepsilon'\right)
      \subset[z-\varepsilon,z+\varepsilon]$.
The thesis follows by Lemma~\ref{t:cruciale} and \cite[Prop.\ 5.1]{[BCO]}. 
\qed

\subsec{On the sparseness of bad sites}{s:bbs}
\par\noindent
In this subsection we state precisely in which sense the bad sites 
in $\cL^{(\wp)}$ are sparse. 
We define the map $\pi:\cM^{(\ell),\wp}\to\{0,1\}^{\cL^{(\wp)}}$ by setting 
for each $n\in\cM^{(\ell),\wp}$ and $t\in\cL^{(\wp)}$ 
\begin{equation}
\label{pizza}
\big(\pi(n)\big)_t:=
\left\{
\begin{array}{ll}
0 & \textrm{if } t\in\cL^{(\wp)}_\delta(n)\\
1 & \textrm{otherwise}\\
\end{array}
\right.
\end{equation}

As a first step we show that the probability that a site
is bad is exponentially small in $\ell$. 
\begin{teo}
\label{t:futuro}
Let the lattice gas potential $U=(z,U_{>1})$ satisfy Condition 
SM$(\ell_0,b,B)$. Then  
there exists a real $C=C(\varepsilon,\ell_0,b,B)>0$ 
such that for any positive integer $\ell$ we have that 
\begin{equation}
\label{ppte}
\sup_{t\in\cL^{(\wp)}}
\nu^{(\ell),\wp}((\pi(n))_t=1)\le\exp\{-C\ell^{(1/3-2\delta)d}\}
\end{equation}
\end{teo}

\noi{\it Proof.}\  
We have
$$
\begin{array}{rcl}
{\displaystyle 
 \sup_{t\in\cL^{(\wp)}}
 \nu^{(\ell),\wp}((\pi(n))_t=1)
}
&\!=\!&
{\displaystyle 
 \sup_{t\in\cL^{(\wp)}}
 \nu^{(\ell)}\big(\exists i\in Q_{\wp/\ell}^{(\ell)}(t):\,
                  |m_i|>\bonta\big)
} \\
&\le&
{\displaystyle
 \cionc d^d
 \sup_{i\in\cL^{(\ell)}}\mu\big(|M_i^{(\ell)}|>\bonta\big)
}\\
\end{array}
$$
We pick $i\in\cL^{(\ell)}$.
To bound the right hand side of the above inequality, we 
recall (\ref{Centr}), consider $L>\ell$ integer, set 
$\Delta_L(i):=\{x\in\cL:\,\dis(x,Q_\ell(i))\le L\}$, 
and use the exponential Chebyshev inequality, with $h\ge0$, as follows
\begin{equation}
\label{hjyt}
\begin{array}{rl}
{\displaystyle 
 \mu\big(M_i^{(\ell)}>\bonta\big)
}
&
{\displaystyle 
 =
 \mu\Big(\sum_{x\in Q_\ell(i)}(\eta_x-\rho)>\ell^{(2/3-\delta)d}\Big)
 \vphantom{\bigg|}
}\\
&
{\displaystyle 
 \!\!\!\!\!\!\!\!\!\!\!\!\!\!\!\!\!\!\!\!\!\le
 e^{-h\ell^{(2/3-\delta)d}}\,
 \mu\Big(\exp\big\{h\sum_{x\in Q_\ell(i)}(\eta_x-\rho)\big\}\Big)
 \vphantom{\bigg|}
}\\
&
{\displaystyle 
 \!\!\!\!\!\!\!\!\!\!\!\!\!\!\!\!\!\!\!\!\!=
 e^{-h\ell^{(2/3-\delta)d}}\,
 \int\!\mu(d\tau)\,
 \mu^\tau_{\Delta_L(i)}\,
  \Big(\exp\big\{h\sum_{x\in Q_\ell(i)}(\eta_x-\rho)\big\}\Big)
 \vphantom{\bigg|}
}\\
&
{\displaystyle 
 \!\!\!\!\!\!\!\!\!\!\!\!\!\!\!\!\!\!\!\!\!=
 e^{-h\ell^{(2/3-\delta)d}}\,
 \int\!\mu(d\tau)\,
  \exp\big\{\log Z_{\Delta_L(i),\ul{z}(i,h)}^\tau
           -\log Z_{\Delta_L(i),\ul{z}(i,0)}^\tau-h\rho\ell^d\big\}
 \vphantom{\bigg|}
}\\
\end{array}
\end{equation}
where we used the DLR equations (\ref{dlr0}) and,
for $\tau\in\cX$ and $\Lambda\subset\subset\cL$,  
we have considered the partition function 
\begin{equation}
\label{uyte}
Z_{\Lambda,\ul{z}(i,h)}^\tau=
  \sum_{\eta\in\cX_\Lambda} 
    \exp\Big\{H_\Lambda(\eta\tau_{\Lambda^\complement})
              +h\sum_{x\in\Lambda\cap Q_\ell(i)}\eta_x\Big\}
\end{equation}
of a lattice gas with not homogeneous activity $\ul{z}(i,h)$ such that 
$z_x(i,h)=z e^h$ for all $x\in Q_\ell(i)$ and 
$z_x(i,h)=z$ otherwise; recall $z$ is the activity of the original lattice gas. 
Note that $Z_{\Lambda,\ul{z}(i,0)}^\tau$ coincides with the partition 
function $Z_\Lambda^\tau$ of the original lattice gas. 

{}From the strong mixing condition SM$(\ell_0,b,B)$ it follows that there 
exist two positive reals $C_1(\ell_0,b,B)<\infty$ and $C_2(\ell_0,b,B)>0$ 
such that for any $L$ multiple of $\ell_0$
and $\tau\in\cX$ we have
\begin{equation}
\label{iuyrp}
\bigg|\bigg(\frac{d\log Z_{\Delta_L(i),\ul{z}(i,h)}^\tau}{dh}
      \bigg)_{h=0}-\rho\ell^d\bigg|=
\bigg|\mu_{\Delta_L(i)}^\tau\Big(\sum_{x\in Q_\ell(i)}\eta_x\Big)-
     \mu\Big(\sum_{x\in Q_\ell(i)}\eta_x\Big)\bigg|\le
C_1\,\ell^d\,e^{-C_2(L-\ell)}
\end{equation}
By Lemma~\ref{t:cruciale} there exist $\varepsilon>0$, $\ell_0'$ 
multiple of $\ell_0$, and the two reals 
$b'=b'(\varepsilon,b,B,\ell_0)$ and $B'=B'(\varepsilon,b,B,\ell_0)$
such that the perturbed lattice gas potential satisfies 
SM$(\ell_0',b',B')$ for all $0\le h\le\varepsilon$. 
Hence, if $L$ is a multiple of $\ell_0'$ 
we have that there exists a real 
$0<C_3=C_3(\varepsilon,\ell_0,b,B)<\infty$ 
such that for any $h\in[0,\varepsilon]$ and $\tau\in\cX$ 
the following bound holds
\begin{equation}
\label{oery}
\bigg|\frac{d^2\log Z_{\Delta_L(i),\ul{z}(i,h)}^\tau}{dh^2}\bigg|=
\bigg|\sum_{x,y\in Q_\ell(i)}
      \mu_{\Delta_L(i),\ul{z}(i,h)}^\tau(\eta_x;\eta_y)\bigg|\le
2C_3\,\ell^d
\end{equation}
where we recall $\mu_{\Lambda,\ul{z}}^\tau$, 
for $\Lambda\subset\subset\cL$ and $\ul{z}\in[0,\infty)^\cL$,
is the finite volume Gibbs measure of the lattice gas with not homogeneous 
activity $\ul{z}$ and boundary condition $\tau\in\cX$. 
By expanding the exponent on the right hand side of (\ref{hjyt}) and
using (\ref{iuyrp}) and (\ref{oery}) we get 
\begin{equation}
\label{ljfhp}
\mu\big(M_i^{(\ell)}>\bonta\big)
\le
\exp\{-h\ell^{(2/3-\delta)d}+C_1\,\ell^d\,e^{-C_2(L-\ell)}
      +h^2C_3\,\ell^d\}
\end{equation}
Taking the limit $L\to\infty$ we finally get 
\begin{equation}
\label{ljfh}
\mu\big(M_i^{(\ell)}>\bonta\big)
\le
\exp\{-h(\ell^{(2/3-\delta)d}-hC_3\,\ell^d)\}
\end{equation}
The bound (\ref{ppte}) follows by choosing $h=\ell^{-(1/3+\delta)d}/(2C_3)$;
indeed the steps in (\ref{hjyt}) can be repeated to 
bound $\mu\big(M_i^{(\ell)}<-\bonta\big)$. 
\qed

In Theorem~\ref{t:partD} below we shall 
state that the bad sites of $\cL^{(\wp)}$ are 
sparse in the following sense. There exists a full measure subset of 
$\cM^{(\ell),\wp}$, such that for each $n$ in such a set there 
exists a gentle disintegration, see Definition~\ref{t:gentle}, 
of the lattice $\cL^{(\wp)}$ with respect to its good part 
$\cL^{(\wp)}_\delta$ and two suitable moderately steep scales 
$\Gamma,\gamma$. The two sequences are chosen as in 
\cite[Remark~2.3]{[BCOabs]} namely, given $\beta\ge9$ we set 
$\Gamma_0=\gamma_0:=0$, 
\begin{equation}
\label{okscale}
\Gamma_k:=e^{(\beta+1)(3/2)^k}\;\;\;\;\;\textrm{ and }\;\;\;\;\;
\gamma_k:=\frac{1}{8}e^{\beta(3/2)^{k+1}}
\;\;\;\;\;\textrm{ for } k\ge1
\end{equation}
Those sequences are steep scales namely, they satisfy 
items~\ref{seq:<}--\ref{seq:covol} in Definition~\ref{t:seq}. 
Moreover, see the remark below Theorem~2.5 in \cite{[BCOabs]}, we 
choose $\beta$ large enough so that the supplementary 
conditions on the steep scales in the hypotheses of
\cite[Theorem~2.5]{[BCOabs]} are satisfied with $\alpha=1$. 
It is easy to prove that for 
\begin{equation}
\label{okscale1}
\varepsilon\in(1/2,1)
\;\;\;\;\textrm{ and }\;\;\;\;
a\ge9d\beta/2
\end{equation}
the steep scales $\Gamma,\gamma$ are moderate namely, 
they also fulfill items \ref{seq:con-ano}--\ref{seq:con-icf} 
in Definition~\ref{t:seq}. The conditions above on $\beta$ and $a$ 
are met  for $\ell$ integer large enough if we set  
\begin{equation}
\label{okscale2}
a\equiv a_\delta(\ell):=\frac{9}{2}d\big[\beta\vee\ell^{(1/3-2\delta)d/2}\big]
\end{equation}
where $\delta\in(0,1/6)$ is the real number which has been picked up 
before (\ref{igood}) to define the good part of the lattice $\cL^{(\wp)}$.

In order to prove Theorem~\ref{t:conv}, we had to choose the parameter $a$ 
diverging with the renormalization scale $\ell$. In fact we shall need 
that the probability that a site of $\cL^{(\wp)}$ belongs to a 
$k$--gentle atom vanishes fast enough as $\ell\to\infty$. 
On the other hand the existence of the gentle disintegration of $\cL^{(\wp)}$
is proven on the basis of Theorem~\ref{t:fm} whose hypotheses are satisfied 
if the probability $p$ for a site to be bad is smaller than 
$\exp\{-a/(1-\varepsilon)\}$. In our application this probability is estimated 
with the stretched exponential in (\ref{ppte}); to ensure that for 
$\ell$ large enough $p$ be smaller than  
$\exp\{-a/(1-\varepsilon)\}$ the function $a_\delta(\ell)$ 
must diverge sufficiently slow. The choice (\ref{okscale2}) meets both the 
above requirements. 

\begin{teo}
\label{t:partD}
Let the lattice gas potential $U$ satisfy Condition SM$(\ell_0,b,B)$.
Consider the two moderately steep scales $\Gamma,\gamma$ defined in 
(\ref{okscale}).
Then for each $\ell$ large enough multiple of $\ell_0$ 
there exists a $\cB^{(\ell),\wp}$--measurable subset 
$\bar\cM^{(\ell),\wp}\subset\cM^{(\ell),\wp}$ 
with $\nu^{(\ell),\wp}(\bar\cM^{(\ell),\wp})=1$
such that
\begin{enumerate}
\item\label{i:partD2}
for each $n\in\bar\cM^{(\ell),\wp}$ there exists a gentle disintegration 
$\cG(n)$, see Definition~\ref{t:gentle},
of $\cL^{(\wp)}$ with respect to $\bG_0(n):=\cL_\delta^{(\wp)}(n)$, 
$\Gamma$, and $\gamma$.
\item\label{i:partD3}
for each $t\in\cL^{(\wp)}$ and  
$X\subset\subset\cL^{(\wp)}$ we have that 
$\left\{n:\,\bG_k(n)\ni t\right\}
 \in\cB^{(\ell),\wp}_{B_{\vartheta_k}^{(\wp)}(t)}$ 
and also 
$\left\{n:\; \cG_k(n)\ni X\right\}\in
\cB^{(\ell),\wp}_{B_{\vartheta_k}^{(\wp)}(X)}$.
\end{enumerate} 
\end{teo}

\noi{\it Proof.}\  
We use the setup of Section~\ref{s:cattivi} with $\bL=\wp^{-1}\cL^{(\wp)}$.
Recall the map $\pi:\cM^{(\ell),\wp}\to\{0,1\}^{\cL^{(\wp)}}$ has been 
defined in (\ref{pizza}).
Note that for each $x,y\in\bL$ we have
$$
\disuno(x,y)=\sum_{i=1}^d|x_i-y_i|\le d\sup_{i\in\{1,\dots,d\}}|x_i-y_i|
=(d/\wp)\dis_\wp(\wp x,\wp y)
$$
From Condition~SM$(\ell_0,b,B)$ it follows that the measure  
$\bQ:=\nu^{(\ell),\wp}\circ\pi^{-1}$ 
on the set $\Omega:=\{0,1\}^{\cL^{(\wp)}}$, 
endowed with its Borel $\sigma$--algebra $\cA$, satisfies the 
bound (\ref{smin}) with constants $b''=\wp b/d$ and $B''=\wp^dB$. 
By taking $\ell$ large enough the scales $\Gamma,\gamma$ in (\ref{okscale}) 
satisfy (\ref{gag}). Moreover by Theorem~\ref{t:futuro} 
$$
p:=\sup_{t\in\cL^{(\wp)}}\bQ\left(\{\omega:\,\omega_t=1\}\right)=
   \sup_{t\in\cL^{(\wp)}}\nu^{(\ell),\wp}\left(\{n:\,(\pi(n))_t=1\}\right)
   \le e^{-C\ell^{(1/3-2\delta)d}}
$$
We can therefore apply Theorem~\ref{t:fm}, we set 
$\bar\cM^{(\ell),\wp}:=\pi^{-1}(\bar\Omega)$. 
The thesis follows by noticing that for each $X\subset\cL^{(\wp)}$
we have $B^{(\wp)}_s\supset \wp O_s(\wp^{-1}X)$ for all $s>0$.
\qed

\sezione{Cluster expansion in the good part of the lattice}{s:cegp}
\par\noindent
In this section we start to compute the renormalized potentials; our 
main technique, as in \cite{[BCO]}, will be the scale adapted
cluster expansion.  

Let $\ell$ be the renormalization scale 
recall $\wp=\cionc d\ell$; for $m\in\cM^{(\ell)}$ the 
set $\cL^{(\wp)}_\delta(\cO_\ell^\wp m)
         \equiv\cL^{(\wp)}_\delta\subset\cL^{(\wp)}$ 
has been defined in (\ref{igood}). 
Pick $\Lambda\subset\subset\cL^{(\wp)}$, a configuration of the renormalized 
variables $m\in\cM^{(\ell)}$, and a boundary condition 
$\tau\in\cX^\ell$; set $J:=\cO_\wp^\ell\Lambda$,
$\Lambda_\delta:=\Lambda\cap\cL_\delta^{(\wp)}$
and $J_\delta:=\cO_\wp^\ell\Lambda_\delta$.
We write
$$
Z_{m,J}^{(\ell)}(\tau) 
 =\sum_{\eta\in\cX_{m,J}^{(\ell)}} 
   \exp\{H_J^{\ell}(\eta\tau_{J^\complement})\} 
 =\sum_{\sigma\in\cX_{m,J\setminus J_\delta}^{(\ell)}} 
   \sum_{\eta\in\cX_{m,J_\delta}^{(\ell)}} 
   \exp\{H_J^{\ell}(\sigma\eta\tau_{J^\complement})\} 
$$
In this section we fix 
$\sigma\in\cX^{(\ell)}_{m,J\setminus J_\delta}$ 
and compute the partition function associated to the good part 
$J_\delta$ of the set $J$ namely, we compute
\begin{equation}
Z^{(\ell)}_{m,J_\delta}\(\sigma\tau_{J^\complement}\)=  
  \sum_{\eta\in\cX^{(\ell)}_{m,J_\delta}}
  \exp\{H_{J_\delta}^{\ell}(\eta\sigma\tau_{J^\complement})\}
\label{Zconstr}
\end{equation}

We rewrite this problem on the scale $\wp$, that is we apply 
the procedure described in Subsection~\ref{s:spm} to the constrained models
introduced in Subsection~\ref{s:sccm} on the scale $\ell$ to group the 
block spin variables on the scale $\wp$. 
We fix a configuration $n\in\cM^{(\ell),\wp}$, the corresponding renormalized 
configuration is $m\equiv m(n):=\cO_\wp^\ell n$. We recall the notion of 
constrained model defined on the configuration space $\cX^{(\ell)}_m$ and, 
via the procedure discussed in Subsection~\ref{s:spm}, we construct  
the configuration space $\cX^{(\ell),\wp}_m$ and its 
$\sigma$--algebra $\cF^{(\ell),\wp}_m$; for $\Lambda\subset\subset\cL^{(\wp)}$
we set 
$\cF^{(\ell),\wp}_{m,\Lambda}:=
 \{\zeta_i\in\cX^{(\ell),\wp}_{m,i},\,i\in\Lambda\}\subset\cF^{(\ell),\wp}_m$.

Finally we consider the potential $U^{(1),\ell,\wp}\equiv U^{\ell,\wp}$, 
obtained by applying the procedure in Subsection~\ref{s:spm} to 
the original lattice gas potential introduced in Subsection~\ref{s:lg},
and, given $\Lambda\subset\subset\cL^{(\wp)}$, 
we consider the Hamiltonian $H^{\ell,\wp}_\Lambda$ 
and the self--interaction
$E^{\ell,\wp}_\Lambda$; we obviously have that for each 
$\zeta\in\cX^{\ell,\wp}$ 
$$
H_\Lambda^{\ell,\wp}(\zeta)=
H_{\cO_\wp^\ell\Lambda}^\ell(\cO_\wp^\ell\zeta)=
H_{\cO_\ell\cO_\wp^\ell\Lambda}(\cO_\ell\cO_\wp^\ell\zeta)
$$
For $\Lambda\subset\subset\cL^{(\wp)}$
we can then write the finite volume Gibbs measure with boundary condition 
$\xi\in\cX^{\ell,\wp}$ as 
$$ 
\mu^{(\ell),\wp,\xi}_{m,\Lambda}(\zeta):=
      \frac{1}{Z^{(\ell),\wp}_{m,\Lambda}(\xi)}\, 
      e^{H^{\ell,\wp}_\Lambda(\zeta\xi_{\Lambda^\complement})}
\quad\quad\quad\quad\zeta\in\cX^{(\ell),\wp}_{m,\Lambda} 
$$
The partition function above is given by 
\begin{equation}
\label{partlp}
Z^{(\ell),\wp}_{m,\Lambda}(\xi)
:=\sum _{\zeta\in\cX^{(\ell),\wp}_{m,\Lambda}} 
    e^{H^{\ell,\wp}_\Lambda(\zeta\xi_{\Lambda^\complement})}
\end{equation}
Note that the boundary condition $\xi\in\cX^{(\ell),\wp}$ is chosen 
independently of the renormalized configuration $m$.
We have that 
$Z^{(\ell),\wp}_{m,\Lambda}(\cdot)\in\cF^{\ell,\wp}_{\Lambda^\complement}$ 
and 
$m\mapsto Z^{(\ell),\wp}_{m,\Lambda}(\xi)\in\cB^{(\ell),\wp}_\Lambda$. 

It is easy to show that 
$Z^{(\ell)}_{m,J_\delta}\(\sigma\tau_{J^\complement}\)=  
Z^{(\ell),\wp}_{m,\Lambda_\delta}(\cO_\ell^\wp(\sigma\tau_{J^\complement}))$.
In the following theorem we shall denote by 
$\xi:=\cO_\ell^\wp\big(\sigma\tau_{J^\complement}\big)$ 
the block spin configuration outside 
$\Lambda_\delta=\Lambda\cap\cL^{(\wp)}_\delta$.

\bteo{t:ip3}
Let the lattice gas potential $U$ satisfy Condition SM$(\ell_0,b,B)$.
Then for each $\ell$ large enough 
multiple of $\ell_0$, $n\in\cM^{(\ell),\wp}$,
and $\Lambda\subset\subset\cL^{(\wp)}$ there exist a family 
of local functions 
$\big\{V^{(\ell),\wp}_{X,\Lambda}(\cdot,n):
  \cX^{\ell,\wp}\to\bR,\,X\subset\subset\cL^{(\wp)}\big\}$,
a real $K^{(\wp)}_\Lambda$, and an integer $\kappa$ such that 
\begin{enumerate}
\item\label{i:ipo03}
for any $\xi\in\cX^{\ell,\wp}$ 
we have the absolutely convergent expansion
\begin{equation}
\label{ipo333}
\log Z_{\cO_\wp^\ell n,\Lambda_\delta}^{(\ell),\wp}(\xi)=
K^{(\wp)}_\Lambda-
\frac{1}{2}\sum_{i\in\cO_\wp^\ell\Lambda}(\cO_\wp^\ell n)_i^2+
\sum_{\newatop{X\subset\subset\cL^{(\wp)}:}
              {X\cap\Lambda\neq\emptyset}}
 V^{(\ell),\wp}_{X,\Lambda}(\xi,n)
\end{equation}
where $V^{(\ell),\wp}_{X,\Lambda}(\cdot,n)$ is constant if 
$X\cap\Lambda_\delta=\emptyset$; moreover,
$V^{(\ell),\wp}_{X,\Lambda}(\cdot,n)=0$ if 
$X\cap\Lambda_\delta=\emptyset$ and $\diam_{\wp}(X)>\kappa$.
\end{enumerate}
For any $X\subset\subset\cL^{(\wp)}$
\begin{enumerate}
\setcounter{enumi}{1}
\item\label{i:ipo13}
we have that 
$V^{(\ell),\wp}_{X,\Lambda}(\cdot,n)
   \in\cF_{X\cap\Lambda_\delta^\complement}^{\ell,\wp}$;
\item\label{i:ipo13.5}
if $X$ is not $\wp$--connected then 
$V^{(\ell),\wp}_{X,\Lambda}(\cdot,n)=0$;
\item\label{i:ipo53}
if $X\cap\Lambda_\delta\neq\emptyset$ we have that 
$X\cap\big(\srclos{\wp}{\kappa}{\Lambda_\delta}\big)^\complement\neq\emptyset$
implies
$V_{X,\Lambda}^{(\ell),\wp}(\cdot,n)=0$;
\end{enumerate}
Moreover
\begin{enumerate}
\setcounter{enumi}{4}
\item\label{i:ipo43}
there exist reals $\alpha_1>0$ and $A_1<\infty$ 
depending on $\ell_0$, $b$, $B$, $\|U\|$, $r$, $d$, and $\delta$ such that 
we have 
\be{ip233}
\sup_{x\in\cL^{(\wp)}}\,
 \sum_{\newatop{X\subset\subset\cL^{(\wp)}:}
               {X\ni x}}
   e^{\alpha_\ell\tree_\wp(X)}
   \sup_{\Lambda\subset\subset\cL^{(\wp)}}
    \|V^{(\ell),\wp}_{X,\Lambda}(\cdot,n)\|_{\infty}
\le A_\ell
\end{equation}
where we have set $\alpha_\ell:=\alpha_1\log(e\ell)$ and 
$A_\ell:=A_1\,\ell^{(\kappa+1)^d\alpha_1+d}$;
\item\label{i:ipo43.5}
we have that 
\be{ip42.5}
\lim_{\ell\to\infty}\,
   \sup_{\Lambda\subset\subset\cL^{(\wp)}}\,
   \sup_{X\subset\Lambda}\,
   \sup_{\newatop{n\in\cM^{(\ell),\wp}:}
                 {\cL^{(\wp)}_\delta(n)\supset X}}\,
    \|V^{(\ell),\wp}_{X,\Lambda}(\cdot,n)\|_{\infty}
=0
\end{equation}
where the limit is taken along a sequence of multiples 
of $\ell_0$;
\item\label{i:ipo23}
for any  
$\Lambda,\Lambda'\subset\subset\cL^{(\wp)}$ and
$X\subset\subset\cL^{(\wp)}$ if $X\cap\Lambda=X\cap\Lambda'$
then $V^{(\ell),\wp}_{X,\Lambda}(\cdot,n)=V^{(\ell),\wp}_{X,\Lambda'}(\cdot,n)$;
\item\label{i:ipo33}
let $X,\Lambda\subset\subset\cL^{(\wp)}$ and $n,n'\in\cM^{(\ell),\wp}$ 
such that $n_X=n'_X$, then
\begin{equation}
\label{ipo33}
V^{(\ell),\wp}_{X,\Lambda}(\cdot,n)=
V^{(\ell),\wp}_{X,\Lambda}(\cdot,n')
\end{equation}
\end{enumerate}
\eteo

We have a situation very similar to the one in \cite{[BCO]}
where we considered the case of a torus; the sole difference is that, now, 
$\Lambda_\delta$ is an arbitrary finite subset of $\cL^{(\wp)}$, hence 
its boundary can be geometrically complicated.
To simplify the exposition, like in \cite{[BCO]}, 
we will treat explicitly only the two--dimensional case.
The general $d$--dimensional case can 
be treated analogously, following the methods of \cite{[OP]}.
We mention that a similar expansion has been used in \cite{[BK1]}
to study coupled maps. 

As in \cite{[BCO]} we will 
transform the constrained system, whose partition function 
is $Z^{(\ell),\wp}_{m,\Lambda_\delta}(\xi)$,
into a small activity polymer system. More precisely,
we shall prove the following formula
\begin{equation}
\label{Zpolym}
Z^{(\ell),\wp}_{m,\Lambda_\delta}(\xi)= 
\bar{Z}^{(\ell),\wp}_{m,\Lambda_\delta}(\xi) 
\Xi^{(\ell),\wp}_{m,\Lambda_\delta}(\xi) 
\end{equation}
where $\bar{Z}^{(\ell),\wp}_{m,\Lambda_\delta}(\xi)$ is 
a product of partition functions on suitable finite volumes; the
dependence on $m$ of the single factors is local. Moreover, 
the {\em reference system} around which we perform the 
perturbative expansion is described by the partition function 
$\bar{Z}^{(\ell),\wp}_{m,\Lambda_\delta}(\xi)$.
On the other hand $\Xi^{(\ell),\wp}_{m,\Lambda_\delta}(\xi)$ is the partition
function of a gas of polymers, see (\ref{Xi}) below.

The expression (\ref{Zpolym}) is well suited to compute
the renormalized potentials; in order to get good estimates on
them we need that the polymer system described by 
$\Xi^{(\ell),\wp}_{m,\Lambda_\delta}(\xi)$ is in the small
activity region thanks to the 
uniform bound in Theorem~\ref{t:bonta}.
In other words, the bound (\ref{CC}) implies that 
the finite size condition of \cite{[OP]} is satisfied
on $\Lambda_\delta$. More precisely, 
recalling the notation introduced in Subsection~\ref{s:wgg},
there exists a real $C<\infty$ 
depending on $\ell_0$, $b$, $B$, $\|U\|_0$, $r$, $d$, and $\delta$ such that 
for $\ell$ multiple of 
$\ell_0$, $m\in\cM^{(\ell)}$, and $i\in\cL^{(\wp)}$ we have
\begin{equation}
\label{CC1}
\sup_{k=1,\dots,d}~
\sup_{I\in P^{i,k}}~
\sup_{\sigma,\zeta,\tau\in\cX^{\ell,\wp}}~ 
\left| {\frac{Z_{m,I\cap\cL^{(\wp)}_\delta}^{(\ell),\wp}\(\s_+\s_-\t_0\) 
              Z_{m,I\cap\cL^{(\wp)}_\delta}^{(\ell),\wp}\(\z_+\z_-\t_0\)} 
             {Z_{m,I\cap\cL^{(\wp)}_\delta}^{(\ell),\wp}\(\s_+\z_-\t_0\)  
              Z_{m,I\cap\cL^{(\wp)}_\delta}^{(\ell),\wp}\(\z_+\s_-\t_0\)}
}
-1\right|
\le \frac{C}{\ell} 
\end{equation}

We start, now, the computation yielding the expansion (\ref{Zpolym}).
We pick $\Lambda\subset\subset\cL^{(\wp)}$ and $n\in\cM^{(\ell),\wp}$; to 
simplify the notation we set $m=\cO_\wp^\ell n$ and 
$\Delta:=\Lambda_\delta=\Lambda\cap\cL^{(\wp)}_\delta(n)$. 
Recall $e^{(\wp)}_1=(\wp,0)$ and $e^{(\wp)}_2=(0,\wp)$;
we  partition $\cL^{(\wp)}$ into the four sub--lattices
$\cA:=\cL^{(2\wp)}$,
$\cB:=\cL^{(2\wp)}+e_2^{(\wp)}$,
$\cC:=\cL^{(2\wp)}+e_1^{(\wp)}+e_2^{(\wp)}$, and
$\cD:=\cL^{(2\wp)}+e_1^{(\wp)}$.
We label the points in those sub--lattices by 
$k\in\cL^{(2\wp)}$ as follows:
$A_k:=k\in\cA$, 
$B_k:=k+e_2^{(\wp)}\in\cB$,
$C_k:=k+e_1^{(\wp)}+e_2^{(\wp)}=B_k+e_1^{(\wp)}\in\cC$,
$D_k:=k+e_1^{(\wp)}=C_k-e_2^{(\wp)}\in\cD$.
It is useful, here and in the sequel, to think to $e^{(\wp)}_1$ as horizontal 
and to $e^{(\wp)}_2$ as vertical. 

Recalling definition (\ref{spiham}) for 
$\xi\in\cX^{\ell,\wp}$, $x\in\cL^{(\wp)}$ we define the function
$E^{\ell,\wp}_{x;\Delta}(\cdot|\xi):\cX^{\ell,\wp}\to\overline{\bR}$ as 
\begin{equation}
\label{porti}
E^{\ell,\wp}_{x;\Delta}(\eta|\xi):=
\left\{
\begin{array}{ll}
E_{\{x\}}^{\ell,\wp}(\eta) & \textrm{if } x\in\Delta \\
0 & \textrm{if } x\not\in\Delta \textrm{ and } \eta_x=\xi_x\\
-\infty & \textrm{if } x\not\in\Delta \textrm{ and } \eta_x\neq\xi_x
\end{array}
\right.
\end{equation}
where we recall that by $E^{\ell,\wp}_{\{x\}}$ we mean 
$E^{(1),\ell,\wp}_{\{x\}}$, see the discussion below (\ref{spihams}). 
We shall understand, below, $\exp\{-\infty\}=0$. We have that 
$E^{\ell,\wp}_{x;\Delta}(\cdot|\xi)\in\cF^{\ell,\wp}_{\{x\}}$; 
we notice here that in the following we will sometimes misuse the notation 
and write 
$E^{\ell,\wp}_{x;\Delta}(\eta_x|\xi)$ instead of 
$E^{\ell,\wp}_{x;\Delta}(\eta|\xi)$.
We define the interaction 
$\tilde W_{X_1,X_2}^{\ell,\wp}:\cX^{\ell,\wp}\to\bR$
between two disjoint sets $X_1,X_2\subset\subset\cL^{(\wp)}$ by setting 
\begin{equation}
\label{interaz}
\tilde W_{X_1,X_2}^{\ell,\wp}:=
E_{X_1\cup X_2}^{\ell,\wp}-
E_{X_1}^{\ell,\wp}-
E_{X_2}^{\ell,\wp}
\end{equation}
Notice that 
$\tilde W^{\ell,\wp}_{X_1,X_2}\in\cF^{\ell,\wp}_{X_1\cup X_2}$.
For $x\in\cL^{(\wp)}$ we define the function 
$W_{x;\Delta}^{\ell,\wp}:\cX^{\ell,\wp}\to\bR$ by setting
\begin{equation}
\label{prtqw}
W_{x;\Delta}^{\ell,\wp}:=
\left\{
\begin{array}{ll}
0 & \textrm{if } x\not\in\Delta\vphantom{\Big\{}\\
\tilde W^{\ell,\wp}_{\{x\},\cA\cup\cB\cup\cC} 
    & \textrm{if } x\in\Delta\cap\cD\vphantom{\Big\{}\\
\tilde W^{\ell,\wp}_{\{x\},\cA\cup\cB\cup(\cD\cap\Delta^\complement)} 
    & \textrm{if } x\in\Delta\cap\cC\vphantom{\Big\{}\\
\tilde W^{\ell,\wp}_{\{x\},\cA\cup[(\cC\cup\cD)\cap\Delta^\complement]} 
    & \textrm{if } x\in\Delta\cap\cB\vphantom{\Big\{}\\
\tilde W^{\ell,\wp}_{\{x\},(\cB\cup\cC\cup\cD)\cap\Delta^\complement} 
    & \textrm{if } x\in\Delta\cap\cA\vphantom{\Big\{}\\
\end{array}
\right.
\end{equation}
By using definitions (\ref{porti}), (\ref{prtqw}), and choosing $\ell$ 
large enough such that $\wp=d\ell>r$ (recall $r$ is the 
range of the original interaction, so that the block spin 
interaction has range one), we have that 
for $\eta,\xi\in\cX^{\ell,\wp}$, such that 
$\eta_{\Delta^\complement}=\xi_{\Delta^\complement}$,
\begin{equation}
\label{Hdecomp}
\begin{array}{rcl}
{\displaystyle
 H^{\ell,\wp}_{\Delta}(\eta)
}
&\!\!=\!\!&
{\displaystyle
\sum_{k\in\cL^{(2\wp)}}\big[
  E^{\ell,\wp}_{A_{k}}(\eta|\xi)+W^{\ell,\wp}_{A_{k};\Delta}(\eta)
 +E^{\ell,\wp}_{B_{k}}(\eta|\xi)+W^{\ell,\wp}_{B_{k};\Delta}(\eta)
}\\
&&
{\displaystyle
\phantom{\sum_{k\in\cL^{(2\wp)}}\big[}
  +E^{\ell,\wp}_{C_{k}}(\eta|\xi)+W^{\ell,\wp}_{C_{k};\Delta}(\eta)
  +E^{\ell,\wp}_{D_{k}}(\eta|\xi)+W^{\ell,\wp}_{D_{k};\Delta}(\eta)
  \big]
}\\
\end{array} 
\end{equation}
For $V\subset\cL^{(\wp)}$ we introduce the set
\begin{equation}
\label{maimai}
\cY^{(\ell),\wp}_{\Delta,m,V}:=
\bigotimes_{x\in\Delta\cap V}\cX^{(\ell),\wp}_{m,\{x\}}
\otimes
\bigotimes_{x\in\Delta^\complement\cap V}\cX^{\ell,\wp}_{\{x\}}
\end{equation}
as usual if $V=\cL^{(\wp)}$ we drop it form the notation.
Hence, we have that for $\xi\in\cX^{\ell,\wp}$ 
the partition function 
in $\Delta$ can be written in the following way 
\begin{equation}
\label{decompZ}
\begin{array}{rcl}
{\displaystyle 
 Z^{(\ell),\wp}_{m,\Delta}(\xi)
}
&\!\!=\!\!& 
{\displaystyle
 \!\!\!\phantom{\times}
 \sum_{\eta\in\cX^{(\ell),\wp}_{m,\Delta}}
 \exp\big\{H^{\ell,\wp}_\Delta(\eta\xi_{\Delta^\complement})\big\}
}\\
&\!\!=\!\!& 
{\displaystyle
 \!\!\!\phantom{\times}
 \sum_{\alpha\in\cY^{(\ell),\wp}_{\Delta,m,\cA}}
 \Big(\prod_{k\in\cL^{(2\wp)}}
 \exp\big\{E^{\ell,\wp}_{A_k;\Delta}(\alpha_{A_k}|\xi)
           +W^{\ell,\wp}_{A_k;\Delta}(\alpha\xi_{\cB\cup\cC\cup\cD})\big\}\Big)
}\\
&& 
{\displaystyle
 \!\!\!\times
 \sum_{\beta\in\cY^{(\ell),\wp}_{\Delta,m,\cB}}
 \Big(\prod_{k\in\cL^{(2\wp)}}
 \exp\big\{E^{\ell,\wp}_{B_k;\Delta}(\beta_{B_k}|\xi)
           +W^{\ell,\wp}_{B_k;\Delta}(\alpha\beta\xi_{\cC\cup\cD})\big\}\Big)
}\\
&& 
{\displaystyle
 \!\!\!\times
 \sum_{\gamma\in\cY^{(\ell),\wp}_{\Delta,m,\cC}}
 \Big(\prod_{k\in\cL^{(2\wp)}}
 \exp\big\{E^{\ell,\wp}_{C_k;\Delta}(\gamma_{C_k}|\xi)
           +W^{\ell,\wp}_{C_k;\Delta}(\alpha\beta\gamma\xi_{\cD})\big\}\Big)
}\\
&& 
{\displaystyle
 \!\!\!\times
 \sum_{\delta\in\cY^{(\ell),\wp}_{\Delta,m,\cD}}
 \Big(\prod_{k\in\cL^{(2\wp)}}
 \exp\big\{E^{\ell,\wp}_{D_k;\Delta}(\delta_{D_k}|\xi)
           +W^{\ell,\wp}_{D_k;\Delta}(\alpha\beta\gamma\delta)\big\}\Big)
}
\end{array}
\end{equation}
Notice that although the sum defining the partition function is 
extended to the volume $\Delta$, it is convenient, for practical reasons, 
to consider the sums extended to the whole lattice $\cL^{(\wp)}$. This 
has been realized in the last step of (\ref{decompZ})
via the definition (\ref{porti}) of the function 
$E^{(\ell),\wp}_{x;\Delta}$.

In order to get (\ref{Zpolym}) we perform a sequence of decimations; we 
fix $\xi$ and sum over $\delta$, $\gamma$, $\beta$, and, finally,
$\alpha$ following this prescribed order. 
At each decimation step, we perform three operations,  
called {\it unfolding}, {\it splitting} and {\it gluing}, 
which will show that the system of variables corresponding to the 
sub--lattice involved in the decimation is weakly coupled. 
This weak coupling is a consequence of the factorization properties
of the partition functions on suitable finite volumes
which follow from (\ref{CC1}). 

We pick a reference configuration $\bar\eta\in\cX^{\ell,\wp}$ and let 
$\bar\xi:=\bar\eta_\Delta\xi_{\Delta^\complement}$. 
By computing the last sum for $\delta\in\cY^{(\ell),\wp}_{\Delta,m,\cD}$ in 
(\ref{decompZ}) and recalling $\wp>r$, we get
\begin{equation}
\label{poggio1}
 \sum_{\delta\in\cY^{(\ell),\wp}_{\Delta,m,\cD}}
 \prod_{k\in\cL^{(2\wp)}}
 \exp\big\{E^{\ell,\wp}_{D_k;\Delta}(\delta_{D_k}|\xi)
           +W^{\ell,\wp}_{D_k;\Delta}(\alpha\beta\gamma\delta)\big\}
=
 \prod_{k\in\cL^{(2\wp)}}
 Z^{(\ell),\wp}_{m,\{D_k\}\cap\Delta}(\alpha\beta\gamma\bar\xi_{\cD})
\end{equation}
where from now on we understand $Z^{(\ell),\wp}_{m,\emptyset}=1$. 
We also note that 
$Z^{(\ell),\wp}_{m,\{D_k\}\cap\Delta}$ depends only on the block spin 
configuration in the boundary of $\{D_k\}\cap\Delta$ namely,
$Z^{(\ell),\wp}_{m,\{D_k\}\cap\Delta}
  \in\cF^{(\ell),\wp}_{\partial^{(\wp)}[\{D_k\}\cap\Delta]}$,
in particular it does not depend on $\bar\xi_{\cD}$. Finally we note that 
by definition (\ref{porti}), when (\ref{poggio1}) is plugged 
into (\ref{decompZ}), the function 
$Z^{(\ell),\wp}_{m,\{D_k\}\cap\Delta}(\cdot)$
will be evaluated in the configuration 
$(\alpha\beta\gamma)_\Delta\bar\xi_{\Delta^\complement\cup\cD}$. 

Given $D_k\in\cD$ we denote by $(\beta\gamma)_\sopra$, resp.\ 
$(\beta\gamma)_\sotto$, the restriction of the configuration $\beta\gamma$ 
to the half--space above, resp.\ below, $D_k$. We now unfold the partition 
function $Z^{(\ell),\wp}_{m,\{D_k\}\cap\Delta}$ in the $e_2^{(\wp)}$ direction 
namely, we write
\begin{equation}
\label{poggio2}
\begin{array}{l}
{\displaystyle
 \!\!\!
 Z^{(\ell),\wp}_{m,\{D_k\}\cap\Delta}(\alpha\beta\gamma\bar\xi_{\cD})
=
 Z^{(\ell),\wp}_{m,\{D_k\}\cap\Delta}
  (\alpha(\beta\gamma)_\sopra(\beta\gamma)_\sotto\bar\xi_{\cD})
  \vphantom{\Bigg\{}
}\\
{\displaystyle
\phantom{i}
=
 \frac{Z^{(\ell),\wp}_{m,\{D_k\}\cap\Delta}
        (\alpha(\beta\gamma)_\sopra(\bar\xi_{\cB\cup\cC})_\sotto\bar\xi_\cD)
       \;
       Z^{(\ell),\wp}_{m,\{D_k\}\cap\Delta}
        (\alpha(\beta\gamma)_\sotto(\bar\xi_{\cB\cup\cC})_\sopra\bar\xi_\cD)}
      {Z^{(\ell),\wp}_{m,\{D_k\}\cap\Delta}
        (\alpha\bar\xi_{\cB\cup\cC\cup\cD})}
\big[1+\Phi_{D_k}(\alpha\beta\gamma,\bar\xi)\big]
}
\end{array}
\end{equation}
where, recall $\bar\xi=\bar\eta_\Delta\xi_{\Delta^\complement}$,
we have defined the function 
$\Phi_{D_k}:\cX^{\ell,\wp}_{\cA\cup\cB\cup\cC}\times\cX^{\ell,\wp}\to\bR$
as follows
\begin{equation}
\label{poggio3}
\Phi_{D_k}(\alpha\beta\gamma,\xi):=
\frac{Z^{(\ell),\wp}_{m,\{D_k\}\cap\Delta}
        (\alpha(\beta\gamma)_\sopra(\beta\gamma)_\sotto\bar\xi_{\cD})
      \;
      Z^{(\ell),\wp}_{m,\{D_k\}\cap\Delta}(\alpha\bar\xi_{\cB\cup\cC\cup\cD})}
     {Z^{(\ell),\wp}_{m,\{D_k\}\cap\Delta}
        (\alpha(\beta\gamma)_\sopra(\bar\xi_{\cB\cup\cC})_\sotto\bar\xi_\cD)
       \;
       Z^{(\ell),\wp}_{m,\{D_k\}\cap\Delta}
        (\alpha(\beta\gamma)_\sotto(\bar\xi_{\cB\cup\cC})_\sopra\bar\xi_\cD)}
-1
\end{equation}
which can be considered as an effective interaction potential 
among the $\alpha$,$\beta$,$\gamma$--variables due to the decimation on 
$\delta$. 
To simplify the notation we do not make explicit the parametric dependence
of $\Phi_{D_k}$ on $\wp$, $\Delta$, and $m$.
From the measurability properties of the partition function 
$Z^{(\ell),\wp}_{m,\{D_k\}\cap\Delta}$ we get
\begin{equation}
\label{misurphi}
\Phi_{D_k}(\cdot,\xi)
  \in\cF^{\ell,\wp}_{\partial^{(\wp)}\{D_k\}\cap(\cA\cup\cB\cup\cC)}=
     \cF^{\ell,\wp}_{\partial^{(\wp)}\{D_k\}}
\;\textrm{ and }\;
\Phi_{D_k}(\alpha\beta\gamma,\cdot)
  \in\cF^{\ell,\wp}_{\partial^{(\wp)}\{D_k\}\cap\Delta^\complement}
\end{equation}
for all $\alpha\in\cX^{\ell,\wp}_{\cA}$, 
$\beta\in\cX^{\ell,\wp}_{\cB}$, 
$\gamma\in\cX^{\ell,\wp}_{\cC}$, and $\xi\in\cX^{\ell,\wp}$,
where we recall the definition of boundary given in Subsection~\ref{s:lat}.  
The bound (\ref{CC1}) implies that 
$\Phi_{D_k}$, as well as similar 
effective interactions that will be defined later on, is uniformly small. 
We note, finally, that $\Phi_{D_k}=0$ if 
$\{D_k\}\cap\Delta=\emptyset$. 

We next split the product of the numerator in (\ref{poggio2}) 
in the $e_2^{(\wp)}$ direction namely, we write 
\begin{equation}
\label{poggio4}
\begin{array}{l}
{\displaystyle
 \prod_{k\in\cL^{(2\wp)}}
 Z^{(\ell),\wp}_{m,\{D_k\}\cap\Delta}
    (\alpha(\beta\gamma)_\sopra(\bar\xi_{\cB\cup\cC})_\sotto\bar\xi_\cD)
     \;
 Z^{(\ell),\wp}_{m,\{D_k\}\cap\Delta}
    (\alpha(\beta\gamma)_\sotto(\bar\xi_{\cB\cup\cC})_\sopra\bar\xi_\cD)
}
\\
{\displaystyle
 \phantom{merdone}
 =
 \prod_{k\in\cL^{(2\wp)}}
 Z^{(\ell),\wp}_{m,\{D_k\}\cap\Delta}
    (\alpha(\beta\gamma)_\sopra(\bar\xi_{\cB\cup\cC})_\sotto\bar\xi_\cD)
     \;
 Z^{(\ell),\wp}_{m,\{D_k+e_2^{(2\wp)}\}\cap\Delta}
    (\alpha(\beta\gamma)_\sotto(\bar\xi_{\cB\cup\cC})_\sopra\bar\xi_\cD)
}
\end{array}
\end{equation}
By (\ref{poggio1}), (\ref{poggio2}) and (\ref{poggio4}) we have 
that 
\begin{equation}
\label{poggio5} 
\begin{array}{l}
{\displaystyle
 \sum_{\gamma\in\cY^{(\ell),\wp}_{\Delta,m,\cC}}
 \prod_{k\in\cL^{(2\wp)}}
 e^{E^{\ell,\wp}_{C_k;\Delta}(\gamma_{C_k}|\xi)
           +W^{\ell,\wp}_{C_k;\Delta}(\alpha\beta\gamma\xi_{\cD})}
 \sum_{\delta\in\cY^{(\ell),\wp}_{\Delta,m,\cD}}
 \prod_{k\in\cL^{(2\wp)}}
 e^{E^{\ell,\wp}_{D_k;\Delta}(\delta_{D_k}|\xi)
           +W^{\ell,\wp}_{D_k;\Delta}(\alpha\beta\gamma\delta)}
} \\
{\displaystyle
 \phantom{m}
 =
 \prod_{k\in\cL^{(\wp)}}
 Z^{(\ell),\wp}_{m,\{D_k\}\cap\Delta}(\alpha\bar\xi_{\cB\cup\cC\cup\cD})^{-1}
}\\
{\displaystyle
 \phantom{mer}
 \times\!\!\!\!\!
 \sum_{\gamma\in\cY^{(\ell),\wp}_{\Delta,m,\cC}}
 \prod_{k\in\cL^{(2\wp)}}
 \Big[
 e^{E^{\ell,\wp}_{C_k;\Delta}(\gamma_{C_k}|\xi)
           +W^{\ell,\wp}_{C_k;\Delta}(\alpha\beta\gamma\xi_{\cD})}
 \,
 (1+\Phi_{D_k}(\alpha\beta\gamma,\bar\xi))
} \\
{\displaystyle
 \phantom{
          mer
          \times\!\!\!\!\!
          \sum_{\gamma\in\cY^{(\ell),\wp}_{\Delta,m,\cC}}
          \prod_{k\in\cL^{(2\wp)}}
          \Big[
         }
 \times
 Z^{(\ell),\wp}_{m,\{D_k\}\cap\Delta}
    (\alpha(\beta\gamma)_\sopra(\bar\xi_{\cB\cup\cC})_\sotto\bar\xi_\cD)
 Z^{(\ell),\wp}_{m,\{D_k+e_2^{(2\wp)}\}\cap\Delta}
    (\alpha(\beta\gamma)_\sotto(\bar\xi_{\cB\cup\cC})_\sopra\bar\xi_\cD)
 \Big]
} \\
{\displaystyle
 \phantom{m}
 =
 \bigg[
 \prod_{k\in\cL^{(\wp)}}
 \frac{
   Z^{(\ell),\wp}_{m,\tilde C_k\cap\Delta}
       ((\alpha\beta)_{\partial^{(\wp)}\{C_k\}\cap(\cA\cup\cB)}
        \bar\xi_{[\partial^{(\wp)}\{C_k\}\cap(\cA\cup\cB)]^\complement})}
  {Z^{(\ell),\wp}_{m,\{D_k\}\cap\Delta}(\alpha\bar\xi_{\cB\cup\cC\cup\cD})}
 \bigg]
}\\
{\displaystyle
 \phantom{mermer}
 \times
 \vphantom{\bigg\}}
 \sum_{\gamma\in\cY^{(\ell),\wp}_{\Delta,m,\cC}}
 \nu_\cC(\gamma|\alpha\beta,\bar\xi)
 \prod_{k\in\cL^{(\wp)}}
 (1+\Phi_{D_k}(\alpha\beta\gamma,\bar\xi))
}\\
\end{array}
\end{equation}
where we have defined 
$\tilde C_k:=\{D_k+e_2^{(2\wp)},C_k,D_k\}\subset\cL^{(\wp)}$, see 
Fig.~\ref{f:cariche} below, and 
introduced the product measure 
\begin{equation}
\label{poggio6}
\nu_\cC(\gamma|\alpha\beta,\bar\xi)
:=
\prod_{k\in\cL^{(2\wp)}}
\nu_{C_k}(\gamma_{C_k}|\alpha\beta,\bar\xi)
\end{equation}
with
\begin{equation}
\label{poggio7}
\begin{array}{rcl}
{\displaystyle
 \nu_{C_k}(\gamma_{C_k}|\alpha\beta,\bar\xi_\cD)
}
&\!\!:=\!\!&
{\displaystyle
 \exp\big\{E^{\ell,\wp}_{C_k;\Delta}(\gamma_{C_k}|\xi)
          +W^{\ell,\wp}_{C_k;\Delta}(\alpha\beta\gamma\xi_{\cD})\big\}
 \vphantom{\Big\{}
} \\
&&
{\displaystyle
 \times
 \frac{
  Z^{(\ell),\wp}_{m,\{D_k\}\cap\Delta}
     (\alpha(\beta\gamma)_\sopra(\bar\xi_{\cB\cup\cC})_\sotto\bar\xi_\cD)
  \,
  Z^{(\ell),\wp}_{m,\{D_k+e_2^{(2\wp)}\}\cap\Delta}
     (\alpha(\beta\gamma)_\sotto(\bar\xi_{\cB\cup\cC})_\sopra\bar\xi_\cD)}
   {Z^{(\ell),\wp}_{m,\tilde C_k\cap\Delta}
       ((\alpha\beta)_{\partial^{(\wp)}\{C_k\}\cap(\cA\cup\cB)}
        \bar\xi_{[\partial^{(\wp)}\{C_k\}\cap(\cA\cup\cB)]^\complement})}
}
\end{array}
\end{equation}
To simplify the notation we do not make explicit the parametric dependence
of $\nu_{C_k}$ on $\wp$, $\Delta$, and $m$.
The definition above is well posed because 
the right hand side depends on the configuration
$\gamma$ only through its restriction to $C_k$. Recalling $\wp>r$ we have that
$Z^{(\ell),\wp}_{m,\tilde C_k\cap\Delta}\in
     \cF^{\ell,\wp}_{\partial^{(\wp)}[\tilde C_k\cap\Delta]}$,
moreover by using definitions (\ref{poggio7}), (\ref{porti}), (\ref{prtqw}),
and the properties of measurability of the partition function 
$Z^{(\ell),\wp}_{m,\{D_k\}\cap\Delta}$ we get
\begin{equation}
\label{misurni}
\nu_{C_k}(\gamma_{C_k}|\cdot,\xi)
  \in\cF^{\ell,\wp}_{\partial^{(\wp)}\{C_k\}\cap(\cA\cup\cB)}
\;\;\;\;\textrm{ and } \;\;\;\;
\nu_{C_k}(\gamma_{C_k}|\alpha\beta,\cdot)
  \in\cF^{\ell,\wp}_{\srclos{\wp}{2}{\{C_k\}}\cap\Delta^\complement}
\end{equation}
for all $\alpha\in\cX^{\ell,\wp}_{\cA}$, 
$\beta\in\cX^{\ell,\wp}_{\cB}$, 
$\gamma\in\cX^{\ell,\wp}_{\cC}$, and $\xi\in\cX^{\ell,\wp}_m$.
Moreover, we remark that 
$\nu_{C_k}$ is a probability measure 
on $\cY^{(\ell),\wp}_{\Delta,m,\{C_k\}}$
since the gluing identity
\begin{equation}
\label{poggio8}
\begin{array}{l}
{\displaystyle
  Z^{(\ell),\wp}_{m,\tilde C_k\cap\Delta}
     ((\alpha\beta)_{\partial^{(\wp)}\{C_k\}\cap(\cA\cup\cB)}
      \bar\xi_{[\partial^{(\wp)}\{C_k\}\cap(\cA\cup\cB)]^\complement})
 \vphantom{\bigg\}}
}\\
\phantom{merdone}
{\displaystyle
 =
 \sum_{\gamma_{C_k}\in\cY^{(\ell),\wp}_{\Delta,m,C_k}}
 \exp\big\{E^{\ell,\wp}_{C_k;\Delta}(\gamma_{C_k}|\xi)
          +W^{\ell,\wp}_{C_k;\Delta}(\alpha\beta\gamma\xi_{\cD})\big\}
 \vphantom{\bigg\{}
} \\
\phantom{merdone=}
{\displaystyle
 \times\;\;\;
  Z^{(\ell),\wp}_{m,\{D_k\}\cap\Delta}
     (\alpha(\beta\gamma)_\sopra(\bar\xi_{\cB\cup\cC})_\sotto\bar\xi_\cD)
  \,
  Z^{(\ell),\wp}_{m,\{D_k+e_2^{(2\wp)}\}\cap\Delta}
     (\alpha(\beta\gamma)_\sotto(\bar\xi_{\cB\cup\cC})_\sopra\bar\xi_\cD)
}
\end{array}
\end{equation}
holds.
We finally remark that 
$\nu_{C_k}(\gamma_{C_k}|\alpha\beta,\bar\xi_\cD)
 =\id_{\{\gamma_{C_k}=\bar\xi_{C_k}\}}$
whenever $C_k\not\in\Delta$. 

By following the procedure of \cite{[BCO]}, with the modifications illustrated 
above, we straightforwardly get (\ref{Zpolym}) with
\begin{equation}
\label{marg}
\bar Z^{(\ell),\wp}_{m,\Delta}(\xi)  
:=
\prod_{k\in\cL^{(2\wp)}}
\frac{Z^{(\ell),\wp}_{m,\tilde A_k\cap\Delta}(\bar\xi)
      \ Z^{(\ell),\wp}_{m,\{D_k\}\cap\Delta}(\bar\xi)}
     {Z^{(\ell),\wp}_{m,F_k\cap\Delta}(\bar\xi)
      \ Z^{(\ell),\wp}_{m,\tilde C_k\cap\Delta}(\bar\xi)}
\end{equation}
where $F_k:=\{C_k-e_1^{(2\wp)},B_k,C_k\}$ and 
$\tilde A_k:=\{A_k\}\cup\partial^{(2\wp)}\{A_k\}$, see Fig.~\ref{f:cariche}, 
and
\begin{equation}
\label{fin}
\begin{array}{rl}
{\displaystyle
\X^{(\ell),\wp}_{m,\Delta}(\xi):=}
&{\displaystyle
 \sum_{\alpha\in\cY^{(\ell),\wp}_{\Delta,m,\cA}}
   \prod_{k\in\cL^{(2\wp)}}
   \nu_{A_k}(\alpha_{A_k}|\bar\xi)\,
    \big(1+\Psi_{D_k}(\alpha,\bar\xi)\big)
    \big(1+\Psi_{A_k}(\alpha,\bar\xi)\big)
    \big(1+\Phi_{B_k}(\alpha,\bar\xi)\big)}\\
&{\displaystyle
 \!\!\!\!\times
 \sum_{\beta\in\cY^{(\ell),\wp}_{\Delta,m,\cB}}
   \prod_{k\in\cL^{(2\wp)}}
   \nu_{B_k}(\beta_{B_k}|\alpha,\bar\xi)\,
    \big(1+\Phi_{C_k}(\alpha\beta,\bar\xi)\big)}\\
&{\displaystyle
 \!\!\!\!\times
 \sum_{\gamma\in\cY^{(\ell),\wp}_{\Delta,m,\cC}}
   \prod_{k\in\cL^{(2\wp)}}
   \nu_{C_k}(\gamma_{C_k}|\alpha\beta,\bar\xi)\,
    \big(1+\Phi_{D_k}(\alpha\beta\gamma,\bar\xi)\big)}\\
\end{array}
\end{equation}
where the $\Psi$'s and $\Phi$'s are error terms similar to the one explicitly
defined in (\ref{poggio3}), and each $\nu_x$ is a probability measures 
on $\cY^{(\ell),\wp}_{\Delta,m,\{x\}}$, for $x\in\cA\cup\cB\cup\cC$, similar 
to the one in (\ref{poggio7}). All these functions can be defined as 
in \cite{[BCO]}, we do not enter here into these details, we just 
recall their measurability properties. 
For each $\alpha\in\cX^{\ell,\wp}_{\cA}$, 
$\beta\in\cX^{\ell,\wp}_{\cB}$, 
$\gamma\in\cX^{\ell,\wp}_{\cC}$, and $\xi\in\cX^{\ell,\wp}$ we have
\begin{equation}
\label{misurphitu}
\begin{array}{lll}
\Phi_{C_k}(\cdot,\xi)
  \in\cF^{\ell,\wp}_{\partial^{(\wp)}\{C_k\}\cap(\cA\cup\cB)} &
\Phi_{C_k}(\alpha\beta,\cdot)
  \in\cF^{\ell,\wp}_{\srclos{\wp}{2}{\{C_k\}}\cap\Delta^\complement} &
\Phi_{B_k}(\cdot,\xi)
  \in\cF^{\ell,\wp}_{\partial^{(\wp)}\{B_k\}\cap\cA}
\\
\Phi_{B_k}(\alpha,\cdot)
  \in\cF^{\ell,\wp}_{\srclos{\wp}{2}{\{B_k\}}\cap\Delta^\complement} &
\Psi_{D_k}(\cdot,\xi)
  \in\cF^{\ell,\wp}_{\partial^{(\wp)}\{D_k\}\cap\cA} &
\Psi_{D_k}(\alpha,\cdot)
  \in\cF^{\ell,\wp}_{\sclos{\wp}{\{D_k\}}\cap\Delta^\complement}
\\
\Psi_{A_k}(\cdot,\xi)
  \in\cF^{\ell,\wp}_{A_k} &
\Psi_{A_k}(\alpha,\cdot)
  \in\cF^{\ell,\wp}_{\srclos{\wp}{2}{\{A_k\}}\cap\Delta^\complement} &
\\
\end{array}
\end{equation}
and
\begin{equation}
\label{misurnutu}
\nu_{B_k}(\beta_{B_k}|\cdot,\xi)
  \in\cF^{\ell,\wp}_{\partial^{(\wp)}\{B_k\}\cap\cA},\;
\nu_{B_k}(\beta_{B_k}|\alpha,\cdot)
  \in\cF^{\ell,\wp}_{\srclos{\wp}{2}{\{B_k\}}\cap\Delta^\complement}
\textrm{ and }
\nu_{A_k}(\alpha_{A_k}|\cdot)
  \in\cF^{\ell,\wp}_{\srclos{\wp}{2}{\{A_k\}}\cap\Delta^\complement}
\end{equation}
We also set 
\begin{equation}
\label{bern}
\nu_\cB(\beta|\xi):=\prod_{B\in\cB}\nu_{B}(\beta_B|\alpha,\xi)
\;\;\textrm{ and }\;\;
\nu_\cA(\alpha|\xi):=\prod_{A\in\cA}\nu_{A}(\alpha_A|\xi)
\end{equation}
for all $\alpha\in\cX^{\ell,\wp}_{\cA}$, 
$\beta\in\cX^{\ell,\wp}_{\cB}$, 
and $\xi\in\cX^{\ell,\wp}$.

We next rewrite the functions $\bar Z^{(\ell),\wp}_{m,\Delta}$ and
$\Xi^{(\ell),\wp}_{m,\Delta}$ having in mind
that our goal is the definition of the family 
$\{V^{(\ell),\wp}_{X,\Lambda},\,X\subset\subset\cL^{(\wp)}\}$ 
whose existence has 
been stated in the theorem. We
first define the collection of subsets of the lattice $\cL^{(\wp)}$
\begin{equation}
\label{gola1}
\cG:=\bigcup_{k\in\cL^{(2\wp)}}\{\{D_k\},\tilde C_k,\,F_k,\,\tilde A_k\}
\end{equation}
and for all $k\in\cL^{(2\wp)}$ we set 
\begin{equation}
\label{gola2}
g(\{D_k\})=+1,\;\;
g(\tilde A_k)=+1,\;\;
g(F_k)=-1,\;\;\textrm{ and }\;\;
g(\tilde C_k)=-1
\end{equation}
{}From (\ref{marg}) we then have 
\begin{equation}
\label{gola3}
\log\bar Z^{(\ell),\wp}_{m,\Delta}(\xi)=
 \sum_{G\in\cG}g(G)\log Z^{(\ell),\wp}_{m,G\cap\Delta}(\bar\xi)
\end{equation}
Recalling that we always understand $Z^{(\ell),\wp}_{m,\emptyset}=1$ and that 
$\Delta$ is a finite subset of the lattice $\cL^{(\wp)}$, we have that the 
sum in (\ref{gola3}) has indeed a finite number of terms. We prove, now,
that 
\begin{equation}
\label{cariche}
\sum_{i\in\cO_\wp^\ell Y}\frac{1}{2}m_i^2
=
 \sum_{G\in\cG}g(G)\sum_{i\in\cO_\wp^\ell(G\cap Y)}\frac{1}{2}m_i^2
\end{equation}
for all $Y\subset\cL^{(\wp)}$. Indeed, we first remark that 
\begin{equation}
\label{cariche1}
 \sum_{G\in\cG}g(G)\sum_{i\in\cO_\wp^\ell(G\cap Y)}\frac{1}{2}m_i^2
=
 \sum_{i\in\cO_\wp^\ell Y}\frac{1}{2}m_i^2
 \sum_{G\in\cG:\,\cO_\wp^\ell G\ni i}g(G)
\end{equation}
The identity (\ref{cariche}) follows from (\ref{cariche1}) once we prove 
that $\sum_{G\in\cG:\,\cO_\wp^\ell G\ni i}g(G)=+1$
for each $i\in\cL^{(\ell)}$.
Pick $i\in\cL^{(\ell)}$ and suppose there exists $k'\in\cL^{(2\wp)}$ such that 
$i\in\cO_\wp^\ell\{D_{k'}\}$. 
The only $G$'s of $\cG$ such that $\cO_\wp^\ell G\ni i$ are 
$\tilde A_{k'}$, $\tilde A_{k'+e^{(2\wp)}_1}$, 
$\tilde C_{k'}$, $\tilde C_{k'-e^{(2\wp)}_2}$, and $\{D_{k'}\}$,  
see Fig.~\ref{f:cariche}. Then
$$
\sum_{G\in\cG:\,\cO_\wp^\ell G\ni i}g(G)=
g(\tilde A_{k'})+g(\tilde A_{k'+e^{(2\wp)}_1})+
g(\tilde C_{k'})+g(\tilde C_{k'-e^{(2\wp)}_2})+g(\{D_{k'}\})=+1
$$
where in the last equality we have used (\ref{gola2}).
The other three cases, where 
$i\in\cO_\wp^\ell\{A_{k'}\}$, $i\in\cO_\wp^\ell\{B_{k'}\}$, or 
$i\in\cO_\wp^\ell\{C_{k'}\}$ for a suitable $k'\in\cL^{(2\wp)}$, 
can be treated similarly. 

We recall that 
$\mu^\tau_X$, for $X\subset\subset\cL$, is the finite volume (grancanonical) 
Gibbs measure of the original lattice gas,
see Subsection~\ref{s:lg}, $\chi$ is the infinite
volume compressibility defined in (\ref{chi}), and that for $i\in\cL^{(\ell)}$
the function $M^{(\ell)}_i$ is defined in (\ref{Centr}). Then for $G\in\cG$ 
we define the $|\cO_\wp^\ell(G\cap\Lambda)|\times|\cO_\wp^\ell(G\cap\Lambda)|$
covariance matrix
\begin{equation}
\label{covma}
\Big(\bV^{(\ell),\bar\eta}_{G\cap\Lambda}\Big)_{i,j}:=
2\pi\chi\ell^d
 \mu^{\cO_\ell\cO^\ell_\wp\bar\eta}_{\cO_\wp(G\cap\Lambda)}
    (M^{(\ell)}_i,M^{(\ell)}_j)
\end{equation}
for $i,j\in\cO_\wp^\ell(G\cap\Lambda)$, 
with $\bar\eta$ the reference  
in $\cX^{(\ell),\wp}$ chosen before (\ref{poggio1}). 
We understand $\bV^{(\ell),\bar\eta}_{\emptyset}$ is
equal to the $1\times1$ matrix with its sole element equal to 1. 
We let, as in Subsection~\ref{s:lg}, 
$Z_X(\tau)$, with $X\subset\subset\cL$ and $\tau\in\cX$, be the 
(grancanonical) partition function of the original lattice gas model, 
on $X$ with boundary condition $\tau$.
Then we define the real 
\begin{equation}
\label{gola5}
K^{(\wp)}_\Lambda:=\sum_{G\in\cG}g(G)
    \log\Big[Z_{\cO_\wp(G\cap\Lambda)}(\cO_\ell\cO_\wp^\ell\bar\eta)/
             \sqrt{\det\bV^{(\ell),\bar\eta}_{G\cap\Lambda}}\Big]
\end{equation}
By using (\ref{gola3}), (\ref{cariche}) for $Y=\Lambda$, 
and (\ref{gola5}) we rewrite 
$\log\bar Z^{(\ell),\wp}_{m,\Delta}(\xi)$ as follows 
\begin{equation}
\label{gola6}
\log\bar Z^{(\ell),\wp}_{m,\Delta}(\xi)
=
K^{(\wp)}_\Lambda
-\sum_{i\in\cO_\wp^\ell\Lambda}\frac{1}{2}m_i^2
+\sum_{G\in\cG}g(G)\Big[
        \log\frac{Z^{(\ell),\wp}_{m,G\cap\Delta}(\bar\xi)
                  \,\sqrt{\det\bV^{(\ell),\bar\eta}_{G\cap\Lambda}}}
                 {Z_{\cO_\wp(G\cap\Lambda)}(\cO_\ell\cO_\wp^\ell\bar\eta)}
       +\sum_{i\in\cO_\wp^\ell(G\cap\Lambda)}\frac{1}{2}m_i^2\Big]
\end{equation}

\setlength{\unitlength}{1.pt}
\begin{figure}
\begin{picture}(200,50)(-20,30)
\thinlines
\multiput(0,0)(0,30){3}{\line(1,0){80}}
\multiput(10,-10)(30,0){3}{\line(0,1){80}}
\multiput(10,0)(30,0){3}{\multiput(0,0)(0,30){3}{\circle*{5}}}
\put(41,33){$\scriptstyle A_k$}
\put(41,63){$\scriptstyle B_k$}
\put(71,63){$\scriptstyle C_k$}
\put(71,33){$\scriptstyle D_k$}
\put(40,8){$\scriptstyle B_{k-e^{(2\wp)}_2}$}
\put(74,-8){$\scriptstyle C_{k-e^{(2\wp)}_2}$}
\put(-15,-8){$\scriptstyle C_{k-e^{(2\wp)}_1-e^{(2\wp)}_2}$}
\put(-15,40){$\scriptstyle D_{k-e^{(2\wp)}_1}$}
\put(-15,70){$\scriptstyle C_{k-e^{(2\wp)}_1}$}
\multiput(110,0)(0,30){3}{\line(1,0){80}}
\multiput(120,-10)(30,0){3}{\line(0,1){80}}
\multiput(150,0)(0,30){3}{\circle*{5}}
\put(151,71){$\scriptstyle D_{k+e^{(2\wp)}_2}$}
\put(151,33){$\scriptstyle C_k$}
\put(151,3){$\scriptstyle D_k$}
\multiput(220,0)(0,30){3}{\line(1,0){80}}
\multiput(230,-10)(30,0){3}{\line(0,1){80}}
\multiput(230,30)(30,0){3}{\circle*{5}}
\put(291,33){$\scriptstyle C_k$}
\put(261,33){$\scriptstyle B_k$}
\put(205,40){$\scriptstyle C_{k-e^{(2\wp)}_1}$}
\multiput(330,0)(0,30){3}{\line(1,0){80}}
\multiput(340,-10)(30,0){3}{\line(0,1){80}}
\put(370,30){\circle*{5}}
\put(371,33){$\scriptstyle D_k$}
\end{picture}
\vskip 2. cm
\caption{From the left to the right the sets 
$\tilde A_k,\tilde C_k,F_k,\{D_k\}\subset\cL^{(\wp)}$ are depicted 
for some $k\in\cL^{(2\wp)}$. Solid circles denote the sites belonging to the 
those subsets; intersections of lines represent sites in $\cL^{(\wp)}$.}
\label{f:cariche}
\end{figure}
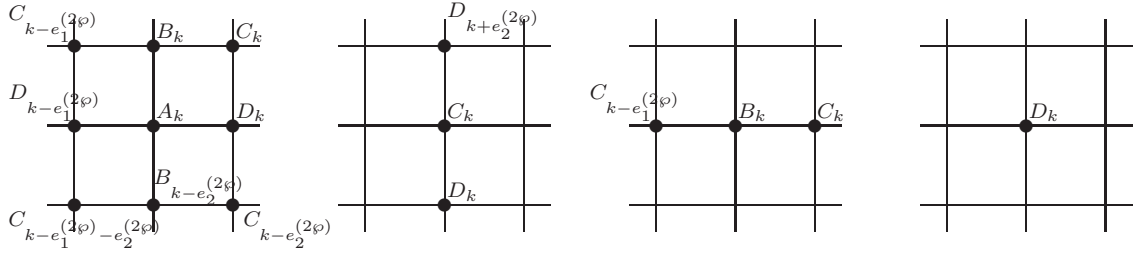

Consider, now, the function $\Xi^{(\ell),\wp}_{m,\Delta}$ defined in 
(\ref{fin}); we show that it can be rewritten 
as the partition function of a gas of polymers. 
We first associate to each error term $\Phi_{D_k},\Phi_{C_k},\dots,\Psi_{D_k}$ 
appearing in (\ref{fin}) a subset of the lattice that will be called 
{\it bond}. More precisely, for the $\Phi$ error terms we set 
\begin{equation}
\label{slitta}
e(\Phi_{D_k})\!:=\partial^{(\wp)}\{D_k\}\cup(\partial^{(\wp),2}\{D_k\}\cap\cA), 
e(\Phi_{C_k})\!:=\partial^{(\wp)}\{C_k\}\cap(\cA\cup\cB), 
e(\Phi_{B_k})\!:=\partial^{(\wp)}\{B_k\}\cap\cA 
\end{equation}
For the $\Psi$ error terms we set 
\begin{equation}
\label{slitta1}
e(\Psi_{D_k}):=\partial^{(\wp)}\{D_k\}\cap\cA
\;\;\;\textrm{ and }\;\;\;
e(\Psi_{A_k}):=\{A_k\}
\end{equation}
see Fig.~\ref{f:legami}. Moreover, in this section we denote by 
\begin{equation}
\label{legami}
\cE:=\bigcup_{k\in\cL^{(2\wp)}}
\{e(\Phi_{D_k}), e(\Phi_{C_k}), e(\Phi_{B_k}), e(\Psi_{D_k}), e(\Psi_{A_k})\}
\end{equation}
the collection of all the bonds.
For each $e\in\cE$ we denote by 
$\Theta_e:\cX^{\ell,\wp}_{\cA\cup\cB\cup\cC}\times\cX^{\ell,\wp}\to\bR$ 
the error term with which the bond $e$ is 
associated and we call it {\it weight} of the bond; for instance if 
$e=e(\Phi_{D_k})$ then $\Theta_e=\Phi_{D_k}$.
We notice that by expanding the products in 
(\ref{fin}) we get also 
addends with a single error term which must be averaged 
against the measures $\nu$'s; the bond have been defined so that 
the infinite volume average in (\ref{fin}) can be replaced by the average 
restricted to the bond itself. More precisely, 
consider the bond $e\in\cE$ and the corresponding error term $\Theta_e$,
by using (\ref{misurphi}), (\ref{misurphitu}),
(\ref{misurni}), and (\ref{misurnutu}) 
we have that for each $\xi\in\cX^{\ell,\wp}$ 
\begin{equation}
\label{ideabond}
\begin{array}{l}
{\displaystyle
 \sum_{\alpha\in\cY^{(\ell),\wp}_{\Delta,m,\cA}}
 \sum_{\beta\in\cY^{(\ell),\wp}_{\Delta,m,\cB}}
 \sum_{\gamma\in\cY^{(\ell),\wp}_{\Delta,m,\cC}}
  \nu_{\cA}(\alpha|\xi)
  \nu_{\cB}(\beta|\alpha,\xi)
  \nu_{\cC}(\gamma|\alpha\beta,\xi)
  \,\Theta_e(\alpha\beta\gamma,\xi)
       \vphantom{._\bigg\{}
}\\
{\displaystyle 
 \;\;\;\;
 =
 \!\!\!\!\!
 \sum_{\alpha\in\cY^{(\ell),\wp}_{\Delta,m,e\cap\cA}}
 \sum_{\beta\in\cY^{(\ell),\wp}_{\Delta,m,e\cap\cB}}
 \sum_{\gamma\in\cY^{(\ell),\wp}_{\Delta,m,e\cap\cC}}
 \!\!\!\!
 \prod_{A\in e\cap\cA}
  \!\!\!
  \nu_{A}(\alpha_{A}|\xi)
  \!\!\!
 \prod_{B\in e\cap\cB}
  \!\!\!
  \nu_{B}(\beta_{B}|\alpha,\xi)
  \!\!\!
 \prod_{C\in e\cap\cC}
  \!\!\!
  \nu_{C}(\gamma_{C}|\alpha\beta,\xi)
       \,\Theta_e(\alpha\beta\gamma,\xi)
}\\
\end{array}
\end{equation}

\setlength{\unitlength}{1.pt}
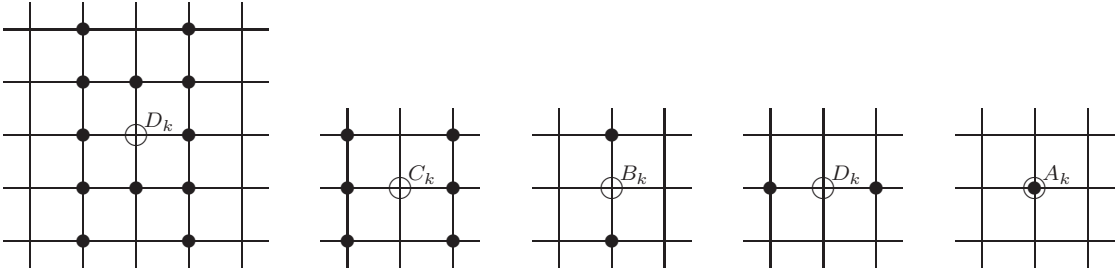
\begin{figure}
\begin{picture}(200,50)(-15,30)
\thinlines
\multiput(0,0)(0,20){5}{\line(1,0){100}}
\multiput(10,-10)(20,0){5}{\line(0,1){100}}
\put(50,40){\circle{8}}
\put(53,43){$\scriptstyle D_k$}
\put(50,60){\circle*{5}} 
\put(50,20){\circle*{5}}
\multiput(30,0)(0,20){5}{\circle*{5}}
\multiput(70,0)(0,20){5}{\circle*{5}}
\multiput(120,0)(0,20){3}{\line(1,0){60}}
\multiput(130,-10)(20,0){3}{\line(0,1){60}}
\put(150,20){\circle{8}}
\put(153,23){$\scriptstyle C_k$}
\multiput(130,0)(0,20){3}{\circle*{5}}
\multiput(170,0)(0,20){3}{\circle*{5}}
\multiput(200,0)(0,20){3}{\line(1,0){60}}
\multiput(210,-10)(20,0){3}{\line(0,1){60}}
\put(230,20){\circle{8}}
\put(233,23){$\scriptstyle B_k$}
\put(230,0){\circle*{5}}
\put(230,40){\circle*{5}}
\multiput(280,0)(0,20){3}{\line(1,0){60}}
\multiput(290,-10)(20,0){3}{\line(0,1){60}}
\put(310,20){\circle{8}}
\put(313,23){$\scriptstyle D_k$}
\put(290,20){\circle*{5}}
\put(330,20){\circle*{5}}
\multiput(360,0)(0,20){3}{\line(1,0){60}}
\multiput(370,-10)(20,0){3}{\line(0,1){60}}
\put(390,20){\circle{8}}
\put(390,20){\circle*{5}}
\put(393,23){$\scriptstyle A_k$}
\end{picture}
\vskip 2. cm
\caption{From the left to the right the bonds 
$e(\Phi_{D_k})$, $e(\Phi_{C_k})$, $e(\Phi_{B_k})$, $e(\Psi_{D_k})$, and 
$e(\Psi_{A_k})$ are depicted for some $k\in\cL^{(2\wp)}$. 
Solid circles denote the sites belonging to the 
bonds; open circles denote the site labelling a bond;
intersections of lines represent sites in $\cL^{(\wp)}$.}
\label{f:legami}
\end{figure}

Consider, now, a collection $\{e_1,\dots,e_k\}$ of pairwise 
different elements of $\cE$, we say that such a collection 
is a polymer if and only if 
for each $i,i'\in\{1,\dots,k\}$ there exists $i_1,\dots,i_s\in\{1,\dots,k\}$ 
such that $e_i=e_{i_1}$, 
$e_{i_1}\cap e_{i_2}\neq\emptyset$, $\cdots$,
$e_{i_{s-1}}\cap e_{i_s}\neq\emptyset$, $e_{i_s}=e_{i'}$. 
We denote by $\cR$ the collection of all polymers and for each $R\in\cR$ we 
set 
\begin{equation}
\label{suppR}
\widetilde R:=\bigcup_{e\in R} e\subset\cL^{(\wp)}
\end{equation}
By expanding the products in (\ref{fin}) and by
standard polymerization, for $\xi\in\cX^{\ell,\wp}$ we have that 
\begin{equation}
\label{Xi}
\X^{(\ell),\wp}_{m,\Delta}(\xi)
=
1+\sum_{k\geq 1}
  \sum_{\newatop{R_1,\dots,R_k\in\cR:}  
                {\tilde R_i\cap\tilde R_j=\emptyset,i\neq j}}  
  \prod_{j=1}^k \zeta^{(\ell),\wp}_{m,R_j,\Delta}(\xi)
\end{equation}
where the \textit{activity} $\zeta^{(\ell),\wp}_{m,R,\Delta}$ 
associated with a polymer $R\in\cR$ is given by 
\begin{equation}
\label{attivita}
\begin{array}{rcl}
{\displaystyle 
 \zeta^{(\ell),\wp}_{m,R,\Delta}(\xi)
}
&\!\!\!:=\!\!\!&
{\displaystyle 
 \sum_{\alpha\in\cY^{(\ell),\wp}_{\Delta,m,\cA}}
 \sum_{\beta\in\cY^{(\ell),\wp}_{\Delta,m,\cB}}
 \sum_{\gamma\in\cY^{(\ell),\wp}_{\Delta,m,\cC}}
  \nu_{\cA}(\alpha|\xi)
  \nu_{\cB}(\beta|\alpha,\xi)
  \nu_{\cC}(\gamma|\alpha\beta,\xi)
  \prod_{e\in R}
  \Theta_e(\alpha\beta\gamma,\bar\xi)
  \vphantom{._{\bigg\{}}
} \\
&\!\!\!=\!\!\!&
{\displaystyle 
 \sum_{\alpha\in\cY^{(\ell),\wp}_{\Delta,m,\tilde R\cap\cA}}
 \sum_{\beta\in\cY^{(\ell),\wp}_{\Delta,m,\tilde R\cap\cB}}
 \sum_{\gamma\in\cY^{(\ell),\wp}_{\Delta,m,\tilde R\cap\cC}}
 \prod_{A\in\tilde R\cap\cA}
  \nu_{A}(\alpha_{A}|\xi)
  \vphantom{._{\bigg\{}}
}\\
&&
{\displaystyle 
  \;\;\;\;\;\;\;\;\;
  \times
 \prod_{B\in\tilde R\cap\cB}
  \nu_{B}(\beta_{B}|\alpha,\xi)
 \prod_{C\in\tilde R\cap\cC}
  \nu_{C}(\gamma_{C}|\alpha\beta,\xi)
 \prod_{e\in R}
  \Theta_e(\alpha\beta\gamma,\xi)
}
\end{array}
\end{equation}
where the last equality holds by the same arguments used to prove 
(\ref{ideabond}). 
We remark that the sum in (\ref{Xi}) is restricted to a finite number of 
``non--intersecting" polymers, indeed the error term $\Theta_e$ associated 
to a bond sufficiently far from $\Delta$ is equal to zero. 
This can be easily checked in the case of $\Phi_{D_k}$: by using  
definition (\ref{poggio3}) and recalling $Z^{(\ell),\wp}_{m,\emptyset}=1$,
we have that $\{D_k\}\cap\Delta=\emptyset$ implies $\Phi_{D_k}=0$. By 
looking at the definitions of the error terms $\Theta_e$, those given 
in \cite{[BCO]} and suitably modified as we did in (\ref{poggio3}),
it is easy to check that for each $e\in\cE$ 
\begin{equation}
\label{zeroerr}
\sclos{\wp}{e}\subset\Delta^\complement
\;\;\Longrightarrow\;\;
\Theta_e=0
\end{equation}
Finally, we note that the activity $\zeta^{(\ell),\wp}_{m,R,\Delta}(\xi)$
of a polymer $R\in\cR$ 
is a local function of $\xi$, indeed by using (\ref{attivita}), 
(\ref{misurnutu}), and (\ref{misurphitu}) we have that 
\begin{equation}
\label{misurz}
\zeta^{(\ell),\wp}_{m,R,\Delta}\in
  \cF^{\ell,\wp}_{(\widetilde R\cup\partial^{(\wp),2}\widetilde R)
                     \cap\Delta^\complement}
\end{equation}

Let us consider a collection of polymers $\{R_1,\dots,R_k\}$, we say 
that it is a {\it cluster of polymers} if and only if 
for each $i,i'\in\{1,\dots,k\}$ there exists $i_1,\dots,i_s\in\{1,\dots,k\}$ 
such that $R_i=R_{i_1}$, 
$\tilde R_{i_1}\cap\tilde R_{i_2}\neq\emptyset$, $\cdots$,
$\tilde R_{i_{s-1}}\cap\tilde R_{i_s}\neq\emptyset$, $R_{i_s}=R_{i'}$. 
We denote by $\underline{\cR}$ the 
collection of all clusters of polymers and for each 
$\underline R\in\underline \cR$ we set 
\begin{equation}
\label{suppcR}
\underline{\tilde R}:=\bigcup_{R\in\underline R}\tilde R\subset\cL^{(\wp)}
\end{equation}
We finally introduce some combinatorial factors as follows:
let $F(R_1,\dots,R_k)$ be the collection of connected subgraphs
with vertex set $\{1,\dots,k\}$ 
of the graph with vertices $\{1,\dots,k\}$ and edges
$\{i,j\}$ corresponding to pairs $R_i,R_j$ such that 
$\widetilde R_i\cap\widetilde R_j\neq \emptyset$, then
\be{inconoscibile}
\f_T(R_1,\dots,R_k):=\frac{1}{k!} \sum_{f\in F(R_1,\dots,R_k)}(-1)^{ \# \;
{\rm edges \;in} \;f}
\end{equation}
we set the sum equal to zero if $F$ is empty and one if $k=1$.
Then, by standard cluster expansion, see for instance \cite{[KP],[GMM]},
under suitable small activity conditions that we shall specify later on,
the polymer gas partition function (\ref{Xi}) can be written as follows
\begin{equation}
\label{exp}
\log \Xi^{(\ell),\wp}_{m,\Delta}(\xi)
=
\sum_{\underline R\in\underline{\tilde R}} 
 \varphi_T(\underline R) \zeta^{(\ell),\wp}_{m,\underline R,\Delta}(\xi)
\end{equation}
where for each $\underline R\in\underline{\tilde R}$ we have set 
\begin{equation}
\label{zetaclu}
\zeta^{(\ell),\wp}_{m,\underline R,\Delta}
:=
\prod_{R\in\underline R} \zeta^{(\ell),\wp}_{m,R,\Delta}
 \in
  \cF^{\ell,\wp}_{(\underline{\widetilde R}
        \cup\partial^{(\wp),2}\underline{\widetilde R})\cap\Delta^\complement}
\end{equation}

As remarked above, see (\ref{zeroerr}), the 
activity of polymers containing at least a bond $e$ such that 
$\sclos{\wp}{e}\subset\Delta^\complement$ 
is equal to zero, so that only polymers with support close 
to $\Delta$ have non--zero activity. Nevertheless, 
the sum on the right--hand side of (\ref{exp}) 
is infinite due to the fact that in a cluster of polymers a given polymer
can be repeated arbitrarily many times. 
We next prove that for $\ell$ large enough multiple of $\ell_0$
the series is absolutely 
convergent. We shall use the technique developed in \cite{[CasO]} 
to get a uniform estimate of the sum of the activity of all the polymers
whose support contains a given site $x\in\cL^{(\wp)}$; such an estimate 
will be then used as the input of the abstract theory developed 
in \cite{[KP]} to estimate the sum (\ref{exp}) which is 
extended to the clusters of polymers whose support intersects $\Delta$. 

Let $e\in\cE$, consider the 
corresponding error term $\Theta_e$. By looking at the 
definition of $\Phi_{D_k}$ given in (\ref{poggio3}) and 
at the similar expressions in \cite{[BCO]} for the other 
error terms we have that (\ref{CC1}) implies
\begin{equation}
\label{bothe}
\sup_{\alpha\in\cX^{\ell,\wp}_{\cA}}\,
\sup_{\beta\in\cX^{\ell,\wp}_{\cB}}\,
\sup_{\gamma\in\cX^{\ell,\wp}_{\cC}}\,
\sup_{\xi\in\cX^{\ell,\wp}}\,
|\Theta_e(\alpha\beta\gamma,\xi)|
 \le \frac{C}{\ell}
\end{equation} 
for $\ell$ multiple of $\ell_0$. 
Consider, now, a polymer and its activity 
$\zeta^{(\ell),\wp}_{m,R,\Delta}$ defined in (\ref{attivita}); from 
(\ref{bothe}) we have the bound 
\begin{equation}
\label{boac}
\|\zeta^{(\ell),\wp}_{m,R,\Delta}\|_\infty
 \le\prod_{e\in R}\frac{C}{\ell}
 \le\varepsilon^{2|\tilde R|}
\end{equation}
where we have set $\varepsilon=\varepsilon(\ell):=(C/\ell)^{1/(2\kappa')}$, 
with $\kappa'=\kappa'(d)$ the maximum 
cardinality of the bonds (equal to 13 in dimension 
two see Fig.~\ref{f:legami}), 
and we are considering $\ell$ so large that $\varepsilon(\ell)<1$. 
We remark that for the current purpose it would have been sufficient to define 
$\varepsilon(\ell)=(C/\ell)^{1/\kappa'}$; the extra factor $2$ will be 
used in the proof of item~\ref{i:ipo43}.

For each polymer $R\in\cR$ we set, now, 
$\bar\zeta_R=\bar\zeta_R(\ell):=[\varepsilon(\ell)]^{|\tilde R|}$ and we prove 
that for $\ell$ large enough 
\begin{equation}
\label{bope}
\sup_{x\in\cL^{(\wp)}}\sum_{R\in\cR:\,\tilde R\ni x}
   \bar\zeta_R(\ell)\,e^{|\tilde R|}\le1
\end{equation}
Indeed, from (\ref{slitta})--(\ref{legami}) 
we have that there exist a real $\kappa''=\kappa''(d)$ such that 
$|\{e\in\cE:\,e\ni x\}|\le\kappa''$
for all $x\in\cL^{(\wp)}$. Moreover by choosing $\ell$ large enough 
we have that 
$\exp\{\kappa''\}\le[e\varepsilon(2-e\varepsilon)]^{-1}$. 
We can now perform the estimate in \cite[Appendix~B]{[CasO]}, 
by replacing there $\zeta(R)$ with $\bar\zeta_R$, $\sigma$ with 
$e\varepsilon$, and $\varphi_e$ with $1$, to obtain 
\begin{equation}
\label{bopef}
\sup_{x\in\cL^{(\wp)}}\sum_{R\in\cR:\,\tilde R\ni x}
   \bar\zeta_R\,e^{|\tilde R|}
  \le
 e\varepsilon\kappa''\Big[1
     +\frac{e^{\kappa''}-1}{1+(e\varepsilon)^2
        e^{\kappa''}-2e\varepsilon e^{\kappa''}}\Big]
\end{equation} 
The bound (\ref{bope}) now follows trivially for $\ell$ large enough. 

We are now ready to apply the abstract theory developed in \cite{[KP]}. 
Given a polymer $S\in\cR$, by using (\ref{bope}), we have that 
\begin{equation}
\label{bokp}
\sum_{\newatop{R\in\cR:}
              {\tilde R\cap\tilde S\neq\emptyset}}\bar\zeta_R\,e^{|\tilde R|}
\le
\sum_{x\in\tilde S}
\sum_{\newatop{R\in\cR:}{\tilde R\ni x}} \bar\zeta_R\,e^{|\tilde R|}
\le|\tilde S|
\;\Longrightarrow\;
\sum_{\newatop{\underline{R}\in\underline{\cR}:}
              {\underline{\tilde R}\cap\tilde S\neq\emptyset}}
     \varphi_T(\underline{R})\prod_{R\in\underline{R}}\bar\zeta_R
\le|\tilde S|
\end{equation}
Where the last bound is a direct consequence of the Theorem in \cite{[KP]}
whenever we choose there $a(R)=|\tilde R|$. 
The absolute convergence of (\ref{exp}) for $\ell$ large enough 
follows easily from (\ref{bokp}) once we recall that the activity 
of a cluster of polymer $\underline{R}$ such that 
$\tilde{\underline{R}}\cap\Delta=\emptyset$ is equal 
to zero and we note that for $\ell$ large enough 
\begin{equation}
\label{pold}
\|\zeta^{(\ell),\wp}_{m,R,\Delta}\|_\infty
 \le(\bar\zeta_R)^2
 \le\bar\zeta_R
\end{equation}
where the first inequality is just a rewriting of (\ref{boac}).

\noi{\it Proof of Theorem \ref{t:ip3}.}\  
{\it Item~\ref{i:ipo03}.}\  
First of all we recall $m=\cO_\wp^\ell n$ and 
define the family $V_{X,\Lambda}^{(\ell),\wp}$ in the following
way: for any $\xi\in\cX^{\ell,\wp}_m$ and $X\subset\subset\cL^{(\wp)}$ 
we set
\begin{equation}
\label{potVb}
V^{(\ell),\wp}_{X,\Lambda}(\xi,n):=
\!\!\!\!
\sum_{\newatop{G\in\cG:}
              {\sclos{\wp}{G}=X}}
        \!\!g(G)\Big[
        \log\frac{Z^{(\ell),\wp}_{m,G\cap\Delta}(\bar\xi)
                  \,\sqrt{\det\bV^{(\ell),\bar\eta}_{G\cap\Lambda}}}
                 {Z_{\cO_\wp(G\cap\Lambda)}(\cO_\ell\cO_\wp^\ell\bar\eta)}
       +
       \!\!\!\!
       \sum_{i\in\cO_\wp^\ell(G\cap\Lambda)}\frac{1}{2}m_i^2\Big] 
       +
       \!\!\!\!\!\!\!\!\!
       \sum_{\newatop{\underline R\in\underline{\tilde R}:}
            {\underline{\tilde R}\cup\partial^{(\wp),2}\underline{\tilde R}=X}} 
         \!\!\!\!\!\!\!\!\!
         \varphi_T(\underline R)\zeta^{(\ell),\wp}_{m,\underline R,\Delta}(\xi)
\end{equation}
We prove, now, that for any $X\subset\subset\cL^{(\wp)}$ 
\begin{equation}
\label{zerofuori}
X\subset\Lambda^\complement 
\;\;\Longrightarrow\;\;
V^{(\ell),\wp}_{X,\Lambda}=0
\end{equation}
Indeed, let $X\subset\Lambda^\complement$. Since
$$
\sclos{\wp}{G}=X\subset\Lambda^\complement 
\;\;\Longrightarrow\;\;
G\subset\Lambda^\complement
\;\;\Longrightarrow\;\;
G\cap\Delta=\emptyset=G\cap\Lambda
$$
we have that the first sum in (\ref{potVb}) is zero. Moreover, since
$
\underline{\tilde R}\cup\partial^{(\wp),2}\underline{\tilde R}=X
  \subset\Lambda^\complement\subset\Delta^\complement
$
definitions (\ref{attivita}), (\ref{zetaclu}), and (\ref{zeroerr})
imply that the second sum in (\ref{potVb}) is zero as well.
The expansion (\ref{ipo333}) finally follows from  
(\ref{Zpolym}), (\ref{gola6}),
(\ref{exp}), and (\ref{potVb}).

Suppose, now, that $X\cap\Delta=\emptyset$, by the same arguments used above 
we can easily prove that 
$$
V^{(\ell),\wp}_{X,\Lambda}(\xi,n)=
\sum_{\newatop{G\in\cG:}
              {\sclos{\wp}{G}=X}}
        g(G)\Big[
        \log\frac{\sqrt{\det\bV^{(\ell),\bar\eta}_{G\cap\Lambda}}}
                 {Z_{\cO_\wp(G\cap\Lambda)}(\cO_\ell\cO_\wp^\ell\bar\eta)}
       +
       \!\!\!\!
       \sum_{i\in\cO_\wp^\ell(G\cap\Lambda)}\frac{1}{2}m_i^2\Big] 
$$
Hence, $V^{(\ell),\wp}_{X,\Lambda}(\cdot,n)$ is constant. 
Finally, we note that if we also have $\diam_\wp(X)>5$ then 
$V^{(\ell),\wp}_{X,\Lambda}(\cdot,n)=0$ since there exists no $G\in\cG$ 
such that $\sclos{\wp}{G}=X$. The proof of the item is completed 
by choosing $\kappa$ large enough; in dimension two $\kappa\ge5$ 
does the job. 

\smallskip\noindent{\it Item~\ref{i:ipo13}.}\  
The statement follows from (\ref{potVb}), the measurability property 
(\ref{zetaclu}), and the following remarks: 
since $\wp>r$, where $r$ is the range of the original lattice gas 
interaction,
$Z^{(\ell),\wp}_{m,Y}(\cdot)\in\cF^{\ell,\wp}_{\partial^{(\wp)}Y}$ for 
all $Y\subset\cL^{(\wp)}$; 
$\partial^{(\wp)}[G\cap\Delta]\subset X$ whenever $\sclos{\wp}{G}=X$;
$\bar\xi_\Delta=\bar\eta_\Delta$.

\smallskip\noindent{\it Item~\ref{i:ipo13.5}.}\ The statement is true by 
construction.   

\smallskip\noindent{\it Item~\ref{i:ipo53}.}\ The statement trivially follows 
from (\ref{potVb}), (\ref{zeroerr}), (\ref{legami}), and (\ref{gola1})
by choosing $\kappa$ large enough. In dimension two it is enough $\kappa\ge8$.

\smallskip\noindent{\it Item~\ref{i:ipo43}.}\  
We first recall that $U$ is the potential of the 
original lattice gas model, $r$ 
its range (see Subsection~\ref{s:lg}), $\ell_0$, $b$, and $B$ the strong 
mixing constants (see Condition SM$^{(s)}(\ell_0,b,B)$ in 
Subsection~\ref{s:gibbs}).
Pick $x\in\cL^{(\wp)}$, and let $\alpha_1>0$ to be chosen later;
by using (\ref{potVb}), the triangular inequality, and the fact 
that $|g(G)|=1$ for all $G\in\cG$, we have 
\begin{equation}
\label{bol1}
\begin{array}{l}
{\displaystyle
\sum_{X\ni x}
 e^{\alpha_\ell\bT_\wp(X)}
  \sup_{\Lambda\subset\subset\cL^{(\wp)}}
    \|V^{(\ell),\wp}_{X,\Lambda}(\cdot,n)\|_\infty
} \\
{\displaystyle
\;\;\;\;\;\;
\le
\sum_{X\ni x}
 e^{\alpha_\ell\bT_\wp(X)}
  \sup_{\Lambda\subset\subset\cL^{(\wp)}}
  \sup_{\xi\in\cX^{\ell,\wp}}
      \sum_{\newatop{G\in\cG:}
                {\sclos{\wp}{G}=X}}
          \!\!\Big|
          \log\frac{Z^{(\ell),\wp}_{m,G\cap\Delta}(\bar\xi)
                    \,\sqrt{\det\bV^{(\ell),\bar\eta}_{G\cap\Lambda}}}
                   {Z_{\cO_\wp(G\cap\Lambda)}(\cO_\ell\cO_\wp^\ell\bar\eta)}
       +
       \!\!\!\!
       \sum_{i\in\cO_\wp^\ell(G\cap\Lambda)}\frac{1}{2}m_i^2\Big|
} \\
{\displaystyle
\;\;\;\;\;\;\;\;\;\;
+
\sum_{X\ni x}
 e^{\alpha_\ell\bT_\wp(X)}
  \sup_{\Lambda\subset\subset\cL^{(\wp)}}
   \bigg\|
       \sum_{\underline R\in\underline{\tilde R}:\,
             \underline{\tilde R}\cup\partial^{(\wp),2}\underline{\tilde R}=X}
     \varphi_T(\underline R)\zeta^{(\ell),\wp}_{m,\underline R,\Delta}(\cdot)
   \bigg\|_\infty
  \vphantom{\bigg\}}
}
\end{array}
\end{equation}
where $\bT_\wp(X)$ has been defined in (\ref{treedec0}), $\alpha_\ell$
is as in the hypothesis of the theorem, and 
$\Delta=\Lambda\cap\cL^{(\wp)}_\delta$.

We now bound the first sum in the right--hand side of (\ref{bol1}). By 
(\ref{gola1}) we have that the terms corresponding to $X\subset\cL^{(\wp)}$ 
such that $\diam_\wp(X)>\kappa$, where $\kappa$ is as in the proof of 
item~\ref{i:ipo03}, are equal to zero. Consider, now, 
$X\subset\subset\cL^{(\wp)}$ such that $\diam(X)\le\kappa$; we have that 
$$
e^{\alpha_\ell\bT_\wp(X)}=(e\ell)^{\alpha_1\bT_\wp(X)}
 \le(e\ell)^{(\kappa+1)^d\alpha_1}
$$
Moreover, since for each $G\in\cG$ one has 
$\diam_\wp(G)\le\kappa-2$, there 
exists a real $C'$ depending on $\ell_0$, $b$, $B$, 
$\|U\|_0$, $r$, and the dimension of the space $d$ such that 
$$
  \sup_{\Lambda\subset\subset\cL^{(\wp)}}
  \sup_{\xi\in\cX^{(\ell),\wp}}
      \sum_{\newatop{G\in\cG:}
                {\sclos{\wp}{G}=X}}
          \Big|
          \log\frac{Z^{(\ell),\wp}_{m,G\cap\Delta}(\bar\xi)
                    \,\sqrt{\det\bV^{(\ell),\bar\eta}_{G\cap\Lambda}}}
                   {Z_{\cO_\wp(G\cap\Lambda)}(\cO_\ell\cO_\wp^\ell\bar\eta)}
       +
       \sum_{i\in\cO_\wp^\ell(G\cap\Lambda)}\frac{1}{2}m_i^2\Big|
\le
C'\ell^d
$$
Indeed, the bound is easy for the logarithm of the 
partition functions, follows from (\ref{Ombi}) for the $m_i^2/2$ 
contribution, and follows from 
the strong mixing condition SM$(\ell_0,b,B)$ and the
result in \cite[Section~4]{[BCO]} for the 
$\det\bV^{(\ell),\bar\eta}_{G\cap\Lambda}$ terms. 
We can therefore conclude that the first sum in the right--hand side of 
(\ref{bol1}) is bounded by 
\begin{equation}
\label{bol1f}
|\{X\subset\subset\cL^{(\wp)}:\,X\ni 0,\,\diam_\wp(X)\le\kappa\}|
	\times(e\ell)^{(\kappa+1)^d\alpha_1}\times C'\ell^d
	=: C''\ell^{(\kappa+1)^d\alpha_1+d}
\end{equation}
where $0$ denotes the origin of the lattice $\cL^{(\wp)}$.

We bound, now, the second sum in the right--hand side of (\ref{bol1}).
Recall $\bar\zeta$ has been defined above (\ref{bope}) and chose $\ell$ so 
large that $\varepsilon(\ell)<1$. Let $X\subset\subset\cL^{(\wp)}$, we claim 
that for each cluster of polymers $\underline{R}$ such that 
$\underline{\tilde R}\cup\partial^{(\wp),2}\underline{\tilde R}=X$  
we have that 
\begin{equation}
\label{morlp}
\prod_{R\in\ul{\cR}}\bar\zeta_R=
\prod_{R\in\ul{\cR}}\varepsilon^{|\tilde R|}\le
\varepsilon^{|\ul{\tilde R}|}\le
\varepsilon^{|X|/5^d}\le
\varepsilon^{\bT_\wp(X)/5^d}= 
e^{(\bT_\wp(X)/5^d)\log\varepsilon} 
\end{equation}
where we have used $\bT_\wp(X)=|X|-1$.
We chose $\alpha_1<1/(2\cdot5^d\kappa')$, recall $\kappa'$ has been 
defined below (\ref{boac}). By taking $\ell$ large enough we have
\begin{equation}
\label{morlp1}
e^{\alpha_\ell\bT_\wp(X)}\,
\prod_{R\in\underline{R}}\bar\zeta_R\le1
\end{equation}
for any $X\subset\subset\cL^{(\wp)}$ and $\underline{R}$ such that 
$\underline{\tilde R}\cup\partial^{(\wp),2}\underline{\tilde R}=X$. 
Therefore, recalling (\ref{boac}), the second term on the r.h.s\ of 
(\ref{bol1}) can be bounded by 
\begin{equation}
\label{morlp2}
\begin{array}{l}
{\displaystyle
\sum_{X\ni x}
 e^{\alpha_\ell\bT_\wp(X)}
 \sum_{\newatop{\underline R\in\underline{\tilde R}:}
           {\underline{\tilde R}\cup\partial^{(\wp),2}\underline{\tilde R}=X}}
         \!\!\!\!\!\!
         |\varphi_T(\underline R)|
         \prod_{R\in\underline{R}}(\bar\zeta_R)^2
} \\
\phantom{merda}
\vphantom{\bigg\{}
{\displaystyle
\le
\sum_{X\ni x}
 \sum_{\newatop{\underline R\in\underline{\tilde R}:}
          {\underline{\tilde R}\cup\partial^{(\wp),2}\underline{\tilde R}=X}}
         \!\!\!\!\!\!
         |\varphi_T(\underline R)|
         \prod_{R\in\underline{R}}\bar\zeta_R
\le
\sum_{\newatop{e\in\cE:}
              {\srclos{\wp}{2}{e}\ni x}}
 \sum_{\newatop{\underline R\in\underline{\tilde R}:}
          {\underline{\tilde R}\cap e\neq\emptyset}}
         |\varphi_T(\underline R)|
         \prod_{R\in\underline{R}}\bar\zeta_R
\le
\sum_{\newatop{e\in\cE:}
              {\srclos{\wp}{2}{e}\ni x}}
|e|=:\kappa'''(d)
}
\end{array}
\end{equation}
where we used (\ref{bokp}).
The bound (\ref{ip233}) follows from 
(\ref{bol1}), (\ref{bol1f}), and (\ref{morlp2}) 
by setting $A_1:=C''+\kappa'''$.

\smallskip\noindent{\it Item~\ref{i:ipo43.5}.}\  
Pick $\Lambda\subset\subset\cL^{(\wp)}$, $X\subset\Lambda$,  
$n\in\cM^{(\ell),\wp}$ such that $\cL^{(\wp)}_\delta(n)\supset X$, and 
set $m=m(n)=\cO_\wp^\ell n$; 
then $\Delta:=\Lambda\cap\cL^{(\wp)}_\delta\supset X$.
Since $\wp>r$, where $r$ is the range of the original lattice gas interaction,
$Z^{(\ell),\wp}_{m,Y}(\cdot)\in\cF^{\ell,\wp}_{\partial^{(\wp)}Y}$ for 
all $Y\subset\cL^{(\wp)}$; 
then for each $G\in\cG$ such that $\sclos{\wp}{G}=X$ we have 
$Z^{(\ell),\wp}_{m,G}\in\cF^{\ell,\wp}_X$. 
Recall, now, that $\bar\eta$ is the reference configuration picked up 
in $\cX^{\ell,\wp}$ before (\ref{poggio1}) and that for each 
$\xi\in\cX^{\ell,\wp}$ we set 
$\bar\xi:=\bar\eta_\Delta\xi_{\Delta^\complement}$. Hence, for $G\in\cG$
\begin{equation}
\label{pol4}
\sclos{\wp}{G}=X\subset\Delta
\;\Longrightarrow\;
\bar\xi_X=\bar\eta_X
\;\Longrightarrow\;
Z^{(\ell),\wp}_{m,G}(\bar\xi)=Z^{(\ell),\wp}_{m,G}(\bar\eta)
\end{equation}
By using 
(\ref{potVb}), (\ref{pol4}), and the triangular inequality we have that 
\begin{equation}
\label{bol3}
\begin{array}{l}
{\displaystyle
    \|V^{(\ell),\wp}_{X,\Lambda}(\cdot,n)\|_\infty
\le
      \sum_{\newatop{G\in\cG:}
                {\sclos{\wp}{G}=X}}
          \!\!\Big|
          \log\frac{Z^{(\ell),\wp}_{m,G}(\bar\eta)
                    \,\sqrt{\det\bV^{(\ell),\bar\eta}_{G}}}
                   {Z_{\cO_\wp G}(\cO_\ell\cO_\wp^\ell\bar\eta)}
       +
       \!\!\!\!
       \sum_{i\in\cO_\wp^\ell G}\frac{1}{2}m_i^2\Big|
} \\
{\displaystyle
\;\;\;\;\;\;\;\;\;\;
\;\;\;\;\;\;\;\;\;\;
\;\;\;\;\;\;\;\;\;\;
\;\;\;\;\;\;\;\;\;\;
\;\;\;\;\;\;\;\;\;\;
\;\;\;\;\;\;\;\;\;\;
+
   \bigg\|
       \sum_{\underline R\in\underline{\tilde R}:\,
             \underline{\tilde R}\cup\partial^{(\wp),2}\underline{\tilde R}=X}
     \varphi_T(\underline R)\zeta^{(\ell),\wp}_{m,\underline R,\Delta}(\cdot)
   \bigg\|_\infty
  \vphantom{\bigg\}^{\big]}}
}
\end{array}
\end{equation}

The estimate (\ref{morlp2}) provides immediately an upper bound 
to the second term on the right--hand side of (\ref{bol3})
vanishing as $\ell\to\infty$.
We consider, now, the first term on the 
right--hand side of (\ref{bol3}): we first notice that (\ref{partlp}), 
(\ref{spihams}), and (\ref{Centr}) imply
$$
Z^{(\ell),\wp}_{m,G}(\bar\eta)
=
\sum _{\zeta\in\cX^{(\ell),\wp}_{m,G}} 
      e^{H^{\ell,\wp}_G(\zeta\bar\eta_{G^\complement})}
=
\sum _{\newatop{\sigma\in\cX_{\cO_\wp G}:}
                     {M^{(\ell)}_i(\sigma_{Q_\ell(i)})
                                    =m_i,\,i\in\cO_\wp^\ell G}}
      e^{H_G(\sigma(\cO_\ell\cO_\wp^\ell\bar\eta)_{G^\complement})}
$$
Hence, we have
\begin{equation}
\label{bol5}
\frac{Z^{(\ell),\wp}_{m,G}(\bar\eta)}
     {Z_{\cO_\wp G}(\cO_\ell\cO_\wp^\ell\bar\eta)}
=
\mu_{\cO_\wp G,z}^{\cO_\ell\cO_\wp^\ell\bar\eta}
          \big(\{M^{(\ell)}_i=m_i,\;i\in\cO_\wp^\ell G\}\big)
\end{equation}
where we recall the notation for the Gibbs measure associated with 
the original lattice gas potential $U$, see Subsection~\ref{s:lg}, with 
activity $z$. Recalling that by hypothesis $U$  
satisfies the strong mixing condition 
SM$(\ell_0,b,B)$, from Lemma~\ref{t:cruciale} we have that there exists 
$\ell_0'$, multiple of $\ell_0$, $b'$, and $B'$ positive reals, such 
that $U$ satisfies SM$(\ell_0',b',B')$ uniformly w.r.t.\ the activity in a 
neighbor of $z$ small enough. We can then apply the local central limit 
theorem \cite[Theorem~4.5]{[BCO]} and (\ref{bol5}) to write 
\begin{equation}
\label{bol7}
\begin{array}{l}
{\displaystyle 
  \Big|
    \log\frac{Z^{(\ell),\wp}_{m,G}(\bar\eta)
             \,\sqrt{\det\bV^{(\ell),\bar\eta}_{G}}}
           {Z_{\cO_\wp G}(\cO_\ell\cO_\wp^\ell\bar\eta)}
     +
    \!\!\!\!
    \sum_{i\in\cO_\wp^\ell G}\frac{1}{2}m_i^2\Big|
}\\
{\displaystyle 
 \;\;\;\;\;\;\;\;\;\;\;\;\;\;\;\;\;
  =\Big|
    \log\Big[\mu_{\cO_\wp G,z}^{\cO_\ell\cO_\wp^\ell\bar\eta}
          \big(\{M^{(\ell)}_i=m_i,\;i\in\cO_\wp^\ell G\}\big)
             \sqrt{\det\bV^{(\ell),\bar\eta}_{G}}\Big]
     +
    \!\!\!\!
    \sum_{i\in\cO_\wp^\ell G}\frac{1}{2}m_i^2\Big|
}\\
{\displaystyle 
 \;\;\;\;\;\;\;\;\;\;\;\;\;\;\;\;\;
  \le\Big|\frac{1}{2}\,
        \sum_{i,j\in\cO_\wp^\ell G}
          m_i\Big(\delta_{ij}
                  -2\pi\chi\ell^d(\bV^{(\ell),\bar\eta}_G)^{-1}_{ij}\Big)m_j
            \Big|
 +
 |\log(1+R_{\cO_\wp G}^{\cO_\ell\cO_\wp^\ell\bar\eta}(m))|
}\\
\end{array}
\end{equation}
where there exist two positive reals 
$\delta'$ and $C_1$ depending on $G$, $\|U\|_0$, and 
$\delta$, such that 
$$
\sup_{\sigma\in\cX}
\sup_{m\in\cM^{(\ell)}:\,\cL^{(\wp)}_\delta(\cO^\wp_\ell m)\supset G}
|R^\sigma_{\cO_\wp G}(m)|\le\frac{C_1}{\ell^{\delta'\,d}}
$$ 
Moreover,  by using the strong mixing condition it is not difficult 
to show, see results in \cite[Subsection~5.2]{[BCO]}, 
that there exists a positive real $C_2$ depending on $\|U\|_0$, such that 
$$
\Big|\delta_{ij}-2\pi\chi\ell^d(\bV^{(\ell),\bar\eta}_G)^{-1}_{ij}\Big|
\le\frac{C_2}{\ell}
$$
By using (\ref{bol3}), (\ref{bol7}), (\ref{morlp2}), and the two above 
estimates we get  
$$
\begin{array}{l}
{\displaystyle 
   \sup_{\Lambda\subset\subset\cL^{(\wp)}}\,
   \sup_{X\subset\Lambda}\,
   \sup_{\newatop{n\in\cM^{(\ell),\wp}:}
                 {\cL^{(\wp)}_\delta(n)\supset X}}\,
    \|V^{(\ell),\wp}_{X,\Lambda}(\cdot,n)\|_{\infty}
}\\
{\displaystyle 
\;\;\;\;\;\;\;\;\;
\le
   \sup_{X\subset\subset\cL^{(\wp)}}\,
   \sup_{\newatop{n\in\cM^{(\ell),\wp}:}
                 {\cL^{(\wp)}_\delta(n)\supset X}}\,
   \sum_{\newatop{G\in\cG:}{\sclos{\wp}{G}=X}}
    \Big(\frac{C_2}{2\ell}|G|^2\sup_{i\in\cO_\wp^\ell G} m_i^2
         +\frac{C_1}{\ell^{\delta'd}}\Big)
   +\kappa'''\,e^{-\alpha_\ell}
}\\
 \vphantom{\bigg\{^\big]}
{\displaystyle 
\;\;\;\;\;\;\;\;\;
\le
    2^{\kappa^d}
    \Big(\frac{C_2}{2\ell}|G|^2\ell^{1/3-2\delta}
         +\frac{C_1}{\ell^{\delta'd}}\Big)
   +\kappa'''\,e^{-\alpha_\ell}
}
\end{array}
$$
By taking the limit $\ell\to\infty$ we complete the proof of (\ref{ip42.5}).

\smallskip\noindent{\it Item~\ref{i:ipo23}.}\  
Suppose $X\cap\Lambda=X\cap\Lambda'$, then we have 
$X\cap\Delta=X\cap\Delta'$ where $\Delta':=\Lambda'\cap\cL^{(\wp)}_\delta(n)$. 
The thesis follows from (\ref{potVb}) and the explicit
expression (\ref{attivita}) of the activity. The key point is that the sums 
in (\ref{potVb}) are extended to subsets of the lattice inside $X$ and to
cluster of polymers $\underline{R}$ such that 
$\underline{\tilde R}\subset X$, and the intersection of 
$\Lambda$ and $\Lambda'$ with $X$ is the same.

\smallskip\noindent{\it Item~\ref{i:ipo33}.}\  
It follows directly from (\ref{potVb}) and (\ref{attivita}).
\qed

\sezione{Construction of the renormalized potential and convergence}{s:diso}
\par\noindent
In this section we construct the renormalized potential and prove the 
main Theorems~\ref{t:wgib} and \ref{t:conv}.

\subsec{Cluster expansion in the bad part of the lattice}{s:cebpl}
\par\noindent
In this subsection we apply the framework in \cite{[BCOabs]} 
to develop a multi--scale cluster expansion for
the {\em constrained model\/} in the {\em bad} part of the lattice 
on the basis of the uniformly 
convergent cluster expansion in the good part
of the lattice proven in Theorem~\ref{t:ip3}. 
Recall that in Section~\ref{s:spm} we 
have introduced $\wp=\cionc d\ell$, with $\ell$ the 
renormalization scale and $d$ the dimension of the lattice, 
on which we have defined the notion of goodness.

We are now ready to evoke \cite[Thm.~2.5]{[BCOabs]}. 
Let $\bar\cM^{(\ell),\wp}\subset\cM^{(\ell),\wp}$ be the set of full 
$\nu^{(\ell),\wp}$--measure in Theorem~\ref{t:partD}. For each 
$x\in\cL^{(\wp)}$ and $n\in\bar\cM^{(\ell),\wp}$ we let 
$k_x(n)<\infty$ be the integer such that item~\ref{gent:cas} in 
Definition~\ref{t:gentle} holds true and set 
\begin{equation}
\label{erre}
\varrho:=
\Big[\frac{1}{1-q}\Big(1
      +\frac{1}{\alpha_\ell}\log A_\ell\Big)\Big]
      \vee 0
\;\;\;\;\textrm{ and }\;\;\;\;
r_x^{(\wp)}(n):=\big[\Gamma_{k_x(n)}+2\vartheta_{k_x(n)}\big]\vee\varrho
\end{equation}
where $q:=2^{-5}3^{-2}$ and $\alpha_\ell$ and $A_\ell$ are as 
in Theorem~\ref{t:ip3}. 

\begin{teo}
\label{t:yteo}
Let the lattice gas potential $U$ satisfy Condition SM$(\ell_0,b,B)$.
Let $\Gamma,\gamma$ be the two moderate steep scales in 
(\ref{okscale}), 
and $\bar\cM^{(\ell),\wp}\subset\cM^{(\ell),\wp}$ be the set of full 
$\nu^{(\ell),\wp}$--measure in Theorem~\ref{t:partD}.
Then for each $\ell$ large enough multiple of $\ell_0$,
each $n\in\bar\cM^{(\ell),\wp}$, and each
$\Lambda\subset\subset\cL^{(\wp)}$ there exist two families of functions
$\{\Psi_{X,\Lambda}^{(\ell),\wp}(\cdot,n):
  \cX^{\ell,\wp}\to\bR,\,X\subset\subset\cL^{(\wp)}\}$ and 
$\{\Phi_{X,\Lambda}^{(\ell),\wp}(\cdot,n):
  \cX^{\ell,\wp}\to\bR,\,X\subset\subset\cL^{(\wp)}\}$
such that
\begin{enumerate}
\item
\label{p:tm1}
for each $\xi\in\cX^{\ell,\wp}$ we have the expansion 
\begin{equation}
\label{tm1}
  \log Z_{\cO_\wp^\ell n,\Lambda}^{(\ell),\wp}(\xi)=
   K^{(\wp)}_\Lambda-
   \frac{1}{2}\sum_{i\in\cO_\wp^\ell\Lambda}(\cO_\wp^\ell n)_i^2+
  \sum_{X\cap\Lambda\neq\emptyset}
     \left[\Psi_{X,\Lambda}^{(\ell),\wp}(\xi,n)
          +\Phi_{X,\Lambda}^{(\ell),\wp}(\xi,n)\right]
\end{equation}
where $K_\Lambda^{(\wp)}$ is as in Theorem~\ref{t:ip3}.
\item
\label{p:tm1.5}
for each $X\subset\subset\cL^{(\wp)}$ we have 
$\Psi_{X,\Lambda}^{(\ell),\wp}(\cdot,n),\Phi_{X,\Lambda}^{(\ell),\wp}(\cdot,n)
 \in\cF^{\ell,\wp}_{\cap\Lambda^\complement}$.
\end{enumerate}
Moreover, for each $n\in\bar\cM^{(\ell),\wp}$ 
\begin{enumerate}
\setcounter{enumi}{2}
\item
\label{p:tm2}
for each $\Lambda,\Lambda'\subset\subset\cL^{(\wp)}$, and each 
$X\subset\subset\cL^{(\wp)}$ we have 
$$
X\cap\Lambda=X\cap\Lambda'
\;\Longrightarrow\;
\Psi_{X,\Lambda}^{(\ell),\wp}(\cdot,n)=\Psi_{X,\Lambda'}^{(\ell),\wp}(\cdot,n)
\;\textrm{ and }\;
\Phi_{X,\Lambda}^{(\ell),\wp}(\cdot,n)=\Phi_{X,\Lambda'}^{(\ell),\wp}(\cdot,n)
$$
\item
\label{p:tm3}
Let $x\in\cL^{(\wp)}$, for any $X\subset\subset\cL^{(\wp)}$ 
if $X\ni x$ and $\diam_\wp(X)>r_x^{(\wp)}(n)$ then 
for each $\Lambda\subset\subset\cL^{(\wp)}$ we have 
$\Psi^{(\ell),\wp}_{X,\Lambda}(\cdot,n)=0$. 
In particular, for each $x\in\cL^{(\wp)}$ there exists a positive 
real $c^{(\wp)}_x(n)<\infty$ such that 
\begin{equation}
\label{tm3}
 \sum_{X\ni x}~
 \sup_{\Lambda\subset\subset\cL^{(\wp)}}~
 \|\Psi^{(\ell),\wp}_{X,\Lambda}(\cdot,n)\|_{\infty}\le c^{(\wp)}_x(n)
\end{equation}
\item
\label{p:tm4}
We have 
\begin{equation}
\label{tm2}
\sup_{x\in\cL^{(\wp)}}
 \sum_{X\ni x}e^{q\alpha_\ell\diam_\wp(X)/d}
 \sup_{\Lambda\subset\subset\cL^{(\wp)}}
  \|\Phi^{(\ell),\wp}_{X,\Lambda}(\cdot,n)\|_{\infty}
\le 
e^{-\alpha_\ell/d}+e^{-q\alpha_\ell\gamma_1/d}
             \Big( 
	          \frac{1+e^{-q\alpha_\ell/(2d^2)}}
		       {1-e^{-q\alpha_\ell/(2d^2)}}
             \Big)^d
\end{equation}
\end{enumerate}
\end{teo}

\smallskip
\noi{\it Proof of Theorem~\ref{t:yteo}.}\ 
By Theorem~\ref{t:partD} for each $n\in\bar\cM^{(\ell),\wp}$ there exists 
a gentle disintegration $\cG(n)$ of $\cL^{(\wp)}$ with respect 
to $\bG_0(n):=\cL_\delta^{(\wp)}(n)$, $\Gamma$, and $\gamma$. 
Moreover, Theorem~\ref{t:ip3} and (\ref{okscale})
ensure that for $\ell$ large enough 
\cite[Condition~2.1]{[BCOabs]} 
is fulfilled with $A$ and $\alpha$ given respectively by 
$A_\ell$ and $\alpha_\ell/d$. 
Note that the factor $1/d$ is due to the fact that here we are using,
as distance on the lattice $\cL^{(\wp)}$, the supremum of the coordinates,
while in \cite{[BCOabs]} we used their sum.
Moreover, we note that items 1--4 in the hypotheses of 
\cite[Theorem~2.5]{[BCOabs]} are satisfied by the scales $\Gamma$, $\gamma$
in (\ref{okscale}).

Items~\ref{p:tm1}--\ref{p:tm4} are, then, a simple restatement of results in 
\cite[Theorem~2.5]{[BCOabs]} once we define the real
\begin{equation}
\label{tm3.1}
\begin{array}{rcl}
{\displaystyle 
 \!\!\!\!\!\!\!\!
 c^{(\wp)}_x(n)
}
&\!:=\!&
 A_\ell
 +k_x(n)(\Gamma_{k_x(n)}+1+2\vartheta_{k_x(n)})^{2d} 
 \vphantom{\Big]}\\
&&
  \times
  \big[\wp^d(\log 2+\|U\|_0)
  +k_x(n)(1\vee A_\ell)(8^d+1)\big]
\end{array}
\end{equation}
for all $n\in\bar\cM^{(\ell),\wp}$ and $x\in\cL^{(\wp)}$.
\qed

\subsec{Locality of the renormalized potential}{s:loc}
\par\noindent
To prove the Gibbsianity of the renormalized measure we need to introduce 
functions of the renormalized variable $n$ which will play the role 
of potentials.
In the subsection we state and prove 
a locality property of the finite volume potentials.

\bteo{t:mep.1} 
Assume the hypotheses of Theorem~\ref{t:yteo} are satisfied.
Let also $X,\Lambda\ssu\cL^{(\wp)}$,  
$n,n'\in\bar\cM^{(\ell),\wp}$ such that $n_X=n'_X$.
Then 
\begin{equation}  
\label{epa.12}  
\Psi_{X,\Lambda}^{(\ell),\wp}(\cdot,n)=
\Psi_{X,\Lambda}^{(\ell),\wp}(\cdot,n')
\;\;\;\textrm{ and }\;\;\; 
\Phi_{X,\Lambda}^{(\ell),\wp}(\cdot,n)=
\Phi_{X,\Lambda}^{(\ell),\wp}(\cdot,n')
\end{equation}
\eteo 

The proof of Theorem~\ref{t:mep.1} needs 
to some extent the details of the recursive construction of 
$\Psi_{X,\Lambda}^{(\ell),\wp}$ and $\Phi_{X,\Lambda}^{(\ell),\wp}$ 
provided in \cite{[BCOabs]} to which we refer for more 
details; we outline here the main idea beneath the computation. 

We pick $n\in\bar\cM^{(\ell),\wp}$ and 
recall the notion of gentle disintegration given 
in Definition~\ref{t:gentle}; for $j\ge 1$ we say 
$G,G'\subset\cG_{\ge j}(n)$ are {\em $j$--connected} iff
$G\cap G'\cap\cG_j(n)\neq\emptyset$.
A system $G_1,\dots,G_k$ with $G_h\subset\cG_{\ge j}(n)$
is said to be {\em $j$--connected} iff
for each $h,h'\in\{1,\dots,k\}$
there exists $h_1,\dots,h_{k'}\in\{1,\dots,k\}$
such that $G_h=G_{h_1}$, $G_{h_{k'}}=G_{h'}$ and 
$G_{h_i}$ is $j$--connected to $G_{h_{i+1}}$ for all $i=1,\dots,k'-1$.

A $j$--polymer is a collection 
$\{(G_1,s_1),\cdots,(G_k,s_k)\}$, with 
$G_h\subset\cG_{\ge j}(n)$ and $s_h\ge 0$ integers for $h=1,\dots,k$, such
that the system $G_1,\dots,G_k$ is $j$--connected. We denote by $\cR_j(n)$
the collection of all the $j$--polymers. 
Given a $j$--polymer $R=\{(G_1,s_1),\dots,(G_k,s_k)\}$ and
$i\ge j$ we set
$R\rest_{i}:=\bigcup_{h=1}^k G_h\cap\cG_i(n)\subset\cG_i(n)$
and 
$R\rest_{\ge i}:=\bigcup_{i'\ge i}R\rest_{i'}\subset\cG_{\ge i}(n)$.
We also introduce the support of the polymer 
\begin{equation}
\label{supol}
\supp{R}:=\bigcup_{h=1}^k \Es_{s_h}(G_h)\subset\cL^{(\wp)}
\;\;\;\textrm{ with }\;\;\;
\Es_s(G_h):=
 \left\{x\in\bL:\,\dis_\wp(x,\env^{(\wp)}(\proj{G_h}))\le \vartheta_j+s\right\}
\end{equation}
for all non--negative integer $s$ and $h=1,\dots,k$, 
where we have set 
where $\proj{G}:=\bigcup_{g\in G}g$ for all $G\subset\cG_{\ge 1}(n)$ and 
we recall $\env^{(\wp)}(\Delta)$ is, for all $\Delta\subset\subset\cL^{(\wp)}$,
the smallest parallelepiped with faces parallel to the coordinate directions 
and containing $\Delta$. 
Moreover for each $s\ge 0$, $h=1,\dots,k$,
we set $\dEs_s(G_h):=\Es_s(G_h)\setminus\Es_{s-1}(G_h)$
where we understand $\Es_{-1}(G_h)=\emptyset$.
We note that the set $\Es_s(G_h)$ will realize, see (\ref{cjp->Ys}),
the volume cutoff mentioned at the end of Subsection~\ref{s:str}.

Given two $j$--polymers $R,S\in\cR_j(n)$
we say they are {\em $j$--compatible} iff
$R\rest_{j} \cap S\rest_{j} =\es$. Conversely we say that $R,S$ are
{\em $j$--incompatible} iff they are not {\em $j$--compatible}.
We say that a collection $\ul{R}=\{R_1,\dots,R_k\}$, where
$R_h\in\cR_j(n)$, $h=1,\dots,k$, of $j$--polymers forms a {\em
cluster of $j$--polymers}
iff it is not decomposable into two non empty
subsets $\ul{R}= \ul{R}_1\cup\ul{R}_2$ such that every pair
$R_1\in\ul{R}_1$, $R_2\in\ul{R}_2$ is $j$--compatible.
We denote by $\ul{\cR}_j(n)$ the collection of all
the clusters of $j$--polymers.
For $i\ge j$, $\ul{R}\in\ul{\cR}_j(n)$ we set
$\ul{R} \rest_{i} := \bigcup_{R\in\ul{R}} R \rest_{i}$,
$\ul{R}\rest_{\ge i}:=\bigcup_{i'\ge i} \ul{R} \rest_{i'}$;
we set $\supp\ul{R}:=\bigcup_{R\in\ul{R}}\supp R$.
We note that $\supp\ul{R}$ is a $\wp$--connected subset of $\cL^{(\wp)}$. 

For any $\Lambda\subset\subset\cL^{(\wp)}$, 
$G\subset\subset\cG_{\ge 1}(n)$, and $s\ge 0$ we define the two 
collection of subsets of the lattice  
\begin{equation}
\label{strangeX}
\begin{array}{rcl}
\Upsilon_\Lambda
&\!:=\!&
\{Y\subset\subset\cL^{(\wp)}:\, Y\cap\Lambda\neq\emptyset
\textrm{ and }
Y\cap\big(\srclos{\wp}{\kappa}{\Lambda}\big)^\complement=\emptyset
\}
\vphantom{\bigg]}\\
\Upsilon_\Lambda(G,s)(n)
&\!:=\!&
\{Y\in\Upsilon_{\Lambda\cap\bG_0(n)}:\, 
\xi(Y)(n)=G,\,Y\subset\Es_s(G),\,Y\cap\dEs_s(G)\neq\emptyset 
\}
\end{array}
\end{equation}
where for each $Y\subset\cL^{(\wp)}$ we have set 
\begin{equation}
\label{xixi}
\xi(Y)(n):=\{g\in\cG_{\ge 1}(n):\,g
 \cap Y\neq\emptyset\}\subset\cG_{\ge 1}(n)
\end{equation}
and $\kappa$ has been introduced in Theorem~\ref{t:ip3}.
Recalling Theorem~\ref{t:ip3}, for
$i\ge 1$, $g\in\cG_i(n)$, $G\ssu\cG_{\ge i}(n)$ such that
$G\cap\cG_i(n)\neq\emptyset$, and $s\ge 0$, we define the $0$--order 
effective potential 
\begin{equation}
\label{autopot}
\begin{array}{rcl}
\Psi^{(i,0)}_{g,\L}(\cdot,n)
&\!:=\!&
{\displaystyle 
 \sum_{Y\in\Upsilon_\Lambda(g,0)(n)} 
   V^{(\ell),\wp}_{Y,\Lambda}(\cdot,n)
}\\
\Phi^{(i,0)}_{G,s,\L}(\cdot,n)
&\!:=\!&
{\displaystyle
 \id_{\{(|G|,s)\neq(1,0)\}}\;
 \sum_{Y\in\Upsilon_\Lambda(G,s)(n)}
  V^{(\ell),\wp}_{Y,\Lambda}(\cdot,n)
}
\end{array}
\end{equation}

We next define by recursion on $j$ the $j$--order effective potentials:
as recursive hypotheses we assume that there exist the families 
$$
\{\Psi^{(i,k)}_{g,\L}(\cdot,n):\cX^{\ell,\wp}\to\bR,
                               \,\Lambda\subset\subset\bL\}
\;\;\;\;\;\textrm{ and }\;\;\;\;\;
\{\Phi^{(i,k)}_{G,s,\L}(\cdot,n):\cX^{\ell,\wp}\to\bR,
                               \,\Lambda\subset\subset\bL\}
$$ 
for any $k=0,\dots,j-1$, any $i\ge k+1$, any $g\in\cG_i(n)$,
any $G\subset\subset\cG_{\ge i}(n)$, such that
$G\cap\cG_i(n)\neq\emptyset$, and any $s\ge 0$. 
We integrate on the scale $j$ and define the $j$--order effective potentials
$\Ps^{(i,j)}_{g,\L}$ and
$\Phi^{(i,j)}_{G,s,\L}$ for $i\ge j+1$,
any $g\in\cG_i(n)$, any $G\subset\subset\cG_{\ge i}(n)$, such that
$G\cap\cG_i(n)\neq\emptyset$, and $s\ge 0$.

Given $g\in\cG_j(n)$, $G\su\su\cG_{\ge j}(n)$ such that $G\cap\cG_j(n)\neq\es$,
and $s\ge 0$ we sum all the lower order contributions, obtained by performing 
the $k$--order cluster expansion with $k=1,\cdots,j-1$, to the effective 
potentials associated with such vertex namely, we define 
\be{intj}
\Psi^{(j)}_{g,\Lambda}(\cdot,n):=
\sum_{k=0}^{j-1}\Psi^{(j,k)}_{g,\Lambda}(\cdot,n)
\;\;\;\;\;\textrm{ and }\;\;\;\;\;
\Phi^{(j)}_{G,s,\L}(\cdot,n):=
\sum_{k=0}^{j-1}\Phi^{(j,k)}_{G,s,\L}(\cdot,n)
\end{equation}

For each vertex $g\in\cG_j(n)$ and block spin configuration 
$\xi\in\cX^{\ell,\wp}$ we define the partition function
\be{ZG}
Z_{g,\Lambda}^{(j)}(\xi,n):=
\sum_{\zeta\in\cX^{(\ell),\wp}_{\cO_\wp^\ell n,g}}
\exp\Big\{
\sum_{\newatop{Y\subset\subset\cL^{(\wp)}:\,
      Y\cap\L\neq\emptyset}{Y\cap\L\subset g\cap\L}}
U^{\ell,\wp}_Y(\zeta\xi_{g^\complement})
+\Psi^{(j)}_{g,\Lambda}(\zeta\xi_{g^\complement},n)
\Big\}
\end{equation}
where $U^{\ell,\wp}_Y$ are the original lattice gas potentials 
rewritten, see the discussion before Theorem~\ref{t:ip3}, for the 
block spin variable in $\cX^{(1),\ell,\wp}\equiv\cX^{\ell,\wp}$, 
and the probability measure $\nu_{g,\Lambda,n,\xi}^{(j)}$ on
$\cX^{(\ell),\wp}_{\cO_\wp^\ell n,g}$ by setting
\be{nuG}
\nu^{(j)}_{g,\Lambda,n,\xi}(\zeta):=
\frac{\delta_\xi(\zeta_{g\cap\Lambda^\complement})}
     {Z_{g,\Lambda}^{(j)}(\xi,n)}
\exp\Big\{
\sum_{\newatop{Y\subset\subset\cL^{(\wp)}:\,
      Y\cap\L\neq\emptyset}{Y\cap\L\subset g\cap\L}}
U^{\ell,\wp}_Y(\zeta\xi_{g^\complement})
+\Psi^{(j)}_{g,\Lambda}(\zeta\xi_{g^\complement},n)
\Big\}
\end{equation}
for any $\zeta\in\cX^{(\ell),\wp}_{\cO_\wp^\ell n,g}$.

We consider, now, a bond $G\subset\subset\cG_{\ge j+1}(n)$, 
such that $G\cap\cG_j(n)\neq\emptyset$, and $s\ge 0$;
our aim is the definition of the $j$--order effective potential associated 
to such a bond and due to the integration over the $j$--gentle sites. 
We set 
\be{cjp->Ys}
\ul{\cR}_j(G,s)(n):=
\lg \ul{R}\in \ul{\cR}_j(n): \
\ul{R}\rest_{\ge j+1} = G\, , \ \supp\ul{R}\su \Es_s(G) ,\
\supp\ul{R} \cap \dEs_s(G) \neq\es
\rg
\end{equation}
We define, now, the activity of a cluster of polymers 
$\ul{R}\in\ul{\cR}_j(G,s)(n)$, whose set of vertices of gentleness order 
greater or equal to $j+1$ is given exactly by $G$, by setting 
\be{accrj}
\zeta_{\ul{R},\L}(\cdot,n):=\prod_{R\in\ul{R}}\zeta_{R,\L}(\cdot,n)
\end{equation}
where for $\xi\in\cX^{\ell,\wp}$ we have set  
\be{acRj}
\zeta_{R,\Lambda}(\xi,n):=
\sum_{\zeta\in\cX^{(\ell),\wp}_{\cO_\wp^\ell n,\proj{R\rest_j}}}
\prod_{g\in R\rest_j}
\nu^{(j)}_{g,\L,n,\xi}(\zeta_g)
\prod_{h=1}^k\[ 
\exp\big\{\Phi^{(j)}_{G_h,s_h,\L}(\zeta\xi_{(\proj{R\rest_j})^\complement})
\big\} -1\]
\end{equation}
for all $R=\{(G_1,s_1),\dots(G_k,s_k)\}\in\ul{R}$.

We are now ready to define the $j$--order effective potentials. Let $i\ge j+1$,
$g\in\cG_i(n)$, $G\ssu\cG_{\ge i}(n)$ such that 
$G\cap\cG_i(n)\neq\es$ and $s\ge 0$, then we set 
\begin{equation}
\label{selfintij}
\begin{array}{rcl}
\Psi^{(i,j)}_{g,\L}(\cdot,n)
&\!:=\!&
{\displaystyle 
 \sum_{\ul{R}\in\ul{\cR}_j(g,0)(n)}
   \f_T\(\ul{R}\)\,\zeta_{\ul{R},\L}(\cdot.n)
}\\
\Phi^{(i,j)}_{G,s,\L}(\cdot,n)
&\!:=\!&
{\displaystyle
 \id_{\{(|G|,s)\neq(1,0)\}}\;
 \sum_{\ul{R}\in\ul{\cR}_j(G,s)(n)}\f_T\(\ul{R}\)\,
                \zeta_{\ul{R},\Lambda}(\cdot,n)
}
\end{array}
\end{equation}
In \cite[Section~4]{[BCOabs]} it is proven that the $j$--order effective 
potentials depend only on those block spins associated to sites 
of order greater than $j$ lying inside the vertices which 
label the function; more precisely 
\begin{equation}
\label{miseff}
\Psi^{(i,j)}_{g,\L}(\cdot,n)
 \in\cF^{\ell,\wp}_{(\Es_0(g)\cap\Lambda^\complement)\cup g}
\;\;\;\;\textrm{ and }\;\;\;\;\;
\Phi^{(i,j)}_{G,s,\L}(\cdot,n)
 \in\cF^{\ell,\wp}_{(\Es_s(G)\cap\Lambda^\complement)\cup\proj{G}}
\end{equation}
where we recall $\proj{G}:=\bigcup_{g\in G}g$. 

We can finally define the functions 
$\Psi_{X,\Lambda}^{(\ell),\wp}$ and $\Phi_{X,\Lambda}^{(\ell),\wp}$ 
whose existence has been stated in Theorem~\ref{t:yteo}. More precisely 
for each $X,\Lambda\subset\subset\cL^{(\wp)}$ and $n\in\bar\cM^{(\ell),\wp}$ 
we 
define
\begin{equation}
\label{slrj.1}
\begin{array}{rcl}
\Psi_{X,\Lambda}^{(\ell),\wp}(\cdot,n)
&\!\!\!\!=\!\!\!\!&
{\displaystyle
 \id_{\{\diam_\wp(X)\le\varrho,X\cap\Lambda\neq\emptyset,\xi(X)(n)=\emptyset\}}
 V^{(\ell),\wp}_{X,\Lambda}(\cdot,n)
 +
 \sum_{j\ge1}
 \sum_{g\in\cG_j(X)(n)}\log Z_{g,\Lambda}^{(j)}(\cdot,n)
}\\
\vphantom{\bigg\{}
\Phi_{X,\Lambda}^{(\ell),\wp}(\cdot,n)
&\!\!\!\!=\!\!\!\!&
{\displaystyle
 \id_{\{\diam_\wp(X)>\varrho,X\cap\Lambda\neq\emptyset,\xi(X)(n)=\emptyset\}}
 V^{(\ell),\wp}_{X,\Lambda}(\cdot,n)
 +
 \sum_{j\ge1}
 \sum_{\ul{R}\in\ul{\cR}_j(X)(n)} 
 \varphi_T\(\ul{R}\)\zeta_{\ul{R},\Lambda}(\cdot,n) 
}\\
\end{array}
\end{equation}
where we have introduced the two sets 
\begin{equation}
\label{GXRX.1} 
\begin{array}{rcl}
{\displaystyle 
\cG_j(X)\,(n)}&:=&{\displaystyle \lg g\in\cG_j(n):\:\Es_0(g)=X\rg  }\\
\vphantom{\bigg\{}
{\displaystyle \ul{\cR}_j(X)\, (n) }&:=&{\displaystyle
     \lg \ul{R}\in \ul{\cR}_j(n):\:  
     \ul{R}\rest_{\ge j+1} =\es,\, \supp\ul{R} = X \rg} 
\end{array} 
\end{equation} 
We remark that the sums over $j$ in (\ref{slrj.1}) are extended to a finite 
number of terms, indeed for $j$ such that $\vartheta_j>\diam_\wp(X)$ the 
sets $\cG_j(X)(n)$ and $\ul{\cR}_j(X)(n)$ are empty for all 
$n\in\bar\cM^{(\ell),\wp}$.
For each $X\subset\subset\cL^{(\wp)}$ and $j\ge 1$ we finally  set
\begin{equation}
\label{Gtild}
\widetilde{\cG}_j(X)(n):= \lg g\in\cG_j(n):\: \Es_0(g)\subset X\rg  
\end{equation}
and $\widetilde{\cG}_{\ge j}(X)(n):=\bigcup_{i\ge j}\widetilde{\cG}_i(X)(n)$. 
 
\blem{t:mGXRX.1} 
Let $X,\Lambda\subset\subset\cL^{(\wp)}$,
$n,n'\in \bar\cM^{(\ell),\wp}$ such that $n_X =n'_X$, 
then 
\begin{enumerate} 
\item
\label{pi}
for each $j\ge 1$ we have
$\widetilde{\cG}_j(X)(n) =\widetilde{\cG}_j(X)(n')$ and 
$\widetilde{\cG}_{\ge j}(X)(n) =\widetilde{\cG}_{\ge j}(X)(n')$;
\item  
\label{pii}
for each $j\ge 1$ we have
$\cG_j(X)(n) =\cG_j(X)(n')$ and   
$\ul{\cR}_j(X) \, (n) = \ul{\cR}_j(X) \, (n')$; 
\item 
\label{piii}
we have
$\Upsilon_\Lambda(G,s)(n)=\Upsilon_\Lambda(G,s)(n')$ for any   
$G\subset\widetilde{\cG}_{\ge 1}(X)(n)=
         \widetilde{\cG}_{\ge 1}(X)(n')$ 
and $s\ge 0$ such that $\Es_s(G)\subset X$; 
\item 
\label{piv}
for each $j\ge 1$ we have that
$\ul{\cR}_j(G,s)(n)=\ul{\cR}_j(G,s)(n')$ for any  
$G\subset\widetilde{\cG}_{\ge j+1}(X)(n)=
         \widetilde{\cG}_{\ge j+1}(X)(n')$
and $s\ge 0$ such that $\Es_s(G)\subset X$;   
\item 
\label{pvi}
for each $j\ge 0$, $i\ge j+1$, $g\in\widetilde{\cG}_i\left(X\right)(n)=
                                    \widetilde{\cG}_i\left(X\right)(n')$,
we have
$\Psi^{(i,j)}_{g,\Lambda}(\cdot,n)=\Psi^{(i,j)}_{g,\Lambda}(\cdot,n')$; 
\item 
\label{pvii}
for each $j\ge 0$, $i\ge j+1$,
$G\subset\widetilde{\cG}_{\ge i}\left(X\right)(n)=
         \widetilde{\cG}_{\ge i}\left(X\right)(n')$ and $s\ge0$ 
such that $G\cap\widetilde{\cG}_i\left(X\right)(n)= 
           G\cap\widetilde{\cG}_i\left(X\right)(n')\neq \emptyset$ 
and $\Es_s(G)\subset X$, 
we have   
$\Phi^{(i,j)}_{G,s,\Lambda}(\cdot,n)=\Phi^{(i,j)}_{G,s,\Lambda}(\cdot,n')$;
\item 
\label{pviii}
for each $j\ge1$, $g\in\cG_j(X)(n)=\cG_j(X)(n')$,
we have   
$Z^{(j)}_{g,\Lambda}(\cdot,n)=Z^{(j)}_{g,\Lambda}(\cdot,n')$;
\item 
\label{pix}
for each $j\ge1$, $\ul{R}\in\ul{\cR}_j(X)(n)=\ul{\cR}_j(X)(n')$,
we have   
$\zeta_{\ul{R},\Lambda}(\cdot,n)=\zeta_{\ul{R},\Lambda}(\cdot,n')$.
\end{enumerate} 
\elem 

\noi{\it Proof of Lemma~\ref{t:mGXRX.1}.}\ 
We first prove items~\ref{pi}--\ref{piv} separately, then \ref{pvi} and 
\ref{pvii} by induction. Items~\ref{pviii} and \ref{pix} will be a byproduct 
of the proof of \ref{pvi} and \ref{pvii}. 

{\it Item~\ref{pi}.}\
Let $g\in\widetilde{\cG}_j(X)(n)$; since 
$X\supset\Es_0(g)\supset 
 B^{(\wp)}_{\vartheta_j}(g)$,  
item~\ref{i:partD3} in Theorem~\ref{t:partD} 
and $n_X=n'_X$ imply $g\in\cG_j(n')$.
Now, $g\in\widetilde{\cG}_j(X)(n')$ follows from
$g\in\cG_j(n')$ and the geometrical property  
$\Es_0(g)\subset X$. Hence  
$\widetilde{\cG}_j(X)(n)\subset\widetilde{\cG}_j(X)(n')$ 
and, by interchanging the role of $n$ and $n'$, 
we get the equality. The second equality follows immediately 
from the first one. 

{\it Item~\ref{pii}.}\
The proof of the first 
equality is similar to the proof of item \ref{pi}. 
Proof of the second equality. 
Let $\ul{R}\in\ul{\cR}_j(X)(n)$ and 
$G=\{g_1,\dots,g_k\}:=\ul{R}\rest_j$ be the  
collection of all the vertices the cluster of polymers  
$\ul{R}$ is built of. By definition  
$g_h\in\cG_{j}(n)$ for any $h=1,\dots,k$. 
The definition of support of a polymer and    
$\supp\ul{R}=X$ imply that  
$\Es_0(g_h)\subset X$ for any  
$h=1,\dots,k$. Hence, $n_X=n'_X$ and 
item~\ref{i:partD3} in Theorem~\ref{t:partD} imply   
$g_h\in\cG_j(n')$ for any $h=1,\dots,k$, which yields 
$\ul{R}\in\ul{\cR}_j(X)(n')$. Hence  
$\ul{\cR}_j(X)(n)\subset\ul{\cR}_j(X)(n')$ and, by interchanging 
the role of $n$ and $n'$, we get the equality. 

{\it Item~\ref{piii}.}\
Recall (\ref{strangeX}), 
let $Y\in\Upsilon_\Lambda(G,s)(n)$. Then we have 
\begin{equation*} 
Y\subset\Es_s(G) 
\;\;\;\textrm{ and }\;\;\; 
Y\cap\dEs_s(G)\neq\emptyset 
\end{equation*} 
Moreover,  
$X\supset\Es_s(G)\supset Y$, 
$n_X=n'_X$, and 
$Y\in\Upsilon_{\Lambda\cap\cL^{(\wp)}_\delta(n)}$  
imply 
$Y\in\Upsilon_{\Lambda\cap\cL^{(\wp)}_\delta(n')}$.  
Finally,  
$\xi(Y)(n)=G$ and $n_Y=n'_Y$ 
imply $\xi(Y)(n')=G$. 
All the properties ensuring $Y\in\Upsilon(G,s)(n')$ have been  
verified, hence we have  
$\Upsilon(G,s)(n)\subset\Upsilon(G,s)(n')$ and, by interchanging the role 
of $n$ and $n'$, we get the equality. 

{\it Item~\ref{piv}.}\
Let $\ul{R}\in\ul{\cR}_j(G,s)(n)$, 
$F=\{f_1,\dots,f_k\}:=\ul{R}\rest_{\ge j}$ the  
collection of all the vertices the cluster of polymers  
$\ul{R}$ is built of (note $G\subset F$) 
and $I=\{i_1,\dots,i_k\}$ the collection of integral 
numbers such that $f_h\in\cG_{i_h}(n)$ for any $h=1,\dots,k$ namely,
$i_h$ is the grade of $f_h$.
Remark that for each $h=1,\dots,k$ either $i_h=j$ or $f_h\in G$. 
We next prove that $f_h\in\tilde\cG_{i_h}(X)(n)$, for $h=1,\dots,k$, 
by showing that $\Es_0(f_h)\subset X$. Indeed,
in the case $f_h\in G$, we have that 
$G\subset\tilde{\cG}_{\ge j+1}(X)(n)$ implies 
$\Es_0(f_h)\subset X$; on the other hand, if $i_h=j$, then,
recall the definition (\ref{supol}) of support of a polymer,   
$X\supset\Es_s(G)\supset\supp\ul{R}\supset\Es_0(f_h)$. 
Now, from item \ref{pi} we get  
$f_h\in\widetilde{\cG}_{i_h}\left(X\right)(n')$ for any $h=1,\dots,k$, which
implies $\ul{R}\in\ul{\cR}_j(n')$. 
We remark, finally, that 
$\ul{R}\in\ul{\cR}_j(G,s)(n)$ 
$\Longrightarrow$ 
$\ul{R}\rest_{\ge j+1}=G$, $\supp\ul{R}\subset\Es_s(G)$
and $\supp\ul{R}\cap\dEs_s(G)\neq \emptyset$.
Hence, 
$\ul{R}\in\ul{\cR}_j(G,s)(n')$. 
We then have  
$\ul{\cR}_j(G,s)(n)\subset\ul{\cR}_j(G,s)(n')$ and,  
by interchanging the role of $n$ and $n'$, we get the equality. 

{\it Items~\ref{pvi}--\ref{pvii}.}\
We proceed by induction on $j$. Let $j=0$. 
For each $i\ge1$ and $g\in\tilde\cG_i(X)(n)=\tilde\cG_i(X)(n')$, by
using (\ref{autopot}), item~\ref{piii},
$Y\in\Upsilon_\Lambda(g,0)(n)=\Upsilon_\Lambda(g,0)(n')$, and 
item~\ref{i:ipo33} in Theorem~\ref{t:ip3}, we have that 
$\Psi^{(i,0)}_{g,\Lambda}(\cdot,n)=\Psi^{(i,0)}_{g,\Lambda}(\cdot,n')$. 
The statement in item~\ref{pvii} for $j=0$ is proven similarly.  

Now, we fix the integer $j$ and suppose that the statements 
in items~\ref{pvi} and \ref{pvii} are verified for all 
$k=0,\dots,j-1$, $i\ge k+1$. From the inductive hypotheses and 
(\ref{intj}) we have that 
the equality 
\begin{equation}
\label{loc1}
\Psi^{(j)}_{g,\Lambda}(\cdot,n)=\Psi^{(j)}_{g,\Lambda}(\cdot,n')
\end{equation}
holds for all $g\in\tilde\cG_j(X)(n)=\tilde\cG_j(X)(n')$.
Hence, recalling (\ref{ZG}) and (\ref{nuG}), we have
\begin{equation}
\label{loc2}
Z^{(j)}_{g,\Lambda}(\cdot,n)=Z^{(j)}_{g,\Lambda}(\cdot,n')
\;\;\;\;\textrm{ and }\;\;\;\;
\nu^{(j)}_{g,\Lambda,n,\xi}=\nu^{(j)}_{g,\Lambda,n',\xi}
\end{equation}
for any $g\in\tilde\cG_j(X)(n)=\tilde\cG_j(X)(n')$
and $\xi\in\cX^{\ell,\wp}$. 

Analogously, from the inductive hypotheses and (\ref{intj}) we have 
\begin{equation}
\label{loc3}
\Phi^{(j)}_{G,s,\Lambda}(\cdot,n)=
\Phi^{(j)}_{G,s,\Lambda}(\cdot,n')
\end{equation}
for any $G\subset\tilde\cG_{\ge j}(X)(n)=\tilde\cG_{\ge j}(X)(n')$ and
$s\ge0$ such that 
$G\cap\tilde\cG_j(X)(n)=G\cap\tilde\cG_j(X)(n')\neq\emptyset$ and
$\Es_s(G)\subset X$.

Consider, now, $i\ge j+1$ and 
$G\subset\widetilde{\cG}_{\ge i}\left(X\right)(n)=
         \widetilde{\cG}_{\ge i}\left(X\right)(n')$,
such that $G\cap\widetilde{\cG}_i\left(X\right)(n)= 
           G\cap\widetilde{\cG}_i\left(X\right)(n')\neq \emptyset$, 
and $s\ge 0$ such that $\Es_s(G)\subset X$. 
Since 
$G\subset\widetilde{\cG}_{\ge i}\left(X\right)(n)=
         \widetilde{\cG}_{\ge i}\left(X\right)(n')$ then
$g\in\tilde\cG_{\ge i}(X)(n)=\tilde\cG_{\ge i}(X)(n')$ for all $g\in G$.
Let $\ul{R}\in\ul{\cR}_j(G,s)(n)=\ul{\cR}_j(G,s)(n')$,
for all $g\in\ul{R}\rest_j$ we have that 
$\Es_0(g)\subset\supp\ul{R}\subset X$ i.e.,
$g\in\tilde\cG_j(X)(n)=\tilde\cG_j(X)(n')$. 
Consider, now, $R=\{(G_1,s_1),\dots,(G_h,s_h)\}\in\ul{R}$,
from definition (\ref{cjp->Ys}) we have that 
$G_l\cap\tilde\cG_j(X)(n)=G_l\cap\tilde\cG_j(X)(n')\neq\emptyset$ and
$\Es_{s_l}(G_l)\subset\supp\ul{R}\subset\Es_s(G)\subset X$ for all 
$l=1,\dots,h$.
Moreover, recalling that for all $l=1,\dots,h$ 
each $g\in G_l$ is either an element of $\ul{R}\rest j$ or an 
element of $G$, we have that 
$\Es_0(g)\subset X$ and, hence, 
$G_l\subset\widetilde{\cG}_{\ge j}(X)(n)=\widetilde{\cG}_{\ge j}(X)(n')$.
Then by using (\ref{loc2}),
(\ref{loc3}), (\ref{accrj}), and (\ref{acRj}) we have that 
\begin{equation}
\label{loc4}
\zeta_{\ul{R},\Lambda}(\cdot,n)=\zeta_{\ul{R},\Lambda}(\cdot,n')
\end{equation}

The inductive proof is completed easily by using (\ref{selfintij}),
(\ref{cjp->Ys}), item~\ref{piv} above and (\ref{loc4}).
\qed 

\smallskip 
\noi{\it Proof of Theorem \ref{t:mep.1}.}\ 
We focus on the 
first of the two identities (\ref{epa.12}), the proof of the 
second can be achieved analogously.
We recall (\ref{slrj.1}) and notice that $n_X=n'_X$ implies 
$$
\id_{\{\diam_\wp(X)\le\varrho,X\cap\Lambda\neq\emptyset,\xi(X)(n)=\emptyset\}}=
\id_{\{\diam_\wp(X)\le\varrho,X\cap\Lambda\neq\emptyset,\xi(X)(n')=\emptyset\}}
$$
Then from item~\ref{i:ipo33} in Theorem~\ref{t:ip3} we get  
\begin{equation}
\label{loc6}
\begin{array}{l}
\id_{\{\diam_\wp(X)\le\varrho,X\cap\Lambda\neq\emptyset,\xi(X)(n)=\emptyset\}}
V^{(\ell),\wp}_{X,\Lambda}(\cdot,n) \\
\;\;\;\;\;\;\;\;\;\;\;\;\;\;\;\;\;\;\;\;\;\;\;\;\;\;\;\;\;\;\;\;\;\;\;
=\id_{\{\diam_\wp(X)\le\varrho,X\cap\Lambda\neq\emptyset,\xi(X)(n')=\emptyset\}}
V^{(\ell),\wp}_{X,\Lambda}(\cdot,n')
\end{array}
\end{equation}
The first of the identities (\ref{epa.12}) finally follows from 
definition (\ref{slrj.1}), the equality (\ref{loc6}), 
and items~\ref{pii} and \ref{pviii} of Lemma~\ref{t:mGXRX.1}.
\qed

\subsec{Proof of Gibbsianity and convergence}{s:gib}
\par\noindent
We notice that for $X\subset\subset\cL^{(\wp)}$
and $n\in\bar\cM^{(\ell),\wp}$ 
item~\ref{p:tm1.5} of Theorem~\ref{t:yteo} implies 
that $\Psi_{X,X}^{(\ell),\wp}(\cdot,n)$ and $\Phi_{X,X}^{(\ell),\wp}(\cdot,n)$ 
are constant functions namely, they do not depend on the first argument. 
In the sequel we shall write 
$\Psi_{X,X}^{(\ell),\wp}(n)$ and $\Phi_{X,X}^{(\ell),\wp}(n)$ 
respectively for 
$\Psi_{X,X}^{(\ell),\wp}(\cdot,n)$ and $\Phi_{X,X}^{(\ell),\wp}(\cdot,n)$. 

We suppose, now, that the hypotheses of Theorem~\ref{t:yteo} are satisfied, 
we pick $\tilde n\in\pi^{-1}(0)$ once for all, recall the map $\pi$ 
has been defined in (\ref{pizza}), and for each 
$X\subset\subset\cL^{(\wp)}$ we define the functions
$\psi_X^{(\ell),\wp}:\cM^{(\ell),\wp}\to\bR$ and 
$\phi_X^{(\ell),\wp}:\cM^{(\ell),\wp}\to\bR$ as follows 
\begin{equation}  
\label{ep.1}  
\psi_X^{(\ell),\wp}(n):=
  \Psi_{X,X}^{(\ell),\wp}(n_X\tilde n_{X^\complement}) 
\;\;\;\textrm{ and }\;\;\; 
\phi_X^{(\ell),\wp}(n):=
  \Phi_{X,X}^{(\ell),\wp}(n_X\tilde n_{X^\complement}) 
\end{equation}
We note that, by definition, 
the functions $\psi_X^{(\ell),\wp}$ and $\phi_X^{(\ell),\wp}$ are local that
is
$\psi_X^{(\ell),\wp},\phi_X^{(\ell),\wp}\in\cB^{(\ell),\wp}_X$; where we recall 
the $\sigma$--algebra $\cB^{(\ell),\wp}$ has been introduced at the 
beginning of Subsection~\ref{s:fs}. 

Let $X,\Lambda\ssu\cL^{(\wp)}$ such that $\Lambda\supset X$ and 
$n\in\bar\cM^{(\ell),\wp}$. The functions 
$\Psi_{X,\Lambda}^{(\ell),\wp}(\cdot,n)$ and 
$\Phi_{X,\Lambda}^{(\ell),\wp}(\cdot,n)$ are constant namely,
$\Psi_{X,\Lambda}^{(\ell),\wp}(\cdot,n),
 \Phi_{X,\Lambda}^{(\ell),\wp}(\cdot,n)
 \in\cF^{\ell,\wp}_{\emptyset}$,
and moreover from Theorem~\ref{t:mep.1} and item~\ref{p:tm2} in 
Theorem~\ref{t:yteo} we get 
\begin{equation}  
\label{epa.1}  
\Psi_{X,\Lambda}^{(\ell),\wp}(\cdot,n)=\psi_X^{(\ell),\wp}(n)
\;\;\;\textrm{ and }\;\;\; 
\Phi_{X,\Lambda}^{(\ell),\wp}(\cdot,n)=\phi_X^{(\ell),\wp}(n)
\end{equation}

\smallskip
\noi{\it Proof of Theorem \ref{t:wgib}.}\ 
To get the renormalized potentials of Theorem~\ref{t:wgib} we next pull 
the $\Psi^{(\ell),\wp}$ and $\Phi^{(\ell),\wp}$ back to the scale $\ell$. 
We define the family
$\{\psi_I^{(\ell)},\phi_I^{(\ell)}:
   \cM^{(\ell)}\to\bR,\,I\subset\subset\cL^{(\ell)}\}$ as follows:
for each $m\in\cM^{(\ell)}$ we set 
\begin{equation}
\label{potf0}
\psi^{(\ell)}_I(m):=\Bigg\{
\begin{array}{ll}
{\displaystyle
 -m_i^2/2
}
&
 \textrm{ if } I=\{i\}\textrm{ with } i\in\cL^{(\ell)}
\\
{\displaystyle
 \psi^{(\ell),\wp}_X(\cO_\ell^\wp m)
}
&
 \textrm{ if } |I|\ge2\textrm{ and } \exists X\subset\cL^{(\wp)}: 
                                   \cO_\wp^\ell X=I
\\
0&\textrm{ otherwise }
\end{array}
\end{equation} 
note that by construction, see (\ref{slrj.1}) and (\ref{ep.1}),
if $|X|\le1$ then $\psi^{(\ell),\wp}_X=0$,
and
\begin{equation}
\label{potf1}
\phi^{(\ell)}_I(m):=\bigg\{
\begin{array}{ll}
{\displaystyle
 \phi^{(\ell),\wp}_X(\cO_\ell^\wp m)
}
&
 \textrm{ if } \exists X\subset\cL^{(\wp)}: \cO_\wp^\ell X=I
\\
0&\textrm{ otherwise }
\end{array}
\end{equation}
Equivalently, for all $I\subset\subset\cL^{(\ell)}$ such that $|I|>2$, 
we can write 
\begin{equation}
\label{potf}
\psi^{(\ell)}_I=\sum_{\newatop{X\subset\cL^{(\wp)}:}
                                  {\cO_\wp^\ell X=I}}
                    \psi^{(\ell),\wp}_X\circ\cO_\ell^\wp
\;\;\;\;\textrm{ and } \;\;\;\;
\phi^{(\ell)}_I=\sum_{\newatop{X\subset\cL^{(\wp)}:}
                                  {\cO_\wp^\ell X=I}}
                    \phi^{(\ell),\wp}_X\circ\cO_\ell^\wp
\end{equation}
we note, indeed, that for each 
$I\subset\subset\cL^{(\ell)}$ there exists at most one $X\subset\cL^{(\wp)}$
such that $\cO_\wp^\ell X=I$.

{\it Item~\ref{i:mtb0}.}\ 
Since for each $X\subset\subset\cL^{(\wp)}$ we have 
$\psi_X^{(\ell),\wp},\phi_X^{(\ell),\wp}\in\cB^{(\ell),\wp}_X$, the 
thesis follows from definition (\ref{potf}) and (\ref{spisigf}). 

{\it Item~\ref{i:mtb1}.}\ 
We note that if we let $X\subset\cL^{(\wp)}$ 
and $I:=\cO_\wp^\ell X\subset\cL^{(\ell)}$, we have that $I$ is 
$\ell$--connected if and only if $X$ is $\wp$--connected.
Then the thesis follows immediately from definitions
(\ref{potf}), (\ref{slrj.1}), (\ref{GXRX.1}), and item~\ref{i:ipo13.5} 
in Theorem~\ref{t:ip3}.

{\it Item~\ref{i:mtb2}.}
Since the original lattice gas potential 
and the algorithmic construction of the gentle atoms in Section~\ref{s:cattivi}
are translationally invariant, 
the statement follows. 

{\it Item~\ref{i:mtb5}.}
Let $\bar\cM^{(\ell),\wp}\subset\cM^{(\ell),\wp}$ as in 
Theorem~\ref{t:partD}. 
We set $\bar\cM^{(\ell)}:=\cO_\wp^\ell\bar\cM^{(\ell),\wp}$, with 
$\cO_\wp^\ell$ the unpacking operator.
Recalling the definition of $\nu^{(\ell),\wp}$ given at the 
end of Subsection~\ref{s:spm}, we have 
\begin{equation}
\label{piur-2}
1=\nu^{(\ell),\wp}(\bar\cM^{(\ell),\wp}) 
 =\nu^{(\ell)}(\cO_\wp^\ell\bar\cM^{(\ell),\wp}) 
 =\nu^{(\ell)}(\bar\cM^{(\ell)}) 
\end{equation}

We recall (\ref{potf0}), (\ref{ep.1}), and that
$\tilde n$ has been picked up above;
for $m\in\bar\cM^{(\ell)}$ and 
$I\subset\subset\cL^{(\ell)}$ such that $|I|\ge2$, we have that if 
there exists $X\subset\subset\cL^{(\ell),\wp}$ such that 
$\cO_\wp^\ell X=I$ we have
\begin{equation}
\label{piur-1}
\psi_I^{(\ell)}(m)=\psi_X^{(\ell),\wp}(\cO_\ell^\wp m)=
\Psi_{X,X}^{(\ell),\wp}((\cO_\ell^\wp m)_X\tilde n_{X^\complement})=
\Psi_{X,X}^{(\ell),\wp}(\cO_\ell^\wp m)
\end{equation}
where the last equality follows from Theorem~\ref{t:mep.1}. 

Recall (\ref{erre}), 
pick $m\in\bar\cM^{(\ell)}$ and $i\in\cL^{(\ell)}$, set 
$r^{(\ell)}_i(m):=2 d r^{(\wp)}_{x(i)}(\cO_\ell^\wp m)$, 
where $x(i)\in\cL^{(\wp)}$ is such that $\{x(i)\}=\cO_\ell^\wp\{i\}$. 
Consider $I\subset\subset\cL^{(\ell)}$ such that 
$I\ni i$ and $\diam_\ell(I)>r_i^{(\ell)}(m)$; from definition (\ref{erre}) 
and $\diam_\ell(I)>r_i^{(\ell)}(m)\ge\cionc d$ we have that $|I|\ge2$.
If it does not exists any $X\subset\subset\cL^{(\ell),\wp}$ such that 
$\cO_\wp^\ell X=I$ we have $\psi_I^{(\ell)}(m)=0$.
On the other hand if 
there exists $X\subset\subset\cL^{(\ell),\wp}$ such that 
$\cO_\wp^\ell X=I$, from (\ref{piur-1}) we have that
$\psi_I^{(\ell)}(m)=0$. Indeed item~\ref{p:tm3}
of Theorem~\ref{t:yteo} implies that 
$\Psi_{X,X}^{(\ell),\wp}(\cO_\ell^\wp m)=0$ once we note that 
$\cO_\ell^\wp m\in\bar\cM^{(\ell),\wp}$, $X\ni x(i)$, and 
$$
\diam_\wp(X)\ge\frac{1}{2d}\diam_\ell(I)>
\frac{1}{2d}r_i^{(\ell)}(m)=
r^{(\wp)}_{x(i)}(\cO_\ell^\wp m)
$$ 

Consider, now, $m\in\bar\cM^{(\ell)}$, $i\in\cL^{(\ell)}$, and 
$x(i)$ as above.
By using (\ref{potf0}), (\ref{piur-1}), and item~\ref{p:tm2} 
in Theorem~\ref{t:yteo}, we then have that
$$
\sum_{I\ni i}\big|\psi^{(\ell)}_I(m)\big|
=
\frac{1}{2}m_i^2+
\sum_{\newatop{I\subset\subset\cL^{(\ell)}:}
              {|I|\ge2,\,I\ni i}}
   \big|\psi^{(\ell)}_I(m)\big|= 
\frac{1}{2}m_i^2+
\sum_{\newatop{X\subset\subset\cL^{(\wp)}:}
              {X\ni x(i)}}
   \big|\Psi_{X,X}^{(\ell),\wp}(\cO_\ell^\wp m)\big|
$$
The statement (\ref{mtb5}) follows from item~\ref{p:tm3} 
in Theorem~\ref{t:yteo} by setting
$$
c^{(\ell)}_i(m):=\frac{1}{2}m_i^2+
c^{(\wp)}_{x(i)}(\cO_\ell^\wp m)
$$
for all $m\in\bar\cM^{(\ell)}$.

{\it Item~\ref{i:mtb4}.}\
By recalling definition (\ref{potf0}) and by 
using the Minkowski inequality we have that 
\begin{equation}
\label{piur0}
\begin{array}{rcl}
{\displaystyle
 \sup_{i\in\cL^{(\ell)}}
  \Big[\nu^{(\ell)}\Big(\Big|
              \sum_{I\ni i}\psi_I^{(\ell)}
              \Big|^q\Big)\Big]^{1/q}
}
&\!=\!&
{\displaystyle
\sup_{i\in\cL^{(\ell)}}
  \Big\{
  \Big[\nu^{(\ell)}\Big(\Big|
              -\frac{1}{2}m_i^2
              \Big|^q\Big)\Big]^{1/q}
  +
  \Big[\nu^{(\ell)}\Big(\Big|
              \sum_{\newatop{I\ni i:}
                            {|I|\ge2}}\psi_I^{(\ell)}
              \Big|^q\Big)\Big]^{1/q}
  \Big\}
}\\
&\!\le\!&
{\displaystyle
 \frac{1}{2\chi}\ell^d
 +
 \sup_{i\in\cL^{(\ell)}}
  \Big[\nu^{(\ell)}\Big(\Big|
              \sum_{\newatop{I\ni i:}
                            {|I|\ge2}}\psi_I^{(\ell)}
              \Big|^q\Big)\Big]^{1/q}
}
\end{array}
\end{equation}
with $\chi$ the infinite volume compressibility defined in (\ref{chi}).

To bound the second term of the right--hand side of the above inequality 
we use (\ref{potf0}), (\ref{piur-2}), (\ref{piur-1}), (\ref{slrj.1}), and 
the Minkowski inequality. We have
\begin{equation}
\label{piur1}
\begin{array}{rcl}
{\displaystyle
 \sup_{i\in\cL^{(\ell)}}
  \Big[\nu^{(\ell)}\Big(\Big|
              \sum_{\newatop{I\ni i:}
                            {|I|\ge2}}\psi_I^{(\ell)}
              \Big|^q\Big)\Big]^{1/q}
 \le
 \sup_{x\in\cL^{(\wp)}}
  \Big[\nu^{(\ell)}\Big(\Big|
              \sum_{X\ni x}[
                \Psi_{X,X}^{(\ell),\wp}\circ\cO_\ell^\wp]
              \Big|^q\Big)\Big]^{1/q}
}
&&\\
&&
\!\!\!\!\!\!\!\!\!\!\!\!\!\!\!\!\!\!\!\!\!\!\!\!\!\!\!\!\!\!\!\!\!\!\!\!\!\!
\!\!\!\!\!\!\!\!\!\!\!\!\!\!\!\!\!\!\!\!\!\!\!\!\!\!\!\!\!\!\!\!\!\!\!\!\!\!
\!\!\!\!\!\!\!\!\!\!\!\!\!\!\!\!\!\!\!\!\!\!\!\!\!\!\!\!\!\!\!\!\!\!\!\!\!\!
\!\!\!\!\!\!\!\!\!\!\!\!\!\!\!\!\!\!\!\!\!\!\!\!\!\!\!\!\!\!\!\!\!\!\!\!\!\!
\!\!\!
{\displaystyle
 \le
 \!\sup_{x\in\cL^{(\wp)}}\Big\{
  \Big[\int_{\bar\cM^{(\ell)}}\!\nu^{(\ell)}(dm)\Big(
              \sum_{X\ni x}
 \id_{\{\diam_\wp(X)\le\varrho,
        \xi(X)(\cO_\ell^\wp m)=\emptyset\}}
 \|V^{(\ell),\wp}_{X,X}(\cdot,\cO_\ell^\wp m)\|_\infty
              \Big)^q\Big]^{1/q}
}\\
&&
\!\!\!\!\!\!\!\!\!\!\!\!\!\!\!\!\!\!\!\!\!\!\!\!\!\!\!\!\!\!\!\!\!\!\!\!\!\!
\!\!\!\!\!\!\!\!\!\!\!\!\!\!\!\!\!\!\!\!\!\!\!\!\!\!\!\!\!\!\!\!\!\!\!\!\!\!
\!\!\!\!\!\!\!\!\!\!\!\!\!\!\!\!\!\!\!\!\!\!\!\!\!\!\!\!\!\!\!\!\!\!\!\!\!\!
\!\!\!\!\!\!\!\!\!\!\!\!\!\!\!\!\!\!\!\!\!\!\!\!\!\!\!\!\!\!\!\!\!\!\!\!\!\!
{\displaystyle
 \phantom{\le
          \!\sup_{x\in\cL^{(\wp)}}\big\{
         }
 \!\!+ 
  \Big[\int_{\bar\cM^{(\ell)}}\!\nu^{(\ell)}(dm)
  \Big(
       \sum_{X\ni x}
       \sum_{j\ge1}
       \sum_{g\in\cG_j(X)(\cO_\ell^\wp m)}
           \|\log Z_{g,X}^{(j)}(\cdot,\cO_\ell^\wp m)\|_\infty
  \Big)^q\Big]^{1/q}\Big\}
}\\
\end{array}
\end{equation}
By using (\ref{ip233}) and the Minkowski inequality we have 
\begin{equation}
\label{piur2}
\begin{array}{rcl}
{\displaystyle
 \sup_{i\in\cL^{(\ell)}}
  \nu^{(\ell)}\Big(\Big|
              \sum_{\newatop{I\ni i:}
                            {|I|\ge2}}\psi_I^{(\ell)}
              \Big|^q\Big)^{1/q}
 \le
 A_\ell
}
&&\\
&&
\!\!\!\!\!\!\!\!\!\!\!\!\!\!\!\!\!\!\!\!\!\!\!\!\!\!\!\!\!\!\!\!\!\!\!\!\!\!
\!\!\!\!\!\!\!\!\!\!\!\!\!\!\!\!\!\!\!\!\!\!\!\!\!\!\!\!\!\!
{\displaystyle
 + 
  \sup_{x\in\cL^{(\wp)}}
  \sum_{j\ge1}
  \Big[\int_{\bar\cM^{(\ell)}}\!\nu^{(\ell)}(dm)
  \Big(
       \sum_{X\ni x}
       \sum_{g\in\cG_j(X)(\cO_\ell^\wp m)}
           \|\log Z_{g,X}^{(j)}(\cdot,\cO_\ell^\wp m)\|_\infty
  \Big)^q\Big]^{1/q}
}
\end{array}
\end{equation}

To bound the second term on the right--hand side of (\ref{piur2}) we recall
(\ref{GXRX.1}) and note that the sum over $g\in\cG_j(X)(\cO_\ell^\wp m)$ 
is zero if there exists no $g\in\cG_j(\cO_\ell^\wp m)$ such that 
$\Es_0(g)\ni x$. Hence this term is estimated from above by 
\begin{equation}
\label{piur3}
  \sup_{x\in\cL^{(\wp)}}
  \sum_{j\ge1}
  \Big[\int_{\bar\cM^{(\ell)}}\!\nu^{(\ell)}(dm)
  \id_{\{\exists g\in\cG_j(\cO_\ell^\wp m):\,\Es_0(g)\ni x\}}
  \Big(
       \sum_{X\ni x}
       \sum_{g\in\cG_j(X)(\cO_\ell^\wp m)}
           \|\log Z_{g,X}^{(j)}(\cdot,\cO_\ell^\wp m)\|_\infty
  \Big)^q\Big]^{1/q}
\end{equation}

Let $j\ge1$,
definition (\ref{supol}) and the bound on the diameter of a 
$j$--gentle atom, see item~\ref{gent:diam} in Definition~\ref{t:gentle},
imply that for all $m\in\bar\cM^{(\ell)}$ and 
$g\in\cG_j(\cO_\ell^\wp m)$ we have that 
$\diam_\wp\Es_0(g)\le\Gamma_j+2\vartheta_j$.
The sum over $X$ in (\ref{piur3}) is then extended only to 
parallelepipedal subsets of the lattice $\cL^{(\wp)}$ whose 
diameter is smaller than $\Gamma_j+2\vartheta_j$; this implies that 
this sum has at most $(\Gamma_j+2\vartheta_j)^{2d}$ terms. 
Moreover given $X$, there exists at most one $g\in\cG_j(\cO_\ell^\wp m)$ 
such that $\Es_0(g)=X$. 
These remarks and the inequality in item~3 of Theorem~3.2 of \cite{[BCOabs]}
imply the expression in (\ref{piur3}) is bounded from above by 
\begin{equation}
\label{piur4}
  c_1\ell^d
  \sup_{x\in\cL^{(\wp)}}
  \sum_{j\ge1}
  \vartheta_j^{4d} 
  \big[\nu^{(\ell)}(
  \{\exists g\in\cG_j(\cO_\ell^\wp m):\,\Es_0(g)\ni x\})
  \big]^{1/q}
\end{equation}
with $c_1$ a suitable real depending on the norm $\|U\|_0$ of the interaction.
By the same estimate as in (\ref{ppppp}), we have 
\begin{equation}
\label{piur5}
  c_1\ell^d
  \sup_{x\in\cL^{(\wp)}}
  \sum_{j\ge1}
  \vartheta_j^{4d} 
  \big[\nu^{(\ell)}(
  \{\exists g\in\cG_j(\cO_\ell^\wp m):\,\Es_0(g)\ni x\})
  \big]^{1/q}
\le
 c_2\ell^d
  \sum_{j\ge1}
  \vartheta_j^{4d+d/q} 
  e^{-a_\delta(\ell)(1+\varepsilon)^j/q}
\end{equation}
where $\varepsilon$ and $a_\delta(\ell)$ are as in (\ref{okscale1}) 
and (\ref{okscale2}), and $c_2$ is a positive real depending on $\|U\|_0$. 
The thesis now follows from (\ref{piur0})--(\ref{piur5}) and 
item~\ref{seq:con-icf} in Definition~\ref{t:seq}. 

{\it Item~\ref{i:mtb3}.}\
We recall that $\wp=\cionc d\ell$, 
$\alpha_1>0$ has been chosen below (\ref{morlp}),
$\alpha_\ell$ have been introduced in 
item~\ref{i:ipo43} of Theorem~\ref{t:ip3}, and 
$q$ has been defined below (\ref{erre}). 
We set $\alpha':=q\alpha_1/(2d^2)$ and 
recall definitions (\ref{potf}) and (\ref{ep.1}). We have
\begin{equation}
\label{stst0}
\begin{array}{l}
{\displaystyle 
 \sup_{m\in\cM^{(\ell)}}
 \sup_{i\in\cL^{(\ell)}}
 \sum_{I\ni i} e^{\alpha'\diam_\ell(I)}
               |\phi^{(\ell)}_I(m)|
}\\
\vphantom{\Big[}
\phantom{merdonemer}
 \le
{\displaystyle 
 \sup_{m\in\cM^{(\ell)}}
 \sup_{i\in\cL^{(\ell)}}
 \sum_{I\ni i} e^{q\alpha_\ell\diam_\ell(I)/(2d^2)}
               |\phi^{(\ell)}_I(m)|
} \\
\vphantom{\Big[}
\phantom{merdonemer}
 \le
{\displaystyle
 \sup_{m\in\cM^{(\ell)}}
 \sup_{i\in\cL^{(\ell)}}
 \sum_{I\ni i} e^{q\alpha_\ell\diam_\ell(I)/(2d^2)}
                \sum_{\newatop{X\subset\cL^{(\wp)}:}
                              {\cO_\wp^\ell X=I}}
      |\Phi^{(\ell),\wp}_{X,X}((\cO_\ell^\wp m)_X\tilde n_{X^\complement})|
}\\
\vphantom{\Big[}
\phantom{merdonemer}
 \le 
{\displaystyle
 \sup_{m\in\cM^{(\ell)}}
 \sup_{i\in\cL^{(\ell)}}
 \sum_{\newatop{X\subset\subset\cL^{(\wp)}:}
               {\cO_\wp^\ell X\ni i}}
     e^{q\alpha_\ell\diam_\wp(X)/d}\,
      |\Phi^{(\ell),\wp}_{X,X}((\cO_\ell^\wp m)_X\tilde n_{X^\complement})|
}\\
\vphantom{\Big[}
\phantom{merdonemer}
\le
{\displaystyle
e^{-\alpha_\ell/d}+e^{-q\alpha_\ell\gamma_1/d}
             \Big( 
	          \frac{1+e^{-q\alpha_\ell/(2d^2)}}
		       {1-e^{-q\alpha_\ell/(2d^2)}}
             \Big)^d
}
\end{array}
\end{equation}
where we have used (\ref{tm2}) and 
$\diam_\ell(\cO_\wp^\ell X)/d\le2\diam_\wp(X)$ for any $X\subset\cL^{(\wp)}$. 

{\it Item~\ref{i:mtb6}.}\
We follow an argument analogous to that in \cite[Section~5.3]{[BCO]}. 
Let $m'\in\cM^{(\ell)}$ and $J\subset\subset\cL^{(\ell)}$, we define 
the following probability kernel on $\cL^{(\ell)}_J$
\begin{equation}
\label{kernq}
q_J(m',m):= 
 \frac{\displaystyle
     \exp\Big\{\sum_{I\cap J\neq\emptyset}
        \big[\psi_I^{(\ell)}(mm'_{J^\complement})
            +\phi_I^{(\ell)}(mm'_{J^\complement})\big]
       \Big\}
      }
 {\displaystyle
  \sum_{m\in\cM^{(\ell)}_J} 
     \exp\Big\{\sum_{I\cap J\neq\emptyset}
        \big[\psi_I^{(\ell)}(mm'_{J^\complement})
            +\phi_I^{(\ell)}(mm'_{J^\complement})\big]
       \Big\}
      }
\end{equation}
where the functions $\psi^{(\ell)}_I$ and $\phi^{(\ell)}_I$ have been defined
in (\ref{potf0}) and (\ref{potf1}). 
Note that, given $m\in\cM^{(\ell)}_J$, we have 
$q_J(\cdot,m)\in\cB^{(\ell)}_{J^\complement}$. 

Pick $J\subset\subset\cL^{(\ell)}$, $f\in\cB^{(\ell)}_J$, recall 
$M^{(\ell)}$ has been defined in (\ref{Centr}), by 
definition of the renormalized measure $\nu^{(\ell)}$ and by 
standard measure theory we have 
$\mu(f(M^{(\ell)}))=\nu^{(\ell)}(f)$ and 
\begin{equation}
\label{poirti}
\int_{\cM^{(\ell)}}
  \nu^{(\ell)}(d m') \,
  \sum_{m\in\cM^{(\ell)}_J}\,
   q_J(m',m)\, f(m)
=
\int_{\cX}
  \mu(d\eta) \,
  \sum_{m\in\cM^{(\ell)}_J}\,
   q_J(M^{(\ell)}(\eta),m)\, f(m)
\end{equation}
The equations (\ref{dlr}) will thus follow from
\begin{equation}
\label{stellina}
\mu(f(M^{(\ell)}))=
\int_{\cX}
  \mu(d\eta) \,
  \sum_{m\in\cM^{(\ell)}_J}\,
   q_J(M^{(\ell)}(\eta),m)\, f(m)
\end{equation}

For $X\su\cL^{(\ell)}$, let us introduce
the family of $\s$--algebras 
$\cE^{(\ell)}_{X}
  :=\s\{M^{(\ell)}_i(\cdot)\,,\: i\in X\}\subset\cF_{\cO_\ell X}$ 
on the configuration space $\cX$.
Now, pick $\Lambda\su\su\cL^{(\wp)}$ so 
that $V:=\cO^\ell_\wp\Lambda\supset J$. For $\eta\in\cX$, we set 
\begin{equation}
\label{sfh}
 G_V(M^{(\ell)}_{V\setminus J}(\eta),\eta_{\cO_\ell V^\complement})
:=
 \mu\big(f(M^{(\ell)})\big|
   \cE^{(\ell)}_{V\setminus J}\otimes\cF_{\cO_\ell V^\complement}\big)(\eta)
\end{equation}
We shall prove that, $\mu$--a.s., 
\begin{equation}
\label{ans5}
\lim_{V\uparrow\cL^{(\ell)}}
 G_V(M^{(\ell)}_{V\setminus J}(\eta),\eta_{\cO_\ell V^\complement})
=
  \sum_{m\in\cM^{(\ell)}_J} 
     q_J(M^{(\ell)}(\eta),m)\, f(m) 
\end{equation}
Therefore, by dominated convergence, we have 
\begin{equation}
\label{plplpl}
\mu(f(M^{(\ell)}))=
\mu(\mu(f(M^{(\ell)}|\cE^{(\ell)}_{V\setminus J}\otimes
             \cF_{\cO_\ell V^\complement}))
\;\stackrel{V\uparrow\cL^{(\ell)}}{\longrightarrow}\;
\int_{\cX}
  \mu(d\eta) \,
  \sum_{m\in\cM^{(\ell)}_J}\,
   q_J(M^{(\ell)}(\eta),m)\, f(m)
\end{equation}
so that (\ref{stellina}) holds. 

We finally prove (\ref{ans5}) in the set of full measure 
$(M^{(\ell)})^{-1}(\bar\cM^{(\ell)})$. 
By the Gibbs property of the original measure $\mu$, for  
$\eta\in\cX$, such that $M^{(\ell)}(\eta)\in\bar\cM^{(\ell)}$, and 
$m'\in\cM_{J^\complement}$, such that 
$m'_{J^\complement}=\big(M^{(\ell)}(\eta)\big)_{J^\complement}$,
we have that
\begin{equation}
\label{ans2}
\begin{array}{l}
{\displaystyle
 G_V(m'_{V\setminus J},\eta_{\cO_\ell V^\complement})
 =
 \mu_{\cO_\ell V}^{\eta} 
   \big(f(M^{(\ell)})\big|M^{(\ell)}_{V\sm J}=m'_{V\setminus J}\big)  
\vphantom{\bigg\}}
}  \\
\phantom{mer}
{\displaystyle 
 =
 \frac{\displaystyle
       \sum_{\sigma\in\cX_{\cO_\ell V}} 
        f(M^{(\ell)}(\sigma))\,
        e^{H_{\cO_\ell V}(\sigma\eta_{\cO_\ell V^\complement})}\,
        \id_{\{M^{(\ell)}_{V\setminus J}(\sigma)=m'_{V\setminus J}\}}}
      {\displaystyle
       \sum_{\sigma\in\cX_{\cO_\ell V}} 
        e^{H_{\cO_\ell V}(\sigma\eta_{\cO_\ell V^\complement})}\,
        \id_{\{M^{(\ell)}_{V\setminus J}(\sigma)=m'_{V\setminus J}\}}}
}\\
\phantom{mer}
{\displaystyle 
 =
 \frac{\displaystyle
       \sum_{m\in\cM^{(\ell)}_J} 
        f(m)\,
       \sum_{\sigma\in\cX_{\cO_\ell V}} 
        e^{H_{\cO_\ell V}(\sigma\eta_{\cO_\ell V^\complement})}\,
        \id_{\{M^{(\ell)}_V(\sigma)=mm'_{V\setminus J}\}}}
      {\displaystyle
       \sum_{m\in\cM^{(\ell)}_J} 
       \sum_{\sigma\in\cX_{\cO_\ell V}} 
        e^{H_{\cO_\ell V}(\sigma\eta_{\cO_\ell V^\complement})}\,
        \id_{\{M^{(\ell)}_V(\sigma)=mm'_{V\setminus J}\}}}
=
 \frac{\displaystyle
       \sum_{m\in\cM^{(\ell)}_J} f(m)\,
                Z^{(\ell)}_{mm',V}(\cO^\ell\eta)}
      {\displaystyle
       \sum_{m\in\cM^{(\ell)}_J} 
               Z^{(\ell)}_{mm',V}(\cO^\ell\eta)}
}\\
\end{array}
\end{equation}
see (\ref{epartf}). 
Recall $V=\cO^\ell_\wp\Lambda$ and set $\xi=\cO^\wp_\ell\cO^\ell\eta$, we have 
$Z^{(\ell)}_{mm',V}(\cO^\ell\eta)=
 Z^{(\ell),\wp}_{mm',\Lambda}(\xi)$ for all $m\in\cM^{(\ell)}_J$.
By using the expansion (\ref{tm1}), this is allowed because 
$mm'\in\bar\cM^{(\ell)}$, we thus get 
\begin{equation}
\label{ans3}
\begin{array}{l}
G_V(m'_{V\setminus J},\eta_{\cO_\ell V^\complement})
\\
\phantom{mer}
= 
 \frac{\displaystyle
  \sum_{m\in\cM^{(\ell)}_J} f(m) \, 
     e^{K^{(\wp)}_\Lambda-\frac{1}{2}\sum_{i\in V}(mm')_i^2  
        + 
        \sum_{X\cap\Lambda\neq\emptyset}
        \big[\Psi_{X,\Lambda}^{(\ell),\wp}(\xi,\cO^\wp_\ell(mm'))
            +\Phi_{X,\Lambda}^{(\ell),\wp}(\xi,\cO^\wp_\ell(mm'))\big]
       }
      }
 {\displaystyle
  \sum_{m\in\cM^{(\ell)}_J} 
     e^{K^{(\wp)}_\Lambda-\frac{1}{2}\sum_{i\in V}(mm')_i^2  
        + 
        \sum_{X\cap\Lambda\neq\emptyset}
        \big[\Psi_{X,\Lambda}^{(\ell),\wp}(\xi,\cO^\wp_\ell(mm'))
            +\Phi_{X,\Lambda}^{(\ell),\wp}(\xi,\cO^\wp_\ell(mm'))\big]
       }
      }
\\
\phantom{mer}
=
 \frac{\displaystyle
  \sum_{m\in\cM^{(\ell)}_J} f(m) \, 
     e^{-\frac{1}{2}\sum_{i\in J} {m_i}^2  
        + 
        \sum_{X\cap\cO^\wp_\ell J\neq\emptyset}
        \big[\Psi_{X,\Lambda}^{(\ell),\wp}(\xi,\cO^\wp_\ell(mm'))
            +\Phi_{X,\Lambda}^{(\ell),\wp}(\xi,\cO^\wp_\ell(mm'))\big]
       }
      }
 {\displaystyle
  \sum_{m\in\cM^{(\ell)}_J} 
     e^{-\frac{1}{2}\sum_{i\in J} {m_i}^2  
        + 
        \sum_{X\cap\cO^\wp_\ell J\neq\emptyset}
        \big[\Psi_{X,\Lambda}^{(\ell),\wp}(\xi,\cO^\wp_\ell(mm'))
            +\Phi_{X,\Lambda}^{(\ell),\wp}(\xi,\cO^\wp_\ell(mm'))\big]
       }
      }
\\
\end{array}
\end{equation}
where in the second step we have used Theorem~\ref{t:mep.1} to simplify
the terms of the potential not intersecting $\cO^\wp_\ell J$. 
By items~\ref{p:tm3} and \ref{p:tm4} in Theorem~\ref{t:yteo} and 
by (\ref{epa.1}) we get 
\begin{equation}
\label{ans4}
\lim_{V\to\cL^{(\ell)}}
 G_V(m'_{V\setminus J},\eta_{\cO_\ell V^\complement})
=
 \frac{\displaystyle
  \sum_{m\in\cM^{(\ell)}_J} 
     \!\!\!
     f(m) \, 
     e^{-\frac{1}{2}\sum_{i\in J} {m_i}^2  
        + 
        \sum_{X\cap\cO^\wp_\ell J\neq\emptyset}
        \big[\psi_X^{(\ell),\wp}(\cO^\wp_\ell(mm'))
            +\phi_X^{(\ell),\wp}(\cO^\wp_\ell(mm'))\big]
       }
      }
 {\displaystyle
  \sum_{m\in\cM^{(\ell)}_J} 
     \!\!\!
     e^{-\frac{1}{2}\sum_{i\in J} {m_i}^2  
        + 
        \sum_{X\cap\cO^\wp_\ell J\neq\emptyset}
        \big[\psi_X^{(\ell),\wp}(\cO^\wp_\ell(mm'))
            +\phi_X^{(\ell),\wp}(\cO^\wp_\ell(mm'))\big]
       }
      }
\end{equation}
By using definitions (\ref{potf0}) and (\ref{potf1}) the above 
expansion reduces to the renormalization scale $\ell$. We thus 
get (\ref{ans5}).
\qed

\medskip
\noi{\it Proof of Theorem \ref{t:conv}.}\ 
{\it Item~\ref{imtc2}.}
Consider (\ref{piur1}) and recall 
(\ref{xixi}); the first 
term on the right--hand side of (\ref{piur1}) 
tends to zero in the limit $\ell\to\infty$ 
by virtue of item~\ref{i:ipo43.5} of Theorem~\ref{t:ip3}. 
The second term is estimated from above by the convergent series in
(\ref{piur5}); it is not difficult to prove that its sum tends to zero 
in the limit $\ell\to\infty$ as a consequence of 
(\ref{okscale})--(\ref{okscale2}).

{\it Item~\ref{imtc1}.}\ The statement is a straightforward consequence 
of (\ref{stst0}) and of the expression of $\alpha_\ell$. 
\qed


 
 
 
\end{document}